\documentclass[12pt]{JHEP3}
\usepackage{amsmath,amssymb,graphics,amscd,amsfonts,mathrsfs,epsf,indentfirst,cite}

\allowdisplaybreaks[1]
\usepackage{epsfig}
\usepackage{bm}
\usepackage{subfigure}

\title{
Electric Field Quench in AdS/CFT
}

\author{Koji Hashimoto$^{1,2}$, Shunichiro Kinoshita$^{3}$, Keiju Murata$^{4}$ and Takashi Oka$^{5}$
\\
$^{1}${\it Department of Physics, Osaka University,
Toyonaka, Osaka 560-0043, Japan}
\\
$^{2}${\it Mathematical Physics Lab., RIKEN Nishina Center,
Saitama 351-0198, Japan}
\\
$^{3}${\it Osaka City University Advanced Mathematical Institute, Osaka 558-8585, Japan}
\\
$^{4}${\it Keio University, 4-1-1 Hiyoshi, Yokohama 223-8521, Japan}
\\
$^{5}${\it Department of Applied Physics, University of Tokyo, 
Tokyo 113-8656, Japan}
\\
\\
{\it E-mails:} \email{koji@phys.sci.osaka-u.ac.jp},
\email{kinosita@sci.osaka-cu.ac.jp},
\email{keiju@phys-h.keio.ac.jp},
\email{oka@ap.t.u-tokyo.ac.jp}
}

\abstract{
An electric field quench, a suddenly applied electric field, 
can induce nontrivial dynamics in confining systems 
which may lead to thermalization as well as
a deconfinement transition. In order to analyze this nonequilibrium transitions,we use the AdS/CFT correspondence for ${\cal N}=2$ supersymmetric 
QCD that has a confining meson sector.
We find that the electric field quench causes the deconfinement transition even when the magnitude of
the applied electric field is smaller than the critical value for the static case (which is
the QCD Schwinger limit for quark-antiquark pair creation).
The time dependence is crucial for this phenomenon, and the gravity dual explains it as an oscillation of a D-brane
in the bulk AdS spacetime. Interestingly, the deconfinement time takes only discrete values as
a function of the magnitude of the electric field. 
We advocate that the new deconfinement phenomenon is analogous to the exciton Mott transition.
}

\preprint{
{\normalsize AP-GR-109} \\
{\normalsize OCU-PHYS-401} \\
{\normalsize OU-HET-819} \\
{\normalsize RIKEN-MP-91}
}


\begin{document}

\maketitle

\section{Introduction: Time-dependent electric field and deconfinement}
\label{sec:intro}

Obviously one of the most important unanswered question in QCD is the mechanism of 
quark confinement. Experimentally, RHIC experiments and subsequent LHC experiments
created a deconfined phase of QCD by heavy ion collisions, which have provided us 
a new perspective of the deconfinement transition. However, the cause of deconfinement
is still a mystery, mainly because we do not know the mechanism of the confinement.

To reach the deconfined phase, we need some external force put into the system.
Heavy ion experiments have two aspects, one is the temperature raise caused by
the thermalized gluons and the other is the strong electromagnetic fields created 
right after the impact of ions \cite{Kharzeev:2007jp,Skokov:2009qp,Voronyuk:2011jd,Bzdak:2011yy,Deng:2012pc}. A high temperature is sufficient for the deconfinement
as lattice simulations of QCD suggest, while  putting strong electric field can make
the QCD vacuum unstable against a creation of quark-antiquark pairs, known as
Schwinger mechanism, which also leads to deconfinement. 

The obstacle in theoretical analysis for this issue of the mechanism of the deconfinement transition 
is apparently the strong coupling and non-perturbative nature of QCD. During the last decade,
the AdS/CFT correspondence \cite{Maldacena:1997re,Gubser:1998bc,Witten:1998qj} turned 
out to be a useful tool for calculating
strongly-coupled gauge theory analytically. The virtue of the AdS/CFT correspondence is 
that it can be applied also to time-dependent system, as opposed to lattice QCD simulations.

\FIGURE[r]{ 
\includegraphics[width=6cm]{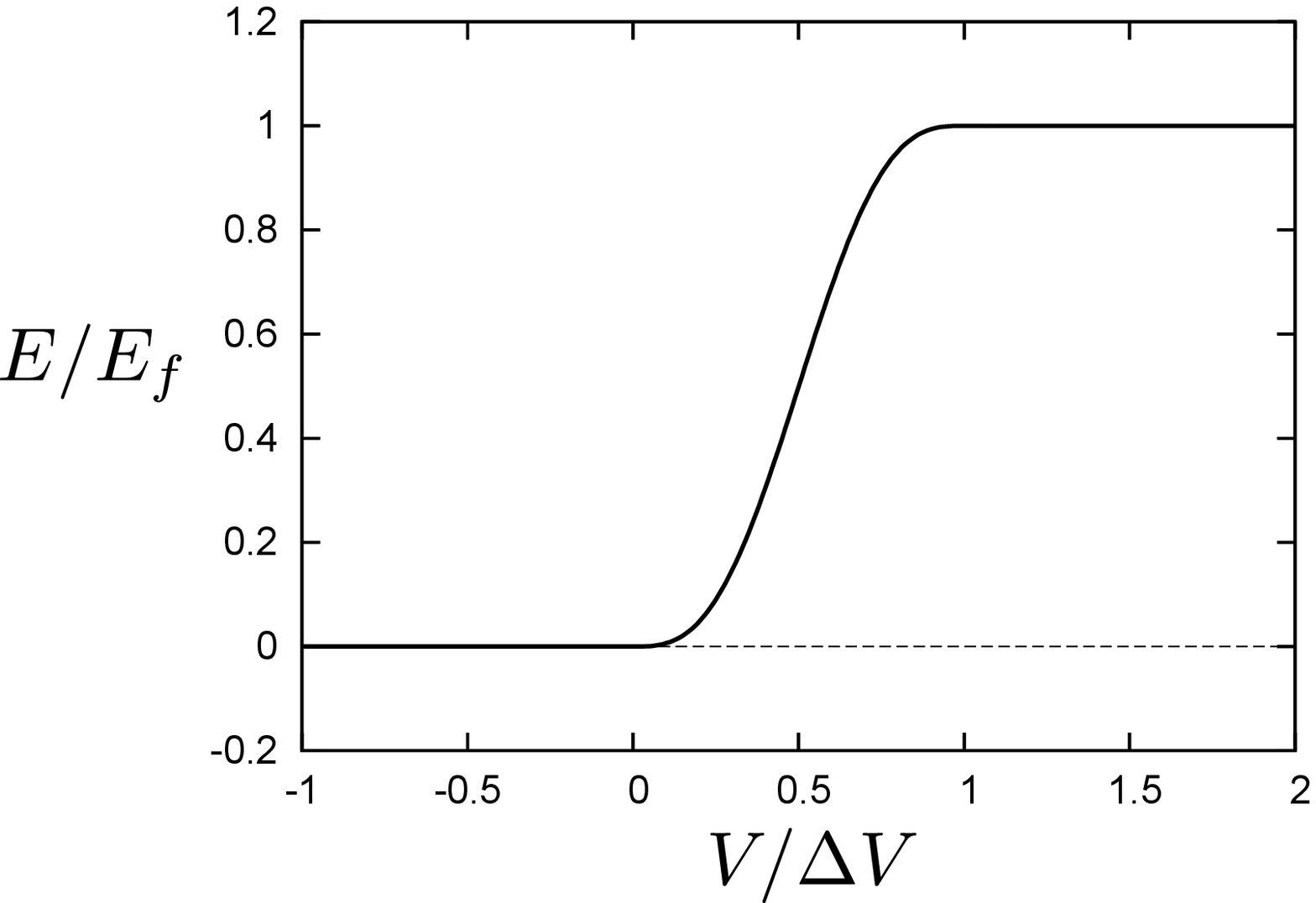}
\caption{The profile of the external electric field applied to the system, 
$E(V)$. $V$ is the time coordinate.}
\label{Eprof}
}

In this paper, we demonstrate an AdS/CFT analysis of an ``electric field quench'' --- a sudden apply
of an electric field --- for a strongly coupled gauge theory.\footnote{
Some AdS/CFT examples of quantum quenches of quark sectors are found in \cite{Das:2010yw,Hashimoto:2010wv,Ishii:2014paa}. 
On the other hand, thermalization due to a
quantum quench on gluonic sector in AdS/CFT were popularly studied (see for
example Refs. \cite{Danielsson:1999fa,Bhattacharyya:2009uu,Janik:2006gp,Ebrahim:2010ra,AbajoArrastia:2010yt,Balasubramanian:2010ce,Buchel:2012gw,Heller:2012km,Balasubramanian:2013rva,Buchel:2013lla}).} 
Figure~\ref{Eprof} shows the characteristic profile
of a time-dependent electric field; starting originally from zero, it is turned on 
with a ramp followed by a constant value. The profile is
a smeared step function whose height is $E_f$ and the duration of the ramp is parametrized by a time period $\Delta V$. 

We shall study the simplest toy model of strongly coupled gauge theory 
in string theory, namely the ${\cal N}=2$ $SU(N_c)$ supersymmetric QCD at large $N_c$ and at strong coupling.
The theory has an ${\cal N}=4$ supersymmetric Yang-Mills (gluonic) sector and an 
${\cal N}=2$ quark hypermultiplet 
\cite{Karch}. When the quark has a mass, the meson spectrum is discrete, while the gluon sector is
completely deconfined. So this serves as a toy model for a quark ``confinement'' occurring only
in the meson sector. The behavior of the system under an external electric field can be
studied by analyzing the dynamics of the probe flavor D7-brane in the $\mathrm{AdS}_5\times S^5$
geometry. 
It is known that, in the static case, 
there exists a critical electric field $E_{\rm crit}$ beyond which the phase 
transition occurs. Beyond the critical electric field $E>E_{\rm crit}$, the confinement is broken
and there appears an electric current carried by the quarks \cite{Karch:2007pd,Albash:2007bq,Erdmenger:2007bn}.\footnote{
Supercritical electric fields can make the QCD vacuum unstable against Schwinger pair production 
of quarks. See \cite{Hashimoto:2013mua,Hashimoto:2014dza} for the evaluation of the Euler-Heisenberg Lagrangian and the instability associated with
the imaginary part of the effective action.
See also \cite{Semenoff:2011ng,Ambjorn:2011wz,Bolognesi:2012gr,Sato:2013pxa,Sato:2013iua,
Sato:2013dwa,Sato:2013hyw,Kawai:2013xya,Sakaguchi:2014gpa,Gorsky:2001up,
Sonner:2013mba,Chernicoff:2013iga} for AdS/CFT calculations of the Schwinger production. 
} 
However, below the critical electric field
$E< E_{\rm crit}$, the system is still a confined phase for mesons.

Interestingly, we find that even if the magnitude of the electric field is 
below the critical electric field,
we can reach the deconfinement phase
 once we apply it in a time-dependent 
manner. 
See our result, Fig.~\ref{bouncenum}. The lines in Fig.~\ref{bouncenum}
divide the $(\Delta V,E_f)$-plane into two regions --- the upper-left region
is a parameter region which leads to the deconfinement. Notice that even for small final value 
of the electric field $E_f$, if the duration $\Delta V$ is sufficiently short,
we can reach the deconfinement.
Our result would imply a novel mechanism which may be working at heavy ion collisions:
the electric field caused by the fast ions can help the deconfinement transition
even if the magnitude of the electric field is small compared to the QCD scale.

Furthermore, we find a strange behavior of the deconfinement timescale: 
The calculated deconfinement time takes only discrete values, as
a function of the magnitude of the final electric field $E_f$ for a fixed $\Delta V$.
See the result shown in Fig.~\ref{td_desc}.

\FIGURE[r]{ 
\includegraphics[width=7cm]{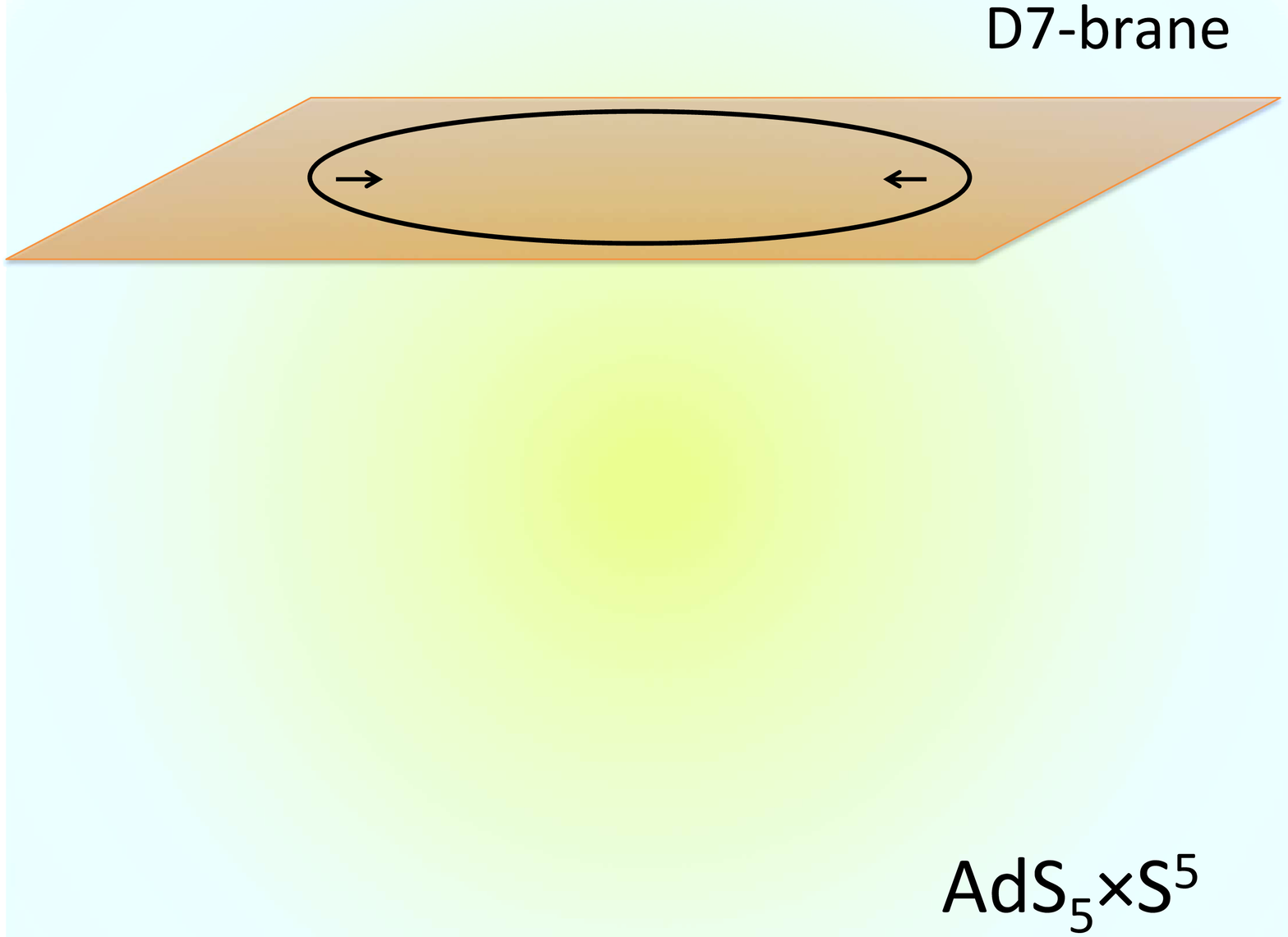}
\caption{A schematic picture of a D7-brane in the $\mathrm{AdS}_5\times S^5$ geometry. 
The D7-brane is shown as a square flat surface. If we turn on the external 
electric field, the fluctuation wave on the D7-brane comes in from the boundary.
The wave is shown as a circle shrinking on the D7-brane.
}
\label{D7wave}
}

Here we briefly describe what is happening in the gravity dual picture to reach the 
conclusions described above.
The AdS/CFT correspondence 
can make the detailed calculation possible in a time-dependent manner.
In the gravity dual, after turning on the electric field, the D7-brane moves in the 
bulk $\mathrm{AdS}_5 \times S^5$ geometry. 
See Fig.~\ref{D7wave} 
for an illustration of the D7-brane motion in the $\mathrm{AdS}_5 \times S^5$ spacetime. 
The motion looks like an oscillation, since
the external input changes only the boundary behavior of the D7-brane, and
the boundary motion propagates into the bulk motion on the D7-brane.
The energy pumped into the D7-brane will create a strongly red-shifted region 
on the D7-brane, which is an indication of the deconfinement in the gravity side.
We find that even if the magnitude of the oscillation is small, if the
energy is pumped in a short duration, the red-shifted region on the D7-brane 
emerges. 
That is the reason why we can make the deconfinement even with
small magnitude of the electric field.

Since the D7-brane fluctuation in the $\mathrm{AdS}_5$ space is similar to a wave in a finite-sized box,
the oscillation caused at the boundary propagates into the bulk but after a while
it is reflected back to the boundary. Repeating this reflection  sharpens the
wave packet and finally creates a naked singularity which causes a 
strongly red-shifted region on the D7-brane.
The number of reflections depends on parameters $E_f$ and $\Delta V$.
The reflection takes place at a multiple of time necessary for the wave to propagate 
from the boundary to the center of the bulk on the D7-brane. 
So, the deconfinement time is discretized from the gravity
viewpoint.
However, how to interpret the AdS/CFT result in the gauge theory side is nontrivial.

In any time-dependent set-up, giving a proper definition of the ``deconfinement'' and 
the ``thermalization''
is a nontrivial issue.
In this paper, we propose a new definition which is universal for any gravity dual setups.
{\it We define the deconfinement time as the time when the redshift factor becomes very large.
Furthermore, we define the thermalization as the time when the redshift factor grows exponentially.}
We are interested in the D7-brane meson sector, so we calculate the redshift factor
of an effective metric on the D7-brane.
There are several reasons for the usefulness of these definitions.
First, if the event horizon is formed, inside of the event horizon and itself 
cannot be known by the boundary observer, while the redshift factor can
be measured. 
Secondly, apparent horizons, which are commonly used for a definition of
the thermalization in AdS/CFT, will not always emerge outside of the
event horizon and they can not capture the universal features of
thermalization for wider gravity duals.
Thirdly, the new definition is directly related to spectrum of the Hawking
radiation to be observed by the boundary observer 
and reduces to the standard Hawking thermal temperature for static cases.

The organization of our paper is as follows. After giving a brief review on the flavor 
D7-brane embedding in the $\mathrm{AdS}_5 \times S^5$ geometry for the static case in section~\ref{sec:static}, we provide our description of the time-dependent D-brane motion
in section~\ref{sec:dynamics}. We explain our coordinate system and the equations of motion,
and the profile of the time-dependent external electric field and the AdS/CFT dictionary to
extract the physical observables. In section~\ref{defs}, we provide careful definitions of
the deconfinement and the thermalization: the deconfinement is defined as the emergence of
a strong redshift factor for the D-brane effective metric, and the thermalization is
defined as a slow settlement of the Hawking temperature given by the effective metric.
Later sections are for the presentation of our numerical results. First, in section~\ref{sec:resultsuper},
we show the thermalization and the deconfinement for the applied electric field
which is greater than the critical value. In section~\ref{sec:resultsub}, we analyze the case with
the electric field smaller than the critical value, and we find that the deconfinement still
takes place. We show that deconfinement time takes only discrete values, and explains
the reason from the AdS bulk viewpoint. Section~\ref{sec:conclusion} 
is devoted for a conclusion and discussions.

\section{A review of static embeddings with electric fields}
\label{sec:static}

\subsection{Basic equation}

In this section, we briefly review results of static embeddings with
electric fields living on the D$7$-brane~\cite{Karch:2007pd,Albash:2007bq,Erdmenger:2007bn}.
We consider AdS$_5\times S^5$ spacetime as the
background solution: 
\begin{equation}
ds^2=\frac{L^2}{z^2}\left[-dV^2-2dVdz+dx_1^2 + dx_2^2 + dx_3^2\right] + L^2(d\phi^2 +
  \cos^2\phi d\Omega_3^2+ \sin^2\phi d\psi^2)\ ,
\label{bulkmetric}
\end{equation}
where $L$ is the AdS radius.
Although one can use the ordinary time
coordinate $dt=dV+dz$ instead of $V$ in static cases,  
we take the ingoing Eddington-Finkelstein coordinates for 
convenience of later dynamical calculations.
The embeddings of D$7$-brane are described by the Dirac-Born-Infeld (DBI) action,
\begin{equation}
 S=-\mu_7g_s^{-1}\int d^8\sigma \sqrt{-\textrm{det}[h_{ab}+2\pi\alpha'F_{ab}]}\
  ,
\label{DBI0}
\end{equation}
where $\mu_7=(2\pi)^{-7}\alpha'{}^{-4}$ and $g_s$ is the string coupling.
$h_{ab}$ is the brane induced metric, which is defined by
$h_{ab}=g_{\mu\nu}\partial_a X^\mu \partial_b X^\nu$. Here $X^\mu$ is
the brane collective coordinate and $g_{\mu\nu}$ is the metric in the
target space.
$F_{ab}$ is the field strength on the brane worldvolume, which is
defined by $F_{ab} = \partial_a A_b - \partial_b A_a$.
As the worldvolume coordinates, we use the target space coordinates
themselves as $\{\sigma^a\}=(V,z,\Omega_3,\vec{x}_3)$.
Assuming time translational symmetry generated by $\partial_V$, 
spherical symmetry of $S^3$,  
translational symmetries generated by $(\partial_{x_1},\partial_{x_2},\partial_{x_3})$, and
rotational symmetry on $(x_2,x_3)$-plane,
the brane position and gauge potential 
are written as
\begin{equation}
 \phi=\Phi(z)\ ,\quad \psi=0, \quad
2\pi\alpha'L^{-2}A_a d\sigma^a=\{-EV+a(z)\}dx_1\ .
\label{ansatz}
\end{equation}
In this paper, since we will not take account of finite baryon number density
in the boundary theory, we have omitted the $V$-components of the gauge
potential, $a_V(z)dV$. 
For static embeddings with non-zero baryon number density, see in~\cite{Karch:2007pd,Kobayashi:2006sb,Kim:2011qh}.
Note that while the gauge potential contains a time dependent component,
$-EVdx_1$, the field strength is time independent and this term gives a 
constant external electric field along $x_1$-direction in the boundary theory. 
Because of the symmetry generated by $\partial_\psi$ in the background
spacetime, we set $\psi=0$ without loss of generality.
Then, the DBI action is written as
\begin{equation}
\begin{split}
&S=-\mu_7 g_s^{-1}V_4 \Omega_3 L^8 \int dz  \frac{\cos^3\Phi(z)}{z^5}\sqrt{\xi}\ ,\\
&\xi\equiv z^2
 \bar{F}(z)\Phi'(z)^2+z^4\{a'(z)^2+2Ea'(z)\}+1\ ,
\end{split}
\end{equation}
where $V_4\equiv \int dVdx_1dx_2dx_3$, $\Omega_3=\textrm{Vol}(S^3)=2\pi^2$
and $\bar{F}(z)\equiv 1-E^2z^4$. 
Equations of motion for $a(z)$ and $\Phi(z)$ are given as
\begin{equation}
 \frac{\cos^3\Phi}{z\sqrt{\xi}}(a'+E)=j\ ,
\quad
\left(\frac{\bar{F}\cos^3\Phi}{z^3\sqrt{\xi}}\Phi'\right)'
+\frac{3\sin\Phi\cos^2\Phi}{z^5}\sqrt{\xi}=0\ .
\label{conserve0}
\end{equation}
Since the action only depends on $a'$ but does not
contain $a$ explicitly, we have obtained
the conservation law as the first equation. 
The constant of motion $j$ will be related to the 
electric current in the boundary theory.
From the first equation in Eq.~(\ref{conserve0}), we have
\begin{equation}
 \xi=\frac{\bar{F}(1+z^2\Phi'{}^2)\cos^6\Phi}{-j^2z^6+\cos^6\Phi}\ .
\label{xi}
\end{equation}
Substituting the above equation into the second equation of
Eq.~(\ref{conserve0}), we obtain a single 
equation for $\Phi$ as
\begin{multline}
\Phi''
=
\frac{1}{2z^8\bar{F}(-j^2+z^{-6}\cos^6\Phi)}\Big[
-6\bar{F}\sin\Phi\cos^5\Phi(1+z^2\Phi'{}^3)\\
- z^4\{
\cos^6\Phi(\bar{F}'-8z^{-1}\bar{F})
-j^2z^{6}(\bar{F}'-2z^{-1}\bar{F})
\}\Phi'{}^3\\
-z^2\{
\cos^6\Phi(\bar{F}'-6z^{-1}\bar{F})
-j^2z^6\bar{F}'
\}\Phi'\Big]\ .
\label{Phieq}
\end{multline}
In practical numerical calculations, 
introducing a new variable $W(z)=z^{-1}\sin\Phi(z)$, we solve the equation for
$W(z)$ obtained by rewriting the above equation in term of $W(z)$.

\subsection{Observables in the boundary theory}

Near the AdS boundary $z=0$, solutions are expanded as
\begin{equation}
 \frac{1}{z}\sin\Phi(z) = m+cz^2+\cdots\ ,\quad
 a(z) =-Ez+\frac{j}{2}z^2+\cdots\ .
\end{equation}
One can easily check that the expansion coefficient $j$ coincides
with the constant of motion appeared in Eq.~(\ref{conserve0}).
The constants $m,E,c$ and $j$ correspond to 
quark mass $m_q$, electric field $\mathcal{E}$, 
quark condensate $\langle \mathcal{O}_m\rangle$, 
and electric current $\langle J^x \rangle$ as
\begin{align}
&m_q=\frac{L^2m}{2\pi\alpha'}=\left(\frac{\lambda}{2\pi^2}\right)^{1/2}m\
 ,\qquad
\mathcal{E}=\frac{L^2 E}{2\pi\alpha'}=\left(\frac{\lambda}{2\pi^2}\right)^{1/2}E\ ,\notag\\
&\langle \mathcal{O}_m\rangle = -\frac{N_c\sqrt{\lambda}}{2^{3/2}\pi^3}\, c\
 ,\qquad
 \langle J^x \rangle
=\frac{N_c\sqrt{\lambda}}{2^{5/2}\pi^3}\, j\ ,
\label{rel}
\end{align}
where $\lambda$ denotes the 't~Hooft coupling.
Ignoring proportional constants, we will refer to $m,E,c$ and
$j$ themselves as quark mass, electric field, 
quark condensate and electric current, hereafter.

\subsection{Effective metric and horizon}

In appendix~\ref{app:EOM_DBI},  
we show that the embedding functions of the brane, which describe the brane
position in the target space, and the gauge field on the brane are governed by non-linear wave
equations on the following effective metric: 
\begin{equation}
 \gamma_{ab} =h_{ab}+(2\pi\alpha')^2h^{cd}F_{ac}F_{bd}\ .
\label{effmetric0}
\end{equation}
Therefore, causality for fluctuations propagating on the brane is
determined by this effective metric~\cite{Seiberg:1999vs,Gibbons:2000xe,Kim:2011qh,Gibbons:2001ck,Gibbons:2002tv}.
Substituting our ansatz~(\ref{ansatz}), 
we obtain the effective metric for the static embedding as 
\begin{multline}
 L^{-2}\gamma_{ab}d\sigma^a d\sigma^b
=-\frac{\bar{F}}{z^2}dV^2
 -\frac{2}{z^2}(1+Ez^4 a')dVdz
 +(\Phi'{}^2+z^2a'{}^2)dz^2\\
 +\frac{z^2\bar{F}\Phi'{}^2+z^4a'(a'+2E)+1}{z^2(1+z^2\Phi'{}^2)}dx_1^2
+\frac{1}{z^2}(dx_2^2+dx_3^2)+\cos^2\Phi d\Omega_3^2\ .
\label{effmetstatic}
\end{multline}
This metric is manifestly singular at $\Phi(z)=\pi/2$, at which the radius of $S^3$
wrapped by the brane goes to zero. 
Thus, 
the domain of the $z$-coordinate is given by $0\le z \le z_\textrm{max}$ where 
$\Phi(z_\textrm{max})=\pi/2$.
The event horizon (Killing horizon) in this metric will appear at 
$z=E^{-1/2}\equiv z_\textrm{eff}$, where $\bar F(z_\textrm{eff}) = 0$, 
if $z_\textrm{eff}<z_\textrm{max}$. 
We refer to the surface $z=z_\textrm{eff}$ as the effective horizon. 
Note that the effective horizon is different from bulk event horizon in
general.
In fact, although the background spacetime is now pure AdS without any
black hole and just the Cauchy horizon is located at $z=\infty$, the
effective horizon can emerge on the D-brane at $z=z_\mathrm{eff}$.
Furthermore, 
the effective horizon is time-like in the view of the bulk metric and can be
seen from the AdS boundary through the bulk null geodesic. 

Based on the effective metric, we can define effective surface gravity.
A Killing vector $\xi_a=(\partial_V)_a$ is the null generator of the effective
horizon. The effective surface gravity $\kappa$ is defined by
$\xi^b\hat{D}_b\xi_a|_{z=z_\textrm{eff}}=-\kappa\xi_a|_{z=z_\textrm{eff}}$ 
where $\hat{D}$ is the covariant
derivative with respect to $\gamma_{ab}$. 
From Eq.~(\ref{effmetstatic}), we
obtain 
\begin{equation}
 \kappa=\frac{2E^{3/2}}{E+a'(z_\textrm{eff})}\ ,
\label{kappa0}
\end{equation}
Quanta of brane fluctuations are emitted from the vicinity of 
the effective horizon 
as Hawking radiation 
with the temperature $\kappa/(2\pi)$.\footnote{
For massless case $m=0$, 
the effective temperature is studied in more general set
up in Ref.~\cite{Nakamura:2013yqa}.
}

\subsection{Boundary conditions at effective horizon and pole}

We can consider two kinds of static embeddings depending on values of 
$z_\textrm{eff}$ and $z_\textrm{max}$. 
When the effective horizon does not emerge on the brane
($z_\textrm{eff}>z_\textrm{max}$),  
the D7-brane solution is called a {\textit{Minkowski embedding}}.
In this case, the brane reaches the pole
($\Phi=\pi/2$) at which the $S^3$ shrinks to zero. 
Now, the first equation in~(\ref{conserve0}) can be rewritten as 
\begin{equation}
 \frac{\cos^6\Phi}{z^2}(a'+E)^2=j^2 
  [z^4(a'+E)^2 + \bar{F}(z)(1+z^2\Phi'^2)] .
\label{eq:first_integral}
\end{equation}
For the Minkowski embeddings, since $\cos\Phi = 0$ should be satisfied
at the pole $z=z_\mathrm{max}$, we have $j=0$ from the above equation.
Furthermore, from the regularity of Eq.~(\ref{Phieq}),
asymptotic solution near the pole becomes 
\begin{equation}
 W(z)=\frac{1}{z_\textrm{max}}-\frac{E^2z_\textrm{max}^2}
{2-E^2z_\textrm{max}^4}(z-z_\textrm{max})+\cdots\ , 
\label{polereg}
\end{equation}
which gives us a boundary condition for the Minkowski embeddings.

When the effective horizon emerges on the brane ($z_\textrm{eff}<z_\textrm{max}$), 
the D7-brane solution is called a {\textit{black hole embedding}}.
In this case, since $\bar{F}(z_\mathrm{eff}) = 0$ at the effective
horizon, (\ref{eq:first_integral}) leads to
\begin{equation}
 j=\frac{\cos^3\Phi(z_\textrm{eff})}{z_\textrm{eff}^3}\ .
\label{reality}
\end{equation}
Here, we have assumed $a'(z_\mathrm{eff}) + E \neq 0$.
Otherwise we obtain $a' + E \sim \sqrt{z_\mathrm{eff}-z}$ which results
in a singular behavior of $a''$ at the effective horizon.
Thus, (\ref{reality}) is a natural condition derived by the equation of
motion for the gauge field.%
\footnote{
Note that it relates to
imposing the reality condition of the D-brane 
action~\cite{Karch:2007pd,Karch:2007br,Erdmenger:2007bn,Albash:2007bq} 
such that the denominator of Eq.~(\ref{xi}) must change the sign at
the effective horizon where $\bar{F}$ changes the sign.
}
From the regularity of Eq.~(\ref{Phieq}),
asymptotic solution near the effective horizon becomes
\begin{equation}
 W(z)=W_\textrm{eff}
-\frac{1-(1-W_\textrm{eff}^2z_\textrm{eff}^2)^{1/2}}{W_\textrm{eff}z_\textrm{eff}^3}
(z-z_\textrm{eff})+\cdots\ .
\label{effreg}
\end{equation}
where $W_\textrm{eff}\equiv W(z_\textrm{eff}) = \sin\Phi(z_\textrm{eff})/z_\textrm{eff}$.

\subsection{Brane solutions}

\begin{figure}
\begin{center}
\includegraphics[scale=0.6]{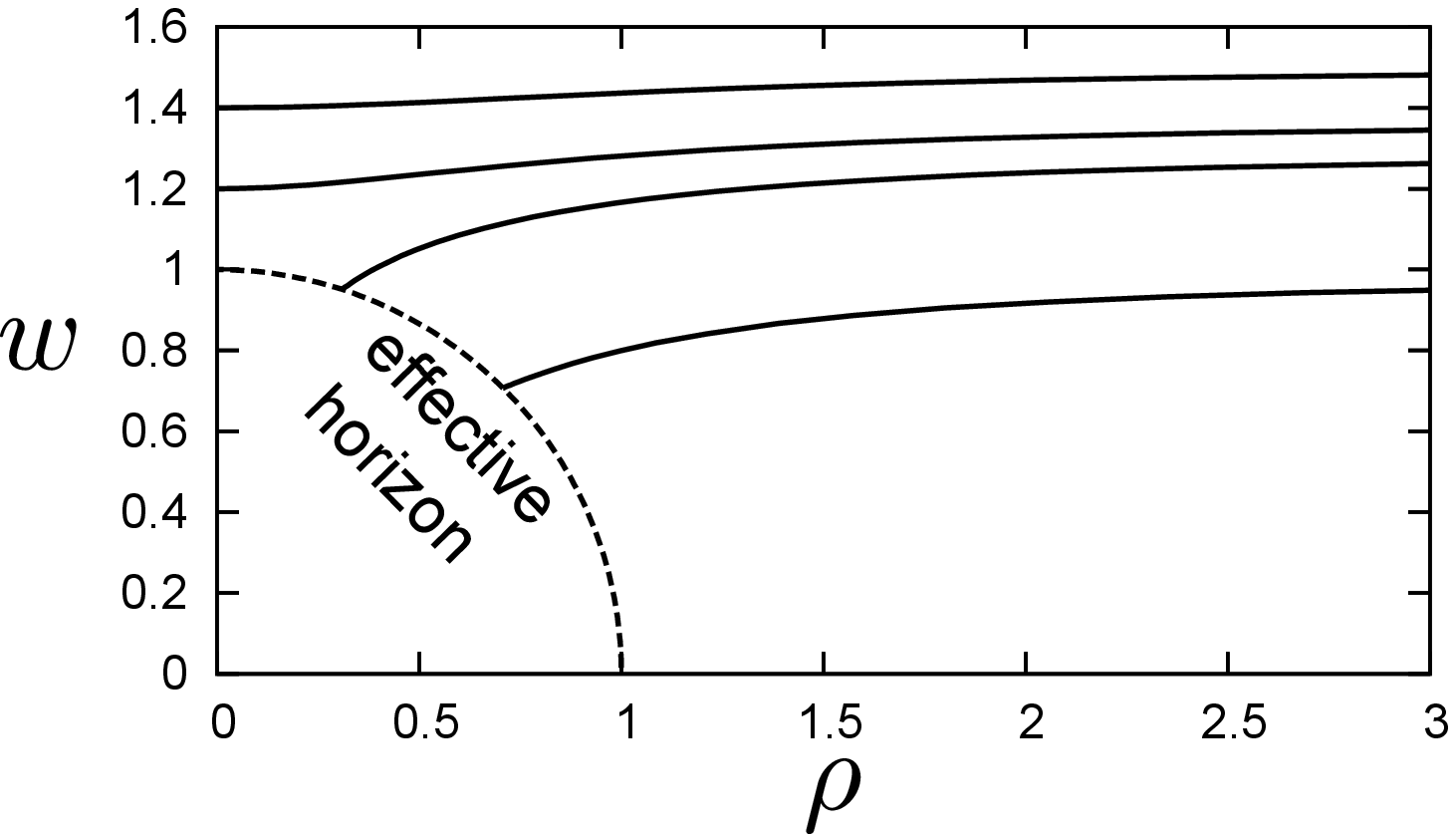}
\end{center}
\caption{
Minkowski and black hole embeddings of D7-brane in the unit of $E=1$. 
The effective horizon is located at $z=1$ which is shown by an unit
circle in this figure.
}
 \label{emb_rp=0_d=0}
\end{figure}

Using Eqs.~(\ref{polereg}) or (\ref{effreg}) as the boundary condition, 
we solve Eq.~(\ref{Phieq}) from the pole or the effective horizon 
to the AdS boundary $z=0$. In Fig.~\ref{emb_rp=0_d=0}, we show profiles of
the D7-brane in the unit of $E=1$. 
As the vertical and the horizontal axes,
we have taken Cartesian-like coordinates $(w,\rho)=(z^{-1}\sin\phi,z^{-1}\cos\phi)$.
In the ($w,\rho$)-plane, the effective horizon is shown by an unit
circle, $w^2+\rho^2=1$. 

\begin{figure}
  \centering
  \subfigure[quark condensate]
  {\includegraphics[scale=0.3]{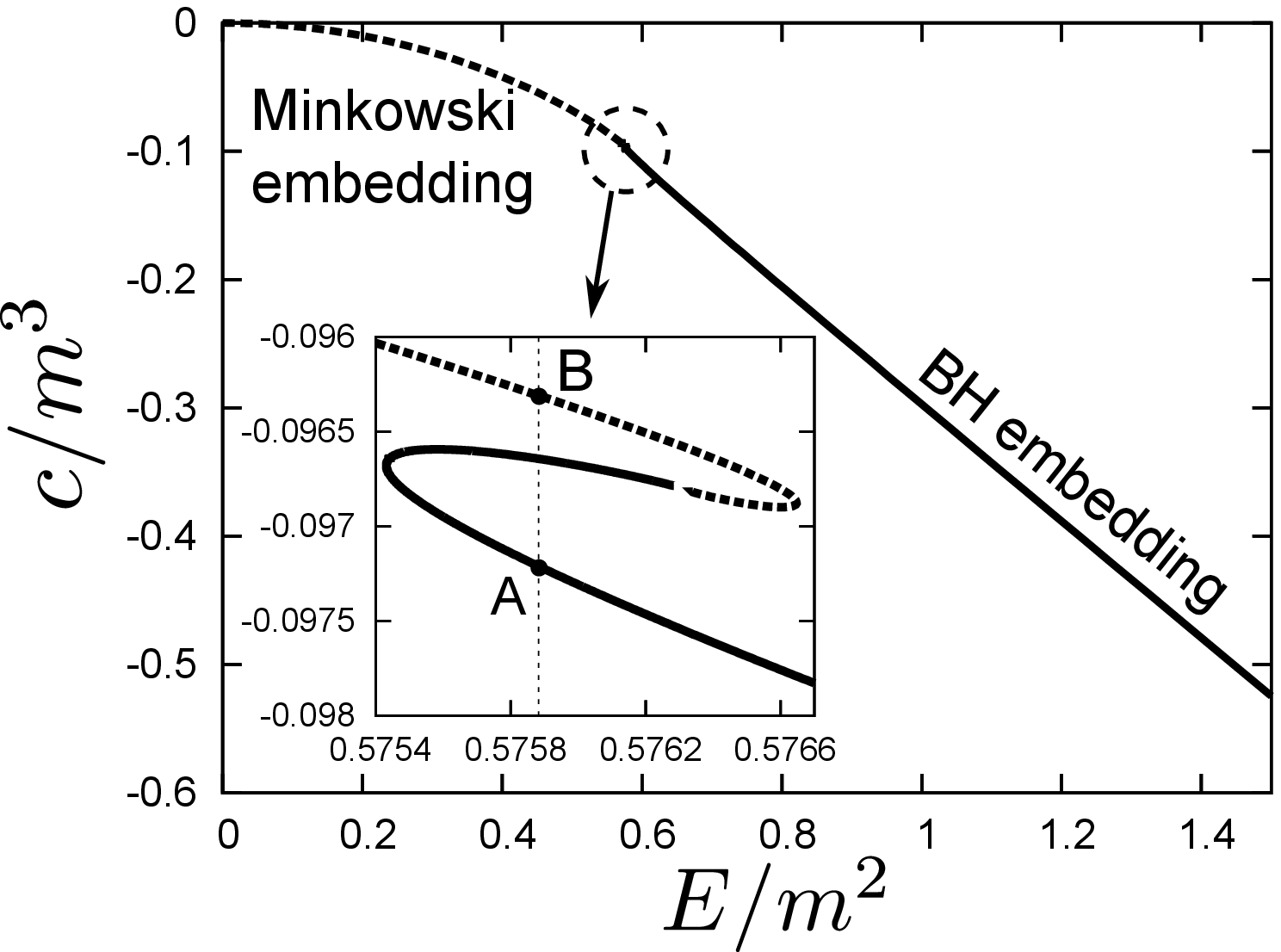}
   }
  \subfigure[electric current]
  {\includegraphics[scale=0.3]{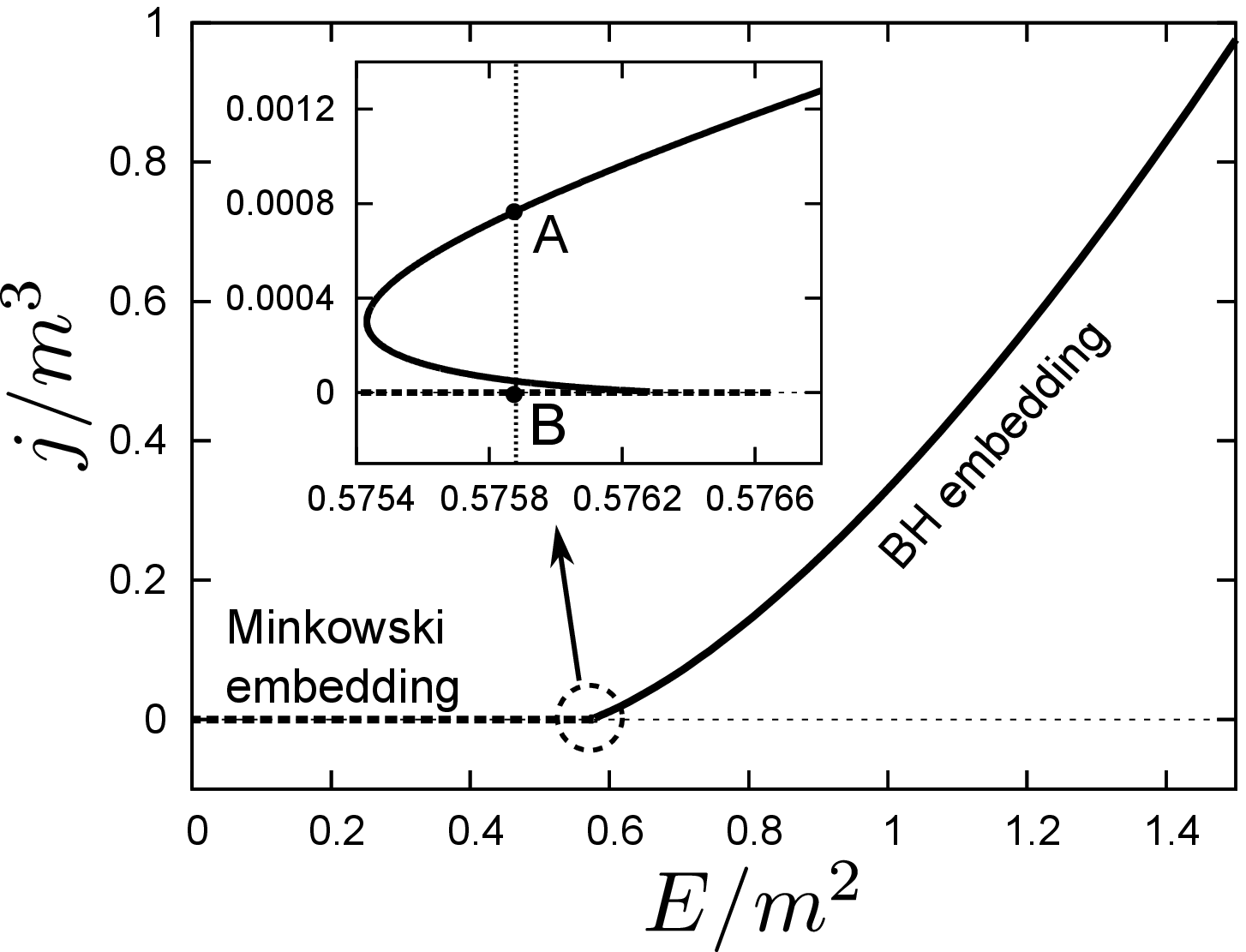}
  }
  \subfigure[effective surface gravity]
  {\includegraphics[scale=0.3]{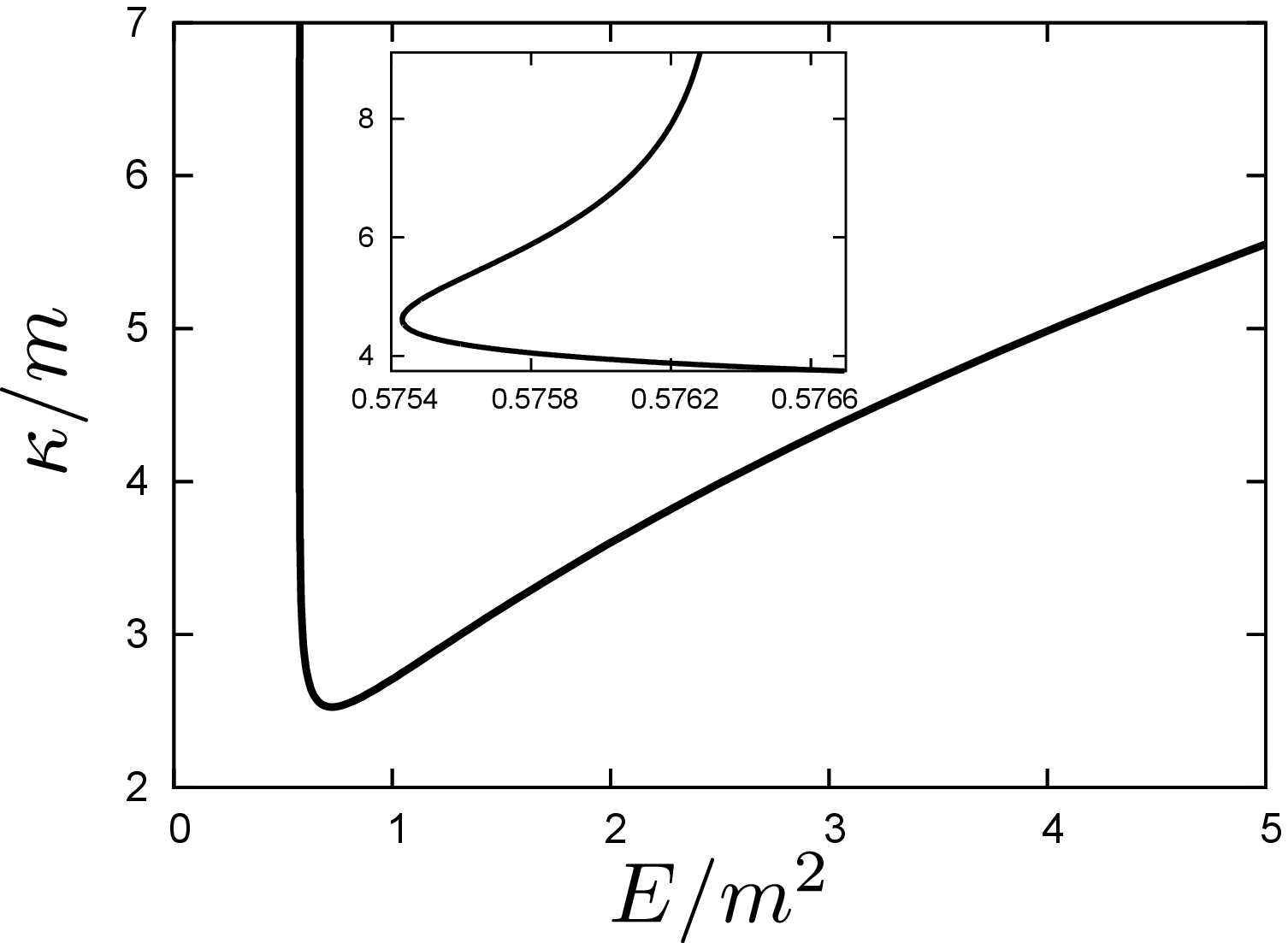} 
  }
  \caption{
Quark condensate $c$, electric current $j$ and effective surface gravity $\kappa$ 
for the static embeddings.
The quark condensate and the electric current 
make finite jump between the points A and B by the first-order phase
 transition.
The effective surface gravity takes minimum value at
 $(E/m^2,\kappa/m)=(0.724,2.527)$.
}
\label{cj_rp=0_d=0}
\end{figure}

From these solutions, we can read off the quark
condensate $c$ and electric current $j$.
In Fig.~\ref{cj_rp=0_d=0}(a) and (b), we plot the  $c$ and $j$
 as functions of electric
field $E$. They are normalized by the quark mass $m$.\footnote{
Throughout this paper, we will nondimensionalize variables by quark mass
$m$ unless otherwise noted.}
(Quark condensate $c(E)$ and electric current $j(E)$ were
computed explicitly in Refs.~\cite{Erdmenger:2007bn,Albash:2007bq}
and Ref.~\cite{Nakamura:2012ae}, respectively.)
They take multiple values in
$0.5754<E/m^2< 0.5766$. This indicates that there is a 
phase transition between Minkowski and black hole embeddings.
In fact, 
in Refs.~\cite{Erdmenger:2007bn,Albash:2007bq},
they found a first-order phase transition at $E/m^2=0.57588$ 
by a thermodynamical argument. We show the transition point by a
vertical line in the figure. The quark condensate and electric current make finite jump 
between points A and B.
Note that, for $E/m^2< 0.5754$, 
we obtain only Minkowski embeddings and, thus, the
electric current is exactly zero. 
For black hole embeddings, 
the differential resistance $dj/dE$ can be negative as pointed out in
Ref.~\cite{Nakamura:2012ae}.

For pure AdS background, 
the effective surface gravity is simply written as 
\begin{equation}
 \kappa=[3E(1+E^{1/2}j^{-1/3})]^{1/2}\ ,
\label{kappaj}
\end{equation}
where we used Eqs.~(\ref{conserve0}), (\ref{xi}), (\ref{kappa0}) and (\ref{reality}).
Since we have already computed the electric current $j$ as a function of $E$, 
we can easily obtain the the effective surface gravity as in
Fig.~\ref{cj_rp=0_d=0}(c).
At the point where both Minkowski and black hole embeddings join,
$\kappa$ diverges. 
For $0.5754<E/m^2<0.724$, 
the surface gravity can be a decreasing function of the electric field
$E$.
It takes minimum value at $(E/m^2,\kappa/m)=(0.724,2.527)$ and, 
for $E/m^2>0.724$, increases monotonically. 
For strong electric field $E/m^2\gg 1$, we have 
$\Phi(z_\textrm{eff})\simeq 0$. 
Thus, we obtain $j\simeq z_\textrm{eff}^{-3} = E^{-3/2}$ from
Eq.~(\ref{reality}). Therefore, for strong limit of the electric field, 
we have analytical expression the the surface gravity as 
$\kappa\simeq (6E)^{1/2}$.

\section{Dynamics of D7-brane with electric fields}
\label{sec:dynamics}

In this paper, we study far-from-equilibrium dynamics of ${\cal N}=2$ supersymmetric 
QCD, which is induced by time dependent external electric fields.
We will turn on a homogeneous electric field from zero to finite
non-zero value and examine the response of the system.
This means we should deal with dynamics of the D$7$-brane and the
gauge field living on the brane.
In this section, we explain our model and formulation for solving the
dynamics numerically.

\subsection{Basic equations}

We use the AdS$_5\times S^5$ spacetime~(\ref{bulkmetric}) as the
background, which means that we focus on zero temperature for the bulk gluon. 
The dynamics of the D7-brane is described by the DBI
action~(\ref{DBI0}). Hereafter, we take the unit where the AdS radius is
unity, $L=1$.

We introduce eight worldvolume coordinates $\{\sigma^a\}$ 
($a=0,1,\cdots,7$) on the brane.
For six of them, we use the target space
coordinates themselves as $(\sigma^2,\cdots,\sigma^7)=(\vec{x}_3,\Omega_3)$.
For the other two coordinates, 
we introduce $(u,v)$-coordinates which are determined by imposing
coordinate conditions later. 
Imposing spherical symmetry of $S^3$,  
translational symmetries generated by $(\partial_{x_1},\partial_{x_2},\partial_{x_3})$, 
and rotational symmetry on $(x_2,x_3)$-plane, the brane collective
coordinates and the gauge potential are written as
\begin{equation}
\begin{split}
&V=V(u,v)\ ,\quad
z=Z(u,v)\ ,\quad
\phi=\Phi(u,v)\ ,\quad
\psi=0\ ,\\
&2\pi\alpha'A_a d\sigma^a =a_u(u,v) du+a_v(u,v)dv + a_x(u,v)dx_1\ .
\label{VZPpara}
\end{split}
\end{equation}
Note that, since ($x_2, x_3$)-components of the gauge potential are
absent, we will denote $x_1$-component of that as $a_x$ briefly. 
Here, because of the $U(1)$-symmetry generated by $\partial_\psi$, we
can set $\psi=0$ without loss of generality. 

Then, the D7-brane action is written as
\begin{equation}
\begin{split}
&S= -\mu_7g_s^{-1} V_3 \Omega_3 \int dudv
 \frac{\cos^3\Phi}{Z^3}\sqrt{\xi}\ ,\\
&\xi\equiv -f_{uv}^2+(h_{uv}+Z^2\partial_u a_x
  \partial_v a_x )^2
-(h_{uu}+Z^2\partial_u a_x^2)(h_{vv}+Z^2\partial_v a_x^2)
\end{split}
\label{DBI1}
\end{equation}
where $V_3\equiv \int dx_1dx_2dx_3$, $\Omega_3=\textrm{Vol}(S^3)=2\pi^2$,
and 
\begin{align}
&f_{uv}=\partial_u a_v-\partial_v a_u\ ,\quad
h_{uv}=-Z^{-2}(V_{,u}V_{,v}+V_{,u}Z_{,v}+V_{,v}Z_{,u})+\Phi_{,u}\Phi_{,v}\ ,\notag\\
&h_{uu}=-Z^{-2}V_{,u}(V_{,u}+2Z_{,u})+\Phi_{,u}^2\ ,\quad
h_{vv}=-Z^{-2}V_{,v}(V_{,v}+2Z_{,v})+\Phi_{,v}^2\ .
\end{align}
From equations of motion for $a_u$ and $a_v$, we obtain
\begin{equation}
 \frac{\cos^3\Phi}{Z^3\sqrt{\xi}}f_{uv}= d\ ,
\label{conserve}
\end{equation}
where $d$ is an integration constant corresponding to the baryon number
density in the boundary theory.
In this paper, we focus on zero baryon number density and assume 
$d=0$, namely $f_{uv}=0$, hereafter. 
(For general cases see appendix~\ref{app:general_EOM}.)

Now, since the action has coordinate freedom of $(u,v)$-coordinates, we
can take a convenient coordinate system for numerically solving dynamics.
As we mentioned, dynamics of the D-brane and the gauge field on the brane
are governed by wave equations on the effective metric.
In order to introduce double-null coordinate system in two-dimensional
part of the effective
metric, 
we impose coordinate conditions:
\begin{align}
&C_1\equiv h_{uu}+Z^2(\partial_u a_x)^2=0\ ,\label{C1}\\
&C_2\equiv h_{vv}+Z^2(\partial_v a_x)^2=0\ ,\label{C2}
\end{align}
which are double-null conditions for the effective metric rather than
the induced metric.
Indeed, under these coordinate conditions, the effective metric is
written as 
\begin{multline}
  \gamma_{ab}d\sigma^a d\sigma^b
=2(h_{uv}+Z^2f_{ux}f_{vx})dudv\\
+\frac{1}{Z^2}\frac{h_{uv}+Z^2f_{ux}f_{vx}}{h_{uv}-Z^2f_{ux}f_{vx}}dx_1^2
+\frac{1}{Z^2}d\vec{x}_2^2+\cos^2\Phi d\Omega_3^2\ ,
\label{effmetric}
\end{multline}
where the effective metric is defined by Eq.~(\ref{effmetric0}).
Note that these coordinate conditions are constraint equations.

Then, the square root in the DBI action~(\ref{DBI1}) can be removed and 
the action is simply written as 
\begin{equation}
S= \mu_7g_s^{-1} V_3 \Omega_3 \int dudv
\frac{\cos^3\Phi}{Z^3}
(h_{uv}+Z^2\partial_u a_x  \partial_v a_x ) .
\end{equation}
Deviating this action, we can obtain evolution equations for $V$, $Z$,
 $\Phi$ and $a_x$. 
For convenience in numerical calculations we introduce a new variable
instead of $\Phi(u,v)$ as  
\begin{equation}
 \Psi(u,v)\equiv \frac{\Phi(u,v)}{Z(u,v)}\ .
\end{equation}
In term of the variables $(V,Z,\Psi,a_x)$, 
the evolution equations are written as 
\begin{align}
&V_{,uv}=
\frac{3}{2}Z(Z\Psi)_{,u}(Z\Psi)_{,v}
+\frac{3}{2}\tan(Z\Psi)
 \{(Z\Psi)_{,u}V_{,v}+(Z\Psi)_{,v}V_{,u}\}
\notag\\
&\hspace{7cm}
-\frac{5}{2Z}V_{,u}V_{,v}
+\frac{Z^3}{2}a_{x,u}a_{x,v}\ ,
\label{Veq}\\
&Z_{,uv}
=
-\frac{3}{2}Z(Z\Psi)_{,u}(Z\Psi)_{,v}
+\frac{3}{2}\tan(Z\Psi)\{(Z\Psi)_{,u}Z_{,v}+(Z\Psi)_{,v}Z_{,u}\}
\notag\\
&\hspace{1cm}
+\frac{5}{2Z}
(V_{,u}V_{,v}+V_{,u}Z_{,v}+V_{,v}Z_{,u})
+\frac{5}{Z}Z_{,u}Z_{,v}
-\frac{Z^3}{2}a_{x,u}a_{x,v}\ ,\\
&\Psi_{,uv}=
\frac{3}{2}\left(\Psi + \frac{\tan(Z\Psi)}{Z}
 \right)(Z\Psi)_{,u}(Z\Psi)_{,v}
\notag\\
&\hspace{2cm}
+ \frac{1}{2Z^2}\{1-3 Z \Psi \tan(Z\Psi)\}
\{(Z\Psi)_{,u}Z_{,v}+(Z\Psi)_{,v}Z_{,u}\}
\notag\\
&
\hspace{3cm}
- \frac{\Psi}{2Z^2}\left(5
- \frac{3\tan(Z\Psi)}{Z\Psi}
\right)
(V_{,u}V_{,v}+V_{,u}Z_{,v}+V_{,v}Z_{,u})\notag\\
&\hspace{4cm}
- \frac{3\Psi}{Z^2}Z_{,u}Z_{,v}
+\frac{Z^2\Psi}{2}\left(
1-\frac{3\tan(Z\Psi)}{Z\Psi}
\right)a_{x,u}a_{x,v}
\ ,\\
&a_{x,uv}
=
\frac{3}{2}\tan(Z\Psi)\{(Z\Psi)_{,u}a_{x,v}+(Z\Psi)_{,v}a_{x,u}\}
+\frac{1}{2Z}(Z_{,u}a_{x,v}+Z_{,v}a_{x,u})
\ .\label{aeq}
\end{align}
They guarantee that 
the coordinate conditions~(\ref{C1}) and (\ref{C2}) are preserved in the
time evolutions as
\begin{equation}
 \partial_u\Big[
\frac{\cos^3\Phi}{Z^5}C_2
\Big]
=
 \partial_v\Big[\frac{\cos^3\Phi}{Z^5}C_1
\Big]=0\ .
\end{equation}
Therefore, once we have imposed the coordinate conditions as initial
conditions and boundary conditions, 
$C_1=0$ and $C_2=0$ are automatically satisfied
and we only have to solve the evolution equations.

It turns out that the form of the equations of motion is quite similar
to that in Ref~\cite{Ishii:2014paa} except for the gauge field $a_x$.
Since a stable numerical method to solve this kind of equations has been
developed there, we will follow the numerical method to solve
Eqs.~(\ref{Veq})-(\ref{aeq}) and skip detail explanations of the
numerics in this paper.

\subsection{Observables at the AdS boundary}

Eliminating $u$ and $v$, we can regard $\Psi$ and $a_x$ as functions of
$V$ and $Z$. 
Near the AdS boundary $Z=0$, these functions are expanded as
\begin{align}
&\Psi(V,Z)=m+\left(c(V)+\frac{m^3}{6}\right)Z^2+\cdots\ ,\label{eq:Psi_asympto}\\
&a_x(V,Z)=\alpha_0(V)+\dot{\alpha}_0(V) Z+\frac{1}{2}j(V)Z^2 +
  \frac{1}{2}\ddot{\alpha}_0(V)Z^2\ln (mZ) +\cdots\ .\label{eq:ax_asympto}
\end{align}
It is convenient to rewrite the leading term of $a_x$ as
\begin{equation}
 \alpha_0(V)\equiv -\int^V dV' E(V')\ .
\label{ax_boundary}
\end{equation}
Here, $m$, $E(V)$, $c(V)$ and $j(V)$ are related to quark mass, electric
field, quark condensate and electric current in the boundary theory 
as in Eq.~(\ref{rel}).
Once we give the leading terms $m$ and $E(V)$ as boundary conditions for
$\Psi$ and $a_x$, 
we can determine $c(V)$ and $j(V)$ by solving the evolution equations.
In our following calculations, we choose a $C^2$ function for $E(V)$ as
\begin{equation}
E(V)=
\begin{cases}
0 & (V<0) \\
E_f[V-\frac{\Delta V}{2\pi}\sin(2\pi V/\Delta V)]/\Delta V & 
 (0 \le V \le \Delta V) \\
E_f & (V>\Delta V) 
\end{cases}
\ ,
\label{Efunc}
\end{equation}
where $E_f$ is a final value of the electric field and $\Delta V$ is a
rise time taken from zero electric field to the final one. 
The profile of the function $E(V)$ is shown in Fig.~\ref{Eprof}.

\subsection{Boundary conditions}
\label{sec:bdrycond}

\begin{figure}
\begin{center}
\includegraphics[scale=0.3]{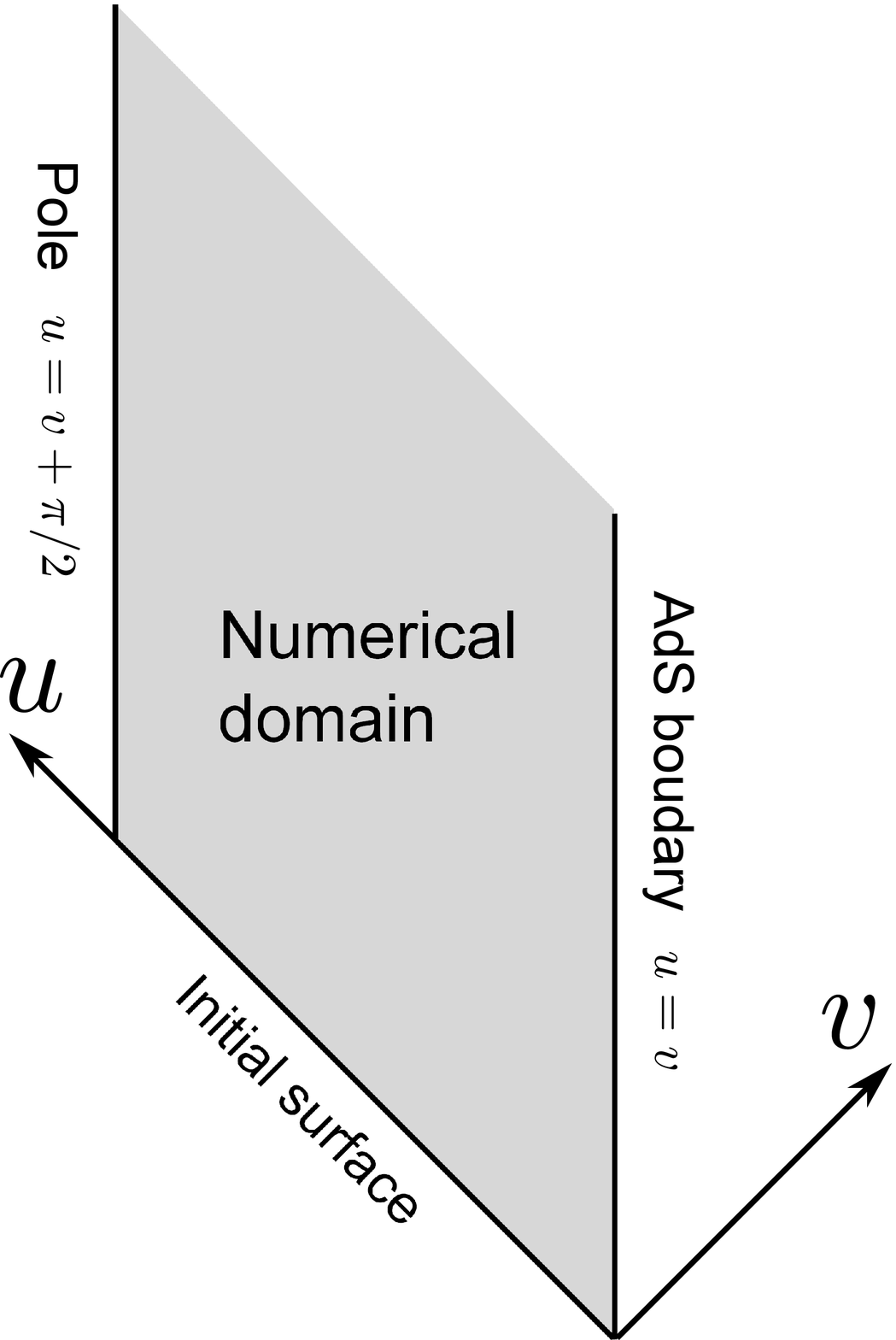}
\end{center}
\caption{
Numerical domain on the worldvolume of the D7-brane.
The AdS boundary and the pole are fixed at $u=v$ and $u=v+\pi/2$, respectively.
}
 \label{Numdom}
\end{figure}

In general, two time-like boundaries will appear in numerical domain on
the brane worldvolume: 
one is the AdS boundary $Z=0$, and the other is the pole
$\Phi=\pi/2$ at which the radius of $S^3$ wrapped by the D7-brane shrinks
to zero. 
For numerical convenience, we should fix the location of each boundary
in the worldvolume $(u,v)$-coordinates if the numerical domain contains 
that boundary.
Note that coordinate conditions~(\ref{C1}) and (\ref{C2}) are invariant under the residual coordinate
transformations, 
\begin{equation}
 \bar{u}=\bar{u}(u)\ ,\qquad \bar{v}=\bar{v}(v)\ ,
\label{residual}
\end{equation}
which generate a conformal transformation in the two-dimensional spacetime. 
Using them, we can fix the location of the
AdS boundary and the pole on the worldvolume coordinates as $u=v$ and
$u=v+\pi/2$, respectively.
In Fig.~\ref{Numdom}, we show our computational domain in $(u,v)$-plane.

Since we are interested in time evolutions on the AdS boundary, the AdS
boundary is always contained in the numerical domain and located at
$u=v$ throughout our calculations. 
Boundary conditions at the AdS boundary for $Z$, $\Psi$ and $a_x$ are
determined by asymptotic behaviors (\ref{eq:Psi_asympto}) and (\ref{eq:ax_asympto}) as $Z|_{u=v}=0$, $\Psi|_{u=v}=m$ and $a_x|_{u=v}=\alpha_0(V)$.
That is, we consider the quark mass is fixed at a non-zero value and the
electric field is time-dependent in the boundary theory.
We can derive the condition for $V$ from regularities of the evolution
equation near the AdS boundary to satisfy the constraint equations. As a result, we obtain 
$V_{,v}|_{u=v}= 2Z_{,u}|_{u=v}$. 
Solving the boundary equation, we can determine boundary value of $V$ at
each time step.

When the pole is contained in the numerical domain, boundary conditions
at the pole are necessary and the pole is fixed at $u=v+\pi/2$. 
Since the pole is located at $\Phi=\pi/2$, one boundary condition is
given by $(Z\Psi)|_{u=v+\pi/2}=\pi/2$. 
The others are obtained from regularities of the evolution equations at the
pole as $V_{,u}=V_{,v}$, $Z_{,u}=Z_{,v}$ and $a_{x,u}=a_{x,v}$.
They are Neumann boundary conditions at the pole.

\subsection{Initial data}

Finally, we explain initial data for our calculations.
Before turning on the electric field $V<0$, 
the brane is static there. In $(u,v)$-coordinates, the static solution
is written as
\begin{equation}
\begin{split}
&V(u,v)=m^{-1}[\phi(u)+\phi(v)-\sin(\phi(u)-\phi(v))] + V_\mathrm{ini}\
 ,\quad a_x(u,v)=0\ ,\\
&Z(u,v)=m^{-1}\sin(\phi(u)-\phi(v))\ ,\quad
\Psi(u,v)=\frac{m(\phi(u)-\phi(v))}{\sin(\phi(u)-\phi(v))}\ ,
\end{split}
\label{exact_static}
\end{equation}
where 
$\phi$ is a free function corresponding to the residual coordinate
freedom on the initial surface, and $V_\mathrm{ini}$ is an integration
constant. 
At an initial surface $v=0$, we set initial data to be the exact solution of the static embedding in pure AdS as
\begin{equation}
\begin{split}
&V(u,0)=m^{-1}(\phi(u)-\sin\phi(u)) + V_\mathrm{ini}\ ,\quad
Z(u,0)=m^{-1}\sin\phi(u)\ ,\\
&\Psi(u,0)=\frac{m\phi(u)}{\sin\phi(u)}\ ,\quad
a_x(u,0)=0
\ .
\end{split}
\label{initial_data}
\end{equation}
We set $\phi(0)=0$, and then $V(0,0)=V_\mathrm{ini}<0$
is the initial time at the AdS boundary.
At the first stage of the time evolution, when the numerical domain
contains the pole, 
we can solve the evolution equations under choosing the free function as
$\phi(u)=u$ simply.
However, if there is a region which causes strong redshift on the brane
such as vicinity of the event horizon, 
the numerical calculations will break down.
To continue the numerical calculation, we pause the numerical calculation
slightly before the breakdown, $v=v_\textrm{int}$.
We refer to the surface $v=v_\textrm{int}$ as intermediate surface.
We define functions at the intermediate surface as
$\bm{f}_\textrm{int}(u)\equiv \bm{f}(u,v_\textrm{int})$ 
where $\bm{f}\equiv (V,Z,\Psi,a_x)$.
We consider the coordinate transformation $u=\phi(\bar{u})$. 
Then, the $v$-coordinate is also transformed as
$v=\phi(\bar{v})$ to locate the AdS boundary at $\bar{u}=\bar{v}$.
Using $\bm{f}_\textrm{int}(\phi(\bar{u}))$ as the initial data,
we restart the numerical calculation from the intermediate surface, 
$\bar{v}=\bar{v}_\textrm{int}$.
We choose this free function $\phi$ so that $V$ and $\bar{v}$ are
synchronized up to a constant at the AdS boundary, i.e.,
$V|_{u=v}=\bar{v}+V_0$. 
For detail of numerical calculation, see~\cite{Ishii:2014paa}.

\section{Thermalization and deconfinement in dynamical
 systems}
\label{defs}
In this section, we attempt to give quantitative definitions of thermalization and deconfinement of mesons 
for dynamical systems in gravity side. 
In the boundary theory, ``thermalization'' means that the distribution
function has settled down in thermal one; 
``deconfinement'' means meson excitations become unstable and dissipate
into the background plasma.
Although both of them will occur at the same point if the systems are in
equilibrium or
steady state, these are physically different notions and may occur at
different times in general time-dependent situations.
Indeed, in the dynamical situation currently discussed, we will see that the system
might not be thermalized but mesons might be deconfined.
Therefore, it is important to explore means of discrimination between
thermalization and dissociation in gravity.

From viewpoint of gravity, in static or stationary cases both of
thermalization and deconfinement
are, also, characterized by a same condition: the existence of the event horizon on the
D-brane effective metric. 
The existence of the event horizon (precisely speaking, the Killing horizon) leads
to thermal spectrum of the Hawking radiation and dissipation of
excitations on the brane, which respectively correspond to thermalization and
deconfinement.
On the other hand, in time-dependent cases, it seems to be ambiguous how one should determine
thermalization time even if event horizons will form.

Of course, black holes and event horizons can be formally defined
without any ambiguities even though systems are dynamical. 
However, by definition, the event horizon cannot be determined unless
global spacetime evolutions have been known.
Temporal observers (or classical fields obeying the equations of motion)
cannot know when the event horizon has formed or whether they have been
inside the black hole, in principle. 
Alternatively, let us consider the apparent horizon instead of the event horizon.
The apparent horizon can be an useful estimator to find the event
horizon 
in dynamical spacetimes. 
In general cases, actually under appropriate energy conditions, apparent
horizons will form inside event horizons.
It means that formation of the apparent horizon does not affect physics
outside the black hole.
It is true that metrics irrelevant to the Einstein equations, such as an effective
metric of the brane, need not satisfy physical energy
conditions. 
(Indeed, in the current case the apparent horizon on the effective
metric can form outside the event horizon.) 
Nevertheless, if we want to discuss black hole formation in the bulk, which correspond to
thermalization of gluon plasma, the above problem is inevitable and the
apparent horizon seems to have trouble with causality.
Hence, appearance of the apparent horizon seems to be not so universal definition
for general thermalization in the boundary theory.

In this paper, to make clear the difference between thermalization and
deconfinement in the gauge/gravity duality, 
we will introduce the redshift factor and the Hawking temperature for
non-stationary spacetime, which are related with the retarded time.
Since these quantities can be determined by the causal past of temporal
observers at the AdS boundary, 
they give us not only practical but also physical manner to
characterize horizon formations.

\subsection{Definition of deconfinement}

In the static case, deconfinement or confinement phases 
is distinguished by seeing whether there is an event horizon in the
effective metric on the brane or not. 
However, such a naive definition cannot be
used in dynamical cases such as a phase transition from confinement to
deconfinement phases.
Since the AdS boundary is not in the causal future of the event horizon, 
boundary observers cannot know ``when'' the event horizon is formed.
To determine when the system is deconfined, 
we focus on the redshift factor instead of the event horizon and give a
practical definition of the deconfinement time.

We consider linear perturbations of the dynamically evolving
D7-brane, which correspond to meson excitations.
In case of sufficiently weak electric field ($E\ll m^2$) 
and slow time dependence ($\Delta V\gg m^{-1}$), 
the linear perturbations are coming and going between AdS boundary and 
the pole $\Phi=\pi/2$ of the brane, namely superpositions of the normal
modes with a discrete spectrum. 
On the other hand, in case of strong electric field ($E\gtrsim m^2$) or 
rapid time dependence ($\Delta V \lesssim m^{-1}$), the brane is strongly
bended and there can be a region which causes strong redshift on the brane.
Strong redshift means that the linear perturbations propagating for the
AdS boundary are trapped and spread out in the region 
and can not come back to the boundary for an extremely long time. 
Then, boundary observers feel that the meson has dissipated into the
background plasma.
Thus, we can identify the existence of strong redshift
on the brane with the deconfinement of mesons.
This definition is nothing but practical and physical notion of black
holes for temporal observers rather than formal and mathematical one.
As we mentioned, temporal observers can never know truly existence of black
holes and event horizons in principle.
They will only observe strong redshift.

Now, we define the {\textit{redshift factor}} which measures strength of
the redshift as follows.
We introduce a time-like vector field on the brane as
$\xi = \partial_V$, where we use coordinates $(V,z)$ defined by 
$V=V(u,v)$ and $z=Z(u,v)$. 
In term of $(u,v)$-coordinates, $\xi$ is written as
$\xi=J^{-1}(Z_{,v}\partial_u-Z_{,u}\partial_v)$, where $J$ is the
Jacobian: $J=V_{,u}Z_{,v}-V_{,v}Z_{,u}$.
The coordinate $V$ becomes ordinary time coordinate in the boundary
metric: $ds^2 = - dV^2 + d\vec{x}^2_3$.
In addition, $\xi$ is a (locally) Killing vector
in the brane effective metric before the electric field quench $V<0$. 
(See Eq.~(\ref{effmetstatic}).)
Therefore, $\xi$ gives us a natural time in both the boundary and
initial stationary regions, while it does not has any specific meaning
but one among time-like vectors in intermediate regions. 

A tangent vector of out-going null geodesic on the effective metric,
which is a null ray described by $u=\text{const.}$, is given by 
\begin{equation}
 k = \frac{d}{ds} = - \frac{C(u)}{\gamma_{uv}(u,v(s))}
  \partial_v ,
\end{equation}
where $C(u)$ is an integration constant associated with each null ray.
Note that $k$ is the eight-momentum of the out-going null ray, 
since this vector is affine parameterized in terms of $s$.  
The energy of the light ray for observers whose natural time is
represented by $\xi$ 
becomes 
\begin{equation}
\omega(v) \equiv - \gamma_{ab} k^a \xi^a
 = \frac{C(u)Z_{,v}(u,v)}{J(u,v)}.
\end{equation}

At an initial time $v=v_0$ and the AdS boundary $v=u$, the energy of the
light ray becomes 
\begin{equation}
 \omega(v_0)=\frac{mC(u)}{\Phi_{,u}(u,v_0)}\ ,\quad
 \omega(u)=\frac{C(u)}{V_{,v}(u,u)}\ .
\end{equation}
We have used for the former equation the static
solution~(\ref{exact_static}) at $v=v_0$ and for the latter equation the
boundary conditions at the AdS boundary: 
$V_{,u}=0$, $V_{,v}=2Z_{,u}$, and $Z_{,u}=-Z_{,v}$.
As a result, the redshift factor, which is the ratio between the energy
observed on the boundary and the initial energy, is
given by%
\footnote{
A past directed null geodesic from the AdS boundary may reach the pole
before the initial surface.
Then, we assume that the null geodesic is reflected at the pole and the AdS
boundary. After several reflections, it reaches the initial surface
eventually.
Taking into account the reflections, 
the expression for the redshift factor is modified as 
\[
R(u)=\frac{m}{2}\frac{V_{,v}(u,u)}{\Phi_{,u}(u-\pi n/2,v_0)}\ ,
\]
where we have used the coordinates defined in section~\ref{sec:bdrycond}.
The integer $n$ is chosen so that $0\leq u-\pi n/2\leq \pi/2$ is satisfied.
}
\begin{equation}
 R(u) 
  \equiv  \frac{\omega(v_0)}{\omega(u)}
  = \frac{m}{2}\frac{V_{,v}(u,u)}{\Phi_{,u}(u,v_0)}
.
\end{equation}
For supersymmetric embedding, $Z=m^{-1}\sin\Phi$, 
since this energy $\omega(v)$ becomes the Killing energy with
respect to the Killing vector $\partial_V$ and it should be conserved, we have $R=1$.
Also, the initial time $v_0$ can be taken arbitrary as far as stationary
regions.
Roughly speaking, this quantity represents how the energy of the light
ray emitted in the infinite past is red-shifted when the ray has arrived at
the boundary.
If $R(u)$ is infinity, such null ray cannot reach the boundary.
When an event horizon will be formed, $R(u)$ will tend to diverse by definition.

In our calculations, if this redshift factor $R(u)$ observed at the AdS
boundary is so large ($R > 100$), we shall say that the system becomes deconfinement phase.

\subsection{Definition of thermalization}

Because of the same reason as the deconfinement, 
the formation of horizons cannot be a good definition of the
thermalization in dynamical cases.
Here, in order to clarify thermalization in gravity side, we use the
Hawking temperature for time-dependent cases based on semi-classical arguments~\cite{Barcelo:2010xk,Kinoshita:2011qs}.
By using the redshift factor, we can define the following quantity  
\begin{equation}
 \kappa(u) \equiv \frac{1}{V_{,v}(u,u)}\frac{d}{du} \log R(u),
\label{eq:def_kappa}
\end{equation}
where $V_{,v}(u,u)$ denotes the normalization in terms of the boundary time.
This describes ``peeling property'' of out-going null geodesics, which
corresponds to ``surface gravity'' for the past horizon when initial state
is finite temperature.%
\footnote{
If the initial state is at a finite temperature, it means that the past
horizon should exist in gravity side.
In such cases we should define the redshift factor by using the Kruskal
time, which is natural initial time on the past horizon, instead of the Killing time.
}
One can find that, if evolutions of $\kappa(u)$ are sufficiently slow,
spectrum of the Hawking radiation becomes approximately thermal with the
temperature determined by $\kappa(u)$. 
When the system settles down in stationary, this temperature eventually 
agrees with the ordinary Hawking temperature associated with the Killing
horizon, of course.
Therefore, we shall define thermalization by saying that $\kappa(u)$ has
been close to the final temperature.

Intuitively, since the redshift factor represents the relation between
the natural times, 
the relation of creation-annihilation operators between initial state
and final one is determined by $\kappa(u)$.
In particular, situations of the current model are quite similar to
considering quantum fluctuations in the so-called moving mirror model.
This is because, for the fluctuations on the brane, the pole can be regard
as a mirror (in fact, we have imposed the Neumann boundary condition
there) and dynamics of the brane will cause this mirror to move effectively.
Thus, this surface gravity $\kappa(u)$ defined above 
is just the quantity which characterizes particle creations caused by
the moving boundary.

We note that $R(u)$ and $\kappa(u)$ are closely related but different
quantities. 
If a system settles down in the final steady state with the horizon,
$R(u)$ becomes exponentially very large and then $\kappa(u)$ becomes a constant value.
This implies that the mesons has been dissociated and the system has
been thermalized in the boundary theory. 
However, even if $R(u)$ becomes so large that the horizon (or naked singularities) would be formed,
$\kappa(u)$ does not always settle down.
Such cases can be interpreted as the phase in
which mesons are dissociated but non-thermalized.

In our calculations, we shall adopt $|\kappa-\kappa_f|/\kappa_f <0.01$ as
criteria for thermalization, where $\kappa_f$
is the final value of the surface gravity.

\section{Results for supercritical electric fields}
\label{sec:resultsuper}

In our setup, time evolutions of the D-brane are characterized by two
model parameters $(E_f/m^2, m\Delta V)$, which are a final value and a rise
time of the homogeneous electric field.
For the static case, there is a critical value of the electric field 
$E_\textrm{crit}=0.5754m^2$
below which only the Minkowski embeddings exist as shown in
section~\ref{sec:static}.
We study the time evolutions of the brane 
dividing the parameter space into two regions: 
supercritical electric field $E_f>E_\textrm{crit}$ and 
subcritical electric field $E_f<E_\textrm{crit}$.
We will show numerical results for supercritical electric fields in this
section and for subcritical ones in the next section.
In appendix~\ref{errorana}, we estimate error in our numerical calculations.

\subsection{Brane motion and boundary observable}

\begin{figure}
  \centering
  \subfigure[snapshots of the brane]
  {\includegraphics[scale=0.5]{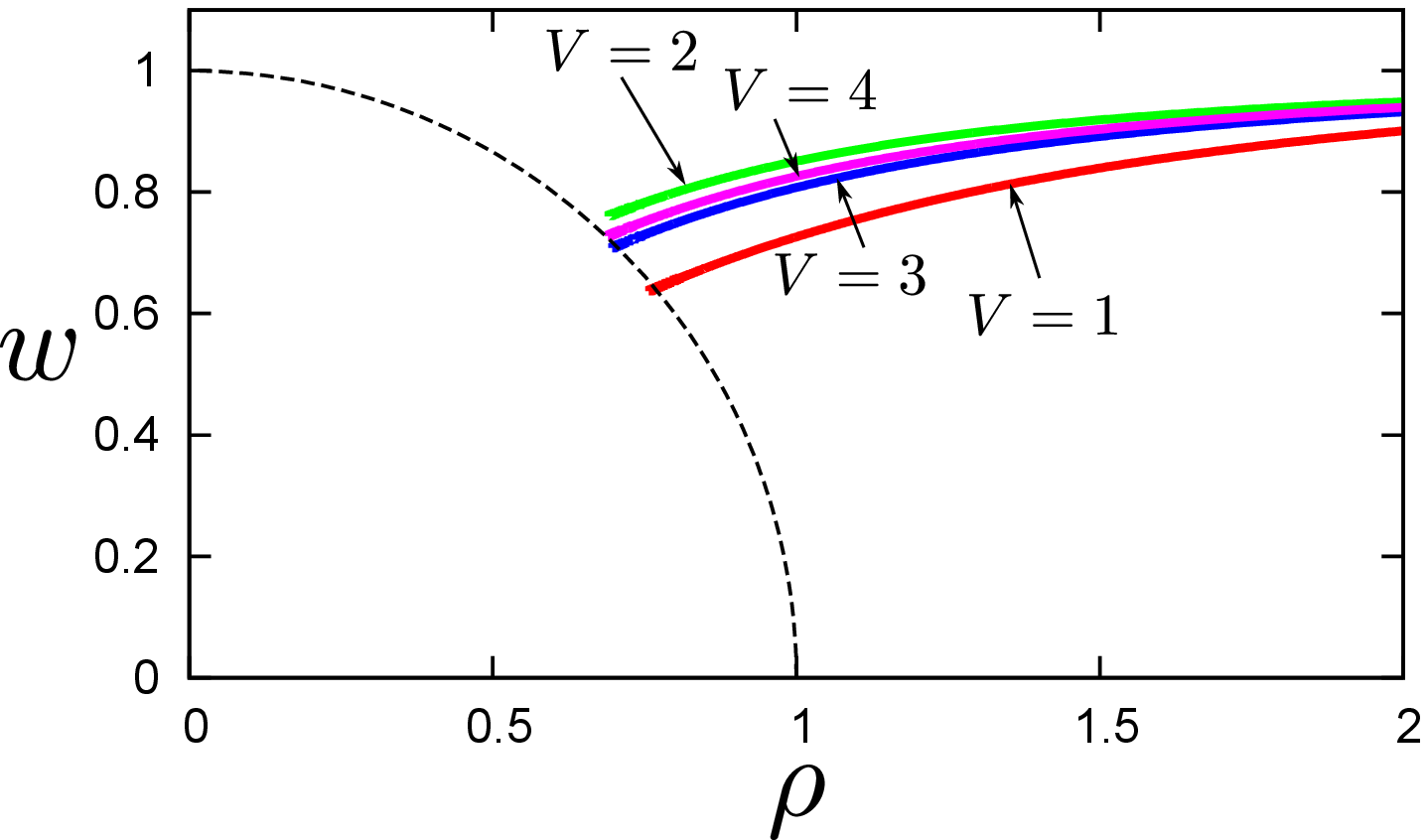}
  \label{fig:motion_snapshot}
  }
  \subfigure[effective horizons]
  {\includegraphics[scale=0.4]{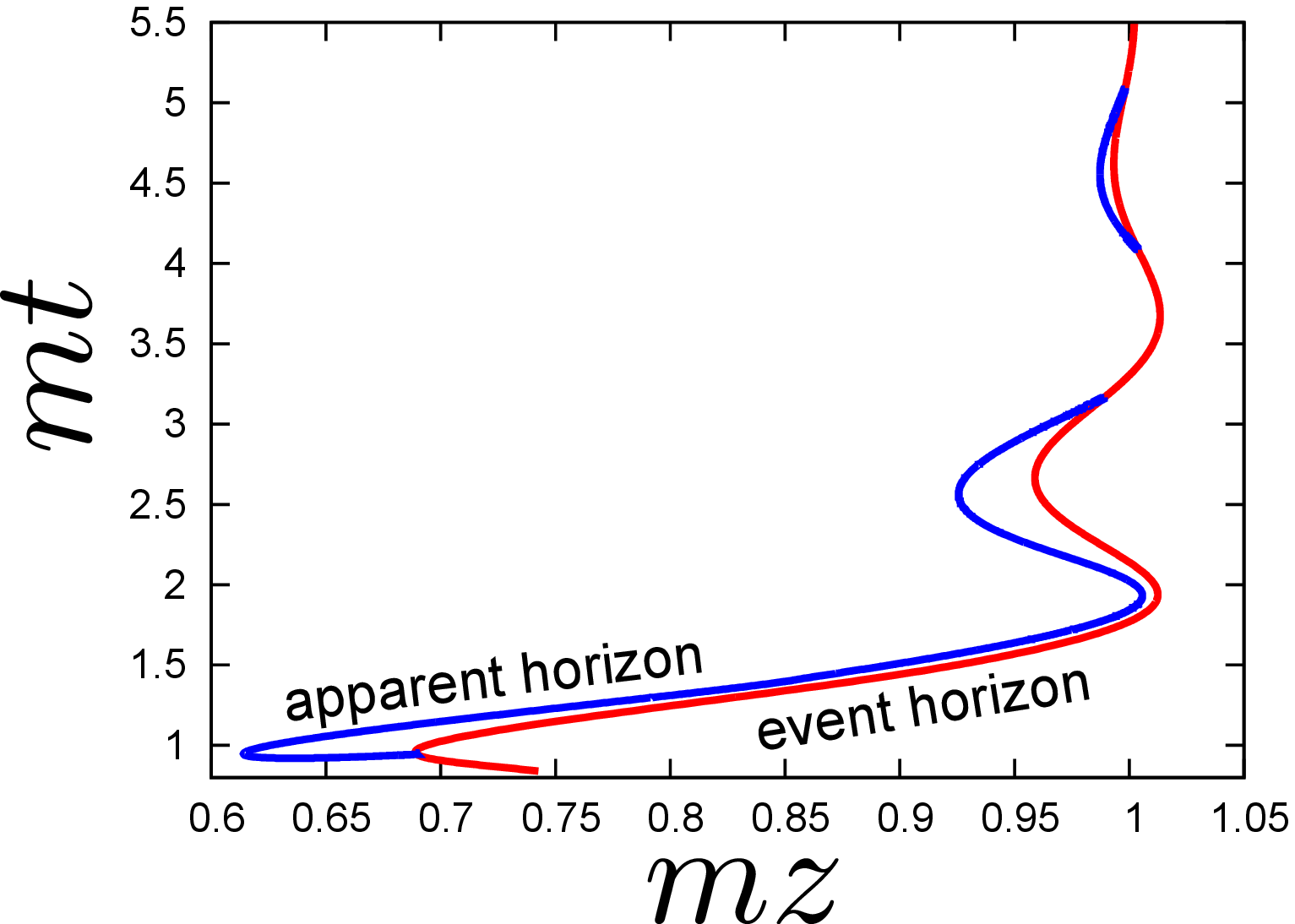}
  \label{fig:motion_horizon}
  }
  \caption{
(a)Snapshots of embeddings of the D7-brane in $(w,\rho)$-plane.
We take the unit of $m=1$ and set parameters as $E_f/m^2=1$ and 
$m \Delta V=0.5$. 
For static embedding, the effective horizon is located at 
$w^2+\rho^2=E=1$, which is shown by a dashed curve in this figure.
(b)Event and apparent horizons. 
The vertical axis is the ordinary time coordinate in the bulk: $t\equiv V+z$. 
}
\label{motion}
\end{figure}

Figure~\ref{fig:motion_snapshot} shows snapshots of the time evolution of the
D7-brane embeddings for
$E_f/m^2=1$ and $m\Delta V=0.5$. 
As vertical and horizontal axes, we have taken Cartesian-like coordinates, 
$w=z^{-1}\sin\phi$ and $\rho=z^{-1}\cos\phi$.
The dashed curve shows 
the effective event horizon for the static embeddings with the parameter
$E/m^2=1$. 
At the late time, the brane configuration tends to be static and
eventually coincides with the static black
hole embedding shown in section~\ref{sec:static}.

From the numerical solution, 
we can find the event and apparent horizons on the effective metric~(\ref{effmetric}). 
Since we are using double-null coordinates, the condition for the
apparent horizon is simply written as
\begin{equation}
 \partial_v\left[
\frac{\cos^3\Phi}{Z^3}\left(\frac{h_{uv}+Z^2f_{ux}f_{vx}}{h_{uv}-Z^2f_{ux}f_{vx}}\right)^{1/2}
\right]=0\ .
\end{equation}
Solving the above equation, we obtain the location of the apparent horizon
$u=u_\textrm{AH}(v)$. The event horizon is defined by the boundary of
the causal past of the AdS boundary. We denote the event horizon as
$u=u_\textrm{EH}$. (The $u_\textrm{EH}$ is a constant since the event
horizon is a null surface.)
In Fig.~\ref{fig:motion_horizon}, we show the locus of event and apparent horizons
in $(t,z)$-coordinates: $(t(u_\textrm{EH},v),z(u_\textrm{EH},v))$ and 
$(t(u_\textrm{AH}(v),v),z(u_\textrm{AH}(v),v))$.%
\footnote{
Strictly speaking, the location of the event horizon can not be
determined unless whole time evolution has been
known by the infinite future on the AdS boundary.
Since, however, we can solve time evolutions only during a
finite time by practical numerical calculations, 
we have approximately estimated the location of this event horizon by using the latest time
of the numerical calculation.
}
Here, $t$ is the ordinary time coordinate: $t\equiv V+z$. 
Note that these effective horizons on the brane worldvolume are different from bulk
ones. (Actually, there is no black hole horizon in the bulk since it is
pure AdS now.)
Especially, the effective event horizon is time like in the view of the
bulk metric and can be seen from the AdS boundary through the bulk null geodesic.
The event and apparent horizons intersect each other and 
the apparent horizon is outside the event horizon in several places.
This implies that the effective metric violates the null energy
condition. 
(Since the effective metric does not obey the Einstein equations, this
condition has just a geometrical meaning.)
Therefore, theorems in general relativity based on the null energy
condition, such as Hawking's area theorem, 
do not hold for brane dynamics.

\begin{figure}
  \centering
  \subfigure[quark condensate]
  {\includegraphics[scale=0.45]{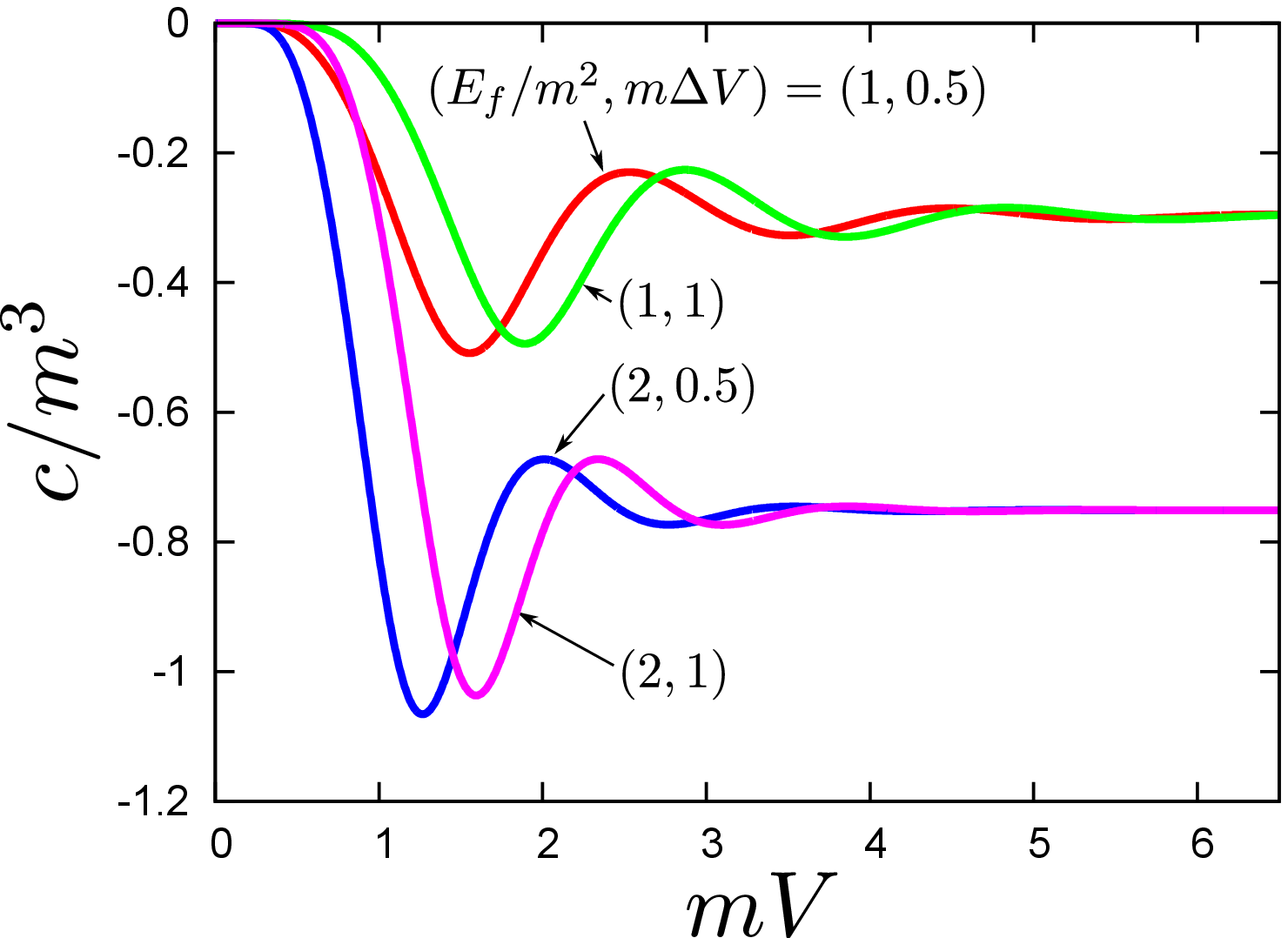}
  }
  \subfigure[electric current]
  {\includegraphics[scale=0.45]{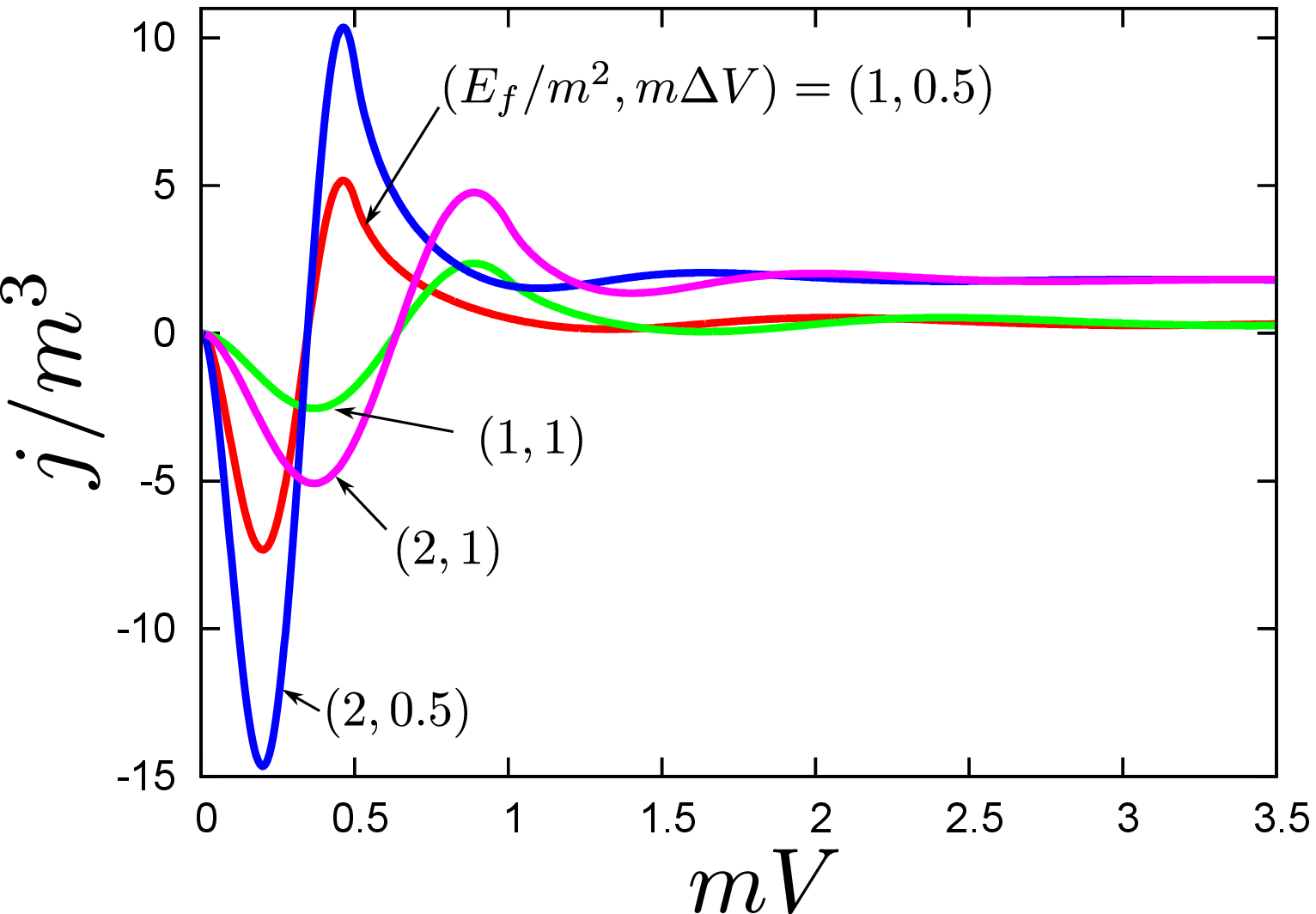} 
  }
  \caption{
Time dependence of quark condensate $c$ and electric current $j$
for $(E_f/m^2,m\Delta V)=(1,1), (1,0.5), (2,0.5)$, and $(2,1)$.
Note that while the electric field is time-dependent (namely 
$0<V<\Delta V$), the obtained value of $j$ has a slight uncertainty.
This is because unhealthy behavior of the equations of motion seems to
 affect numerical error.
\label{strongE}
}
\end{figure}

Now, we turn to quantities on the boundary theory.
Figure~\ref{strongE} shows quark condensate $c$ and
electric current $j$ as functions of boundary time $V$.
As typical examples for supercritical electric fields, 
we show our numerical results for 
$(E_f/m^2,m\Delta V)=(1,1), (1,0.5), (2,0.5)$, and $(2,1)$.
Although the quark condensate and the electric current oscillate at the first stage of the time evolutions, 
they approach constant
values at the late time.\footnote{It would be interesting to compare our maximum oscillation with
the universal scaling found in \cite{Das:2014jna} (see also \cite{Buchel:2013lla}).
However, our case is not conformal as we have the mass scale 
$m$.}
This means that the fluctuations on the brane have dissipated in the
effective event horizon.
For the static embeddings, 
the quark condensate and the electric current are given by 
$(c/m^3,j/m^3)=(-0.297,0.331)$ and $(-0.751,1.81)$ 
for $E/m^2=1$ and $2$, respectively.
We can confirm that these values coincide with the asymptotic values of
$c$ and $j$ for the dynamical cases.

\subsection{Thermalization and deconfinement time}

\begin{figure}
  \centering
  \subfigure[redshift factor]
  {\includegraphics[scale=0.45]{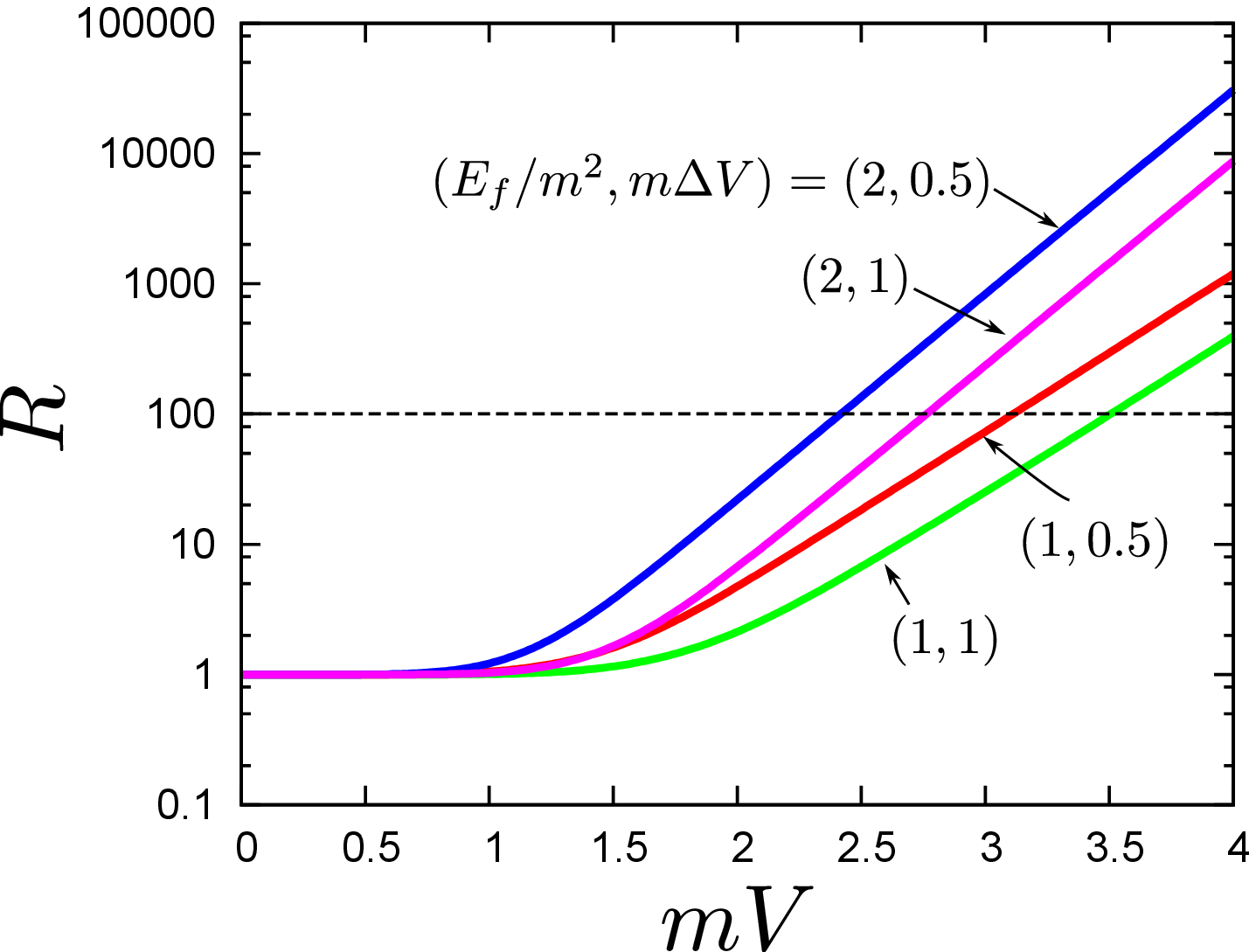}
  }
  \subfigure[surface gravity]
  {\includegraphics[scale=0.45]{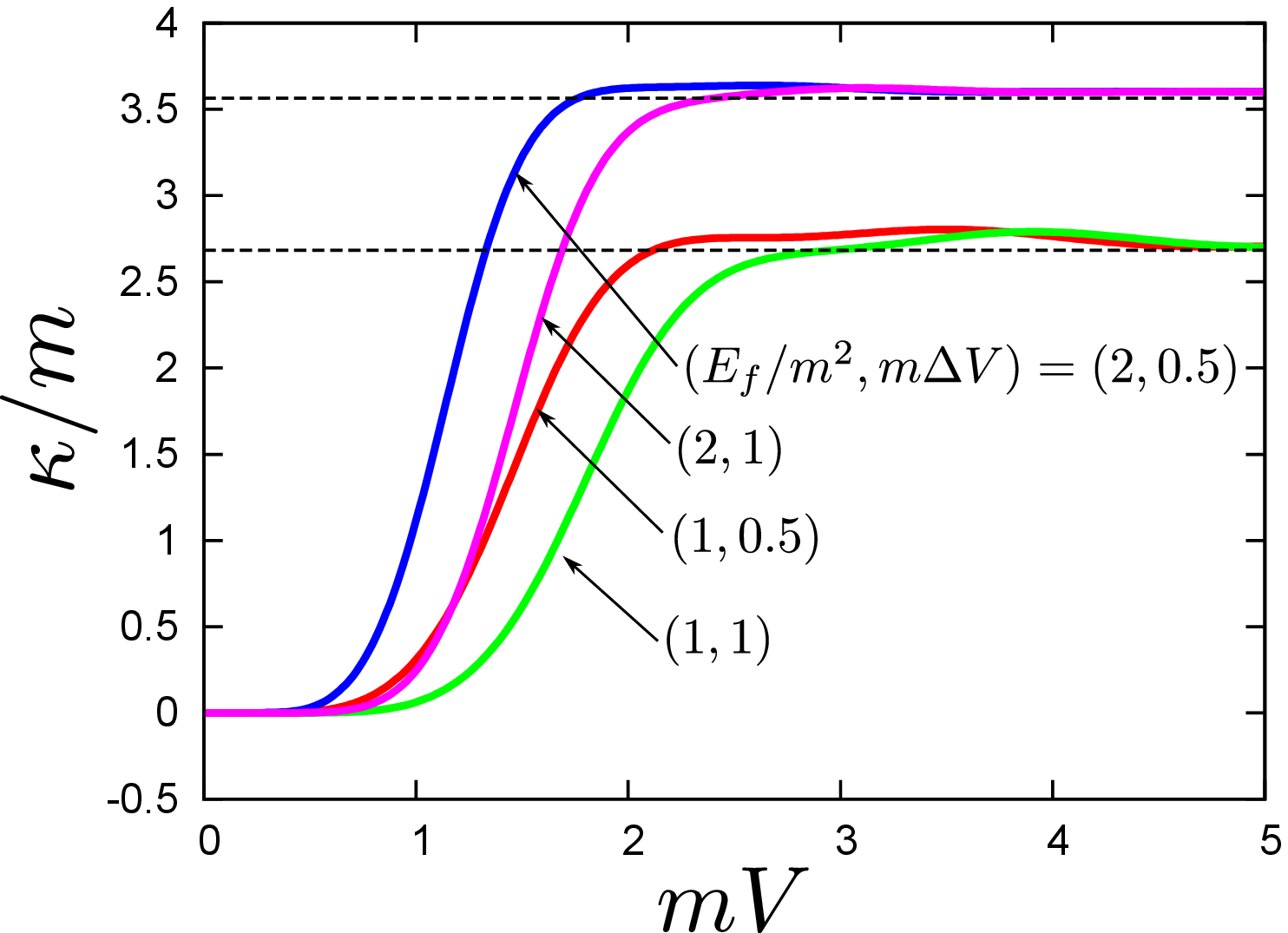} 
  }
  \caption{
Time dependence of redshift factor $R$ and surface gravity $\kappa$ 
for $(E_f/m^2,m\Delta V)=(1,1), (1,0.5), (2,0.5)$, and $(2,1)$.
\label{strongE_Rk}
}
\end{figure}

In this subsection, we study thermalization and deconfinement based on
the definitions introduced in section~\ref{defs}.
In Fig.~\ref{strongE_Rk}, we show the 
redshift factor $R$ and the surface gravity $\kappa$ as the boundary time $V$
for several parameters $(E_f/m^2,m\Delta V)=(1,1), (1,0.5), (2,0.5)$, and $(2,1)$.
In section~\ref{defs}, 
we have defined a criterion of the deconfinement as $R=100$ which is shown
by the horizontal line in Fig.~\ref{strongE_Rk}(a).
Since the redshift factors increase exponentially at the late time, 
they exceed the criterion and the systems change to deconfinement phases.
On the other hand, we have defined the thermalization by 
$|\kappa-\kappa_f|/\kappa_f <0.01$ where $\kappa_f$ is the final value of the
surface gravity.\footnote{
We evaluated $\kappa_f$ from static embeddings.} 
The criteria for $E_f/m^2=1,2$ are shown by 
horizontal lines in Fig.~\ref{strongE_Rk}(b).
We see that the systems have been thermalized at the late time.

\begin{figure}
  \centering
  \subfigure[thermalization time]
  {\includegraphics[scale=0.45]{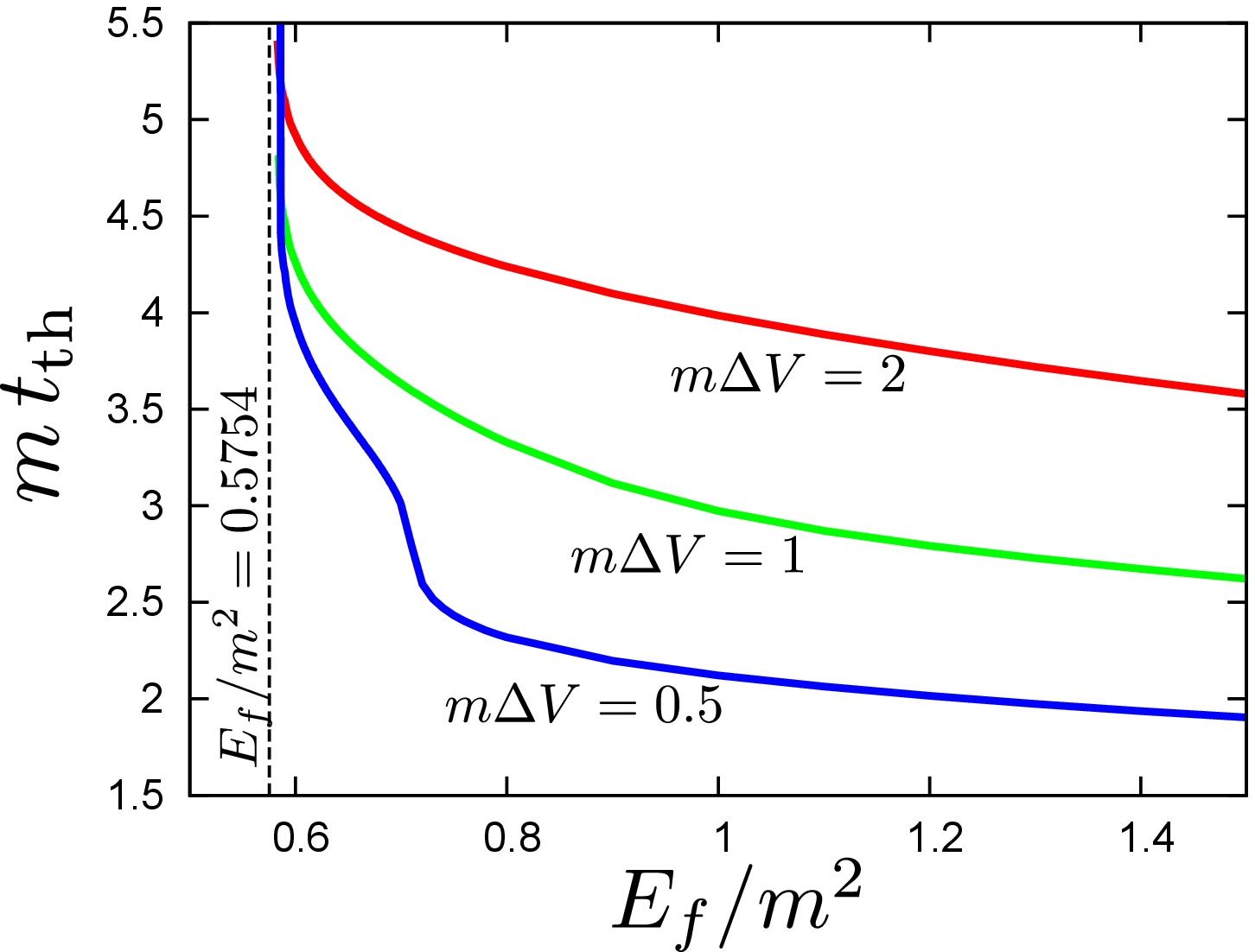}
  }
  \subfigure[deconfinement time]
  {\includegraphics[scale=0.45]{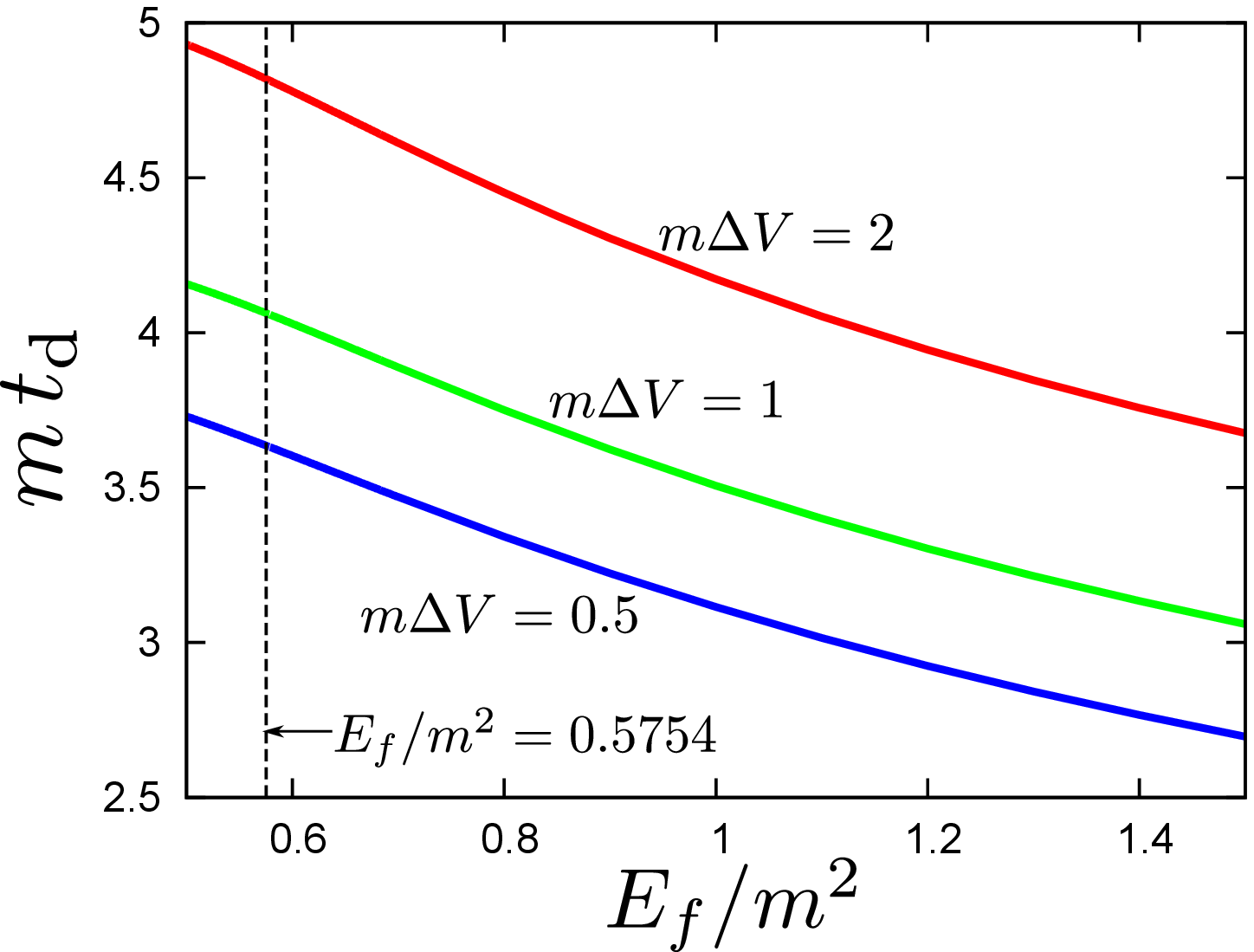} 
  }
  \caption{
Thermalization and deconfinement time as functions of $E_f$.
\label{strongE_tthd}
}
\end{figure}

In Fig.~\ref{strongE_tthd}(a), we show the thermalization time $t_\textrm{th}$
as a function of final value of
the electric field $E_f$.
At the critical value of the electric field $E_f=0.5754m^2 \equiv E_\textrm{crit}$, 
the thermalization time appears to diverge. 
This is natural behavior because, for $E_f<E_\textrm{crit}$,  
there are no static black hole embeddings and   
the system has never been thermalized.
In contrast, the thermalization time becomes small as $E_f$ increases.
This is because the brane fluctuations are damped by 
$\sim e^{-\kappa V}$ as general features for quasi-normal modes and, thus, 
we can estimate the thermalization time as 
$t_\textrm{th}\sim 1/\kappa \sim 1/\sqrt{E_f}$.\footnote{
Below Eq.~(\ref{kappaj}), we showed that 
the surface gravity is given by $\kappa\simeq (6E)^{1/2}$ for $E\to \infty$.
}
This is nothing but the Plankian thermalization time pointed out in Ref.~\cite{Hashimoto:2013mua}.

In Fig.~\ref{strongE_tthd}(b), we show the deconfinement time $t_\textrm{d}$
as a function of $E_f$.
The $t_\textrm{d}$ is finite at the critical electric field $E_f=E_\textrm{crit}$.
Furthermore, even for $E<E_\textrm{crit}$, it is conceivable that the system becomes deconfinement
phase if the system is dynamical. 
We will discuss the deconfinement below the critical electric field in detail
in the next section. 

\section{Results for subcritical electric fields}
\label{sec:resultsub}

\subsection{Quark condensate and electric current}
\label{subQJ}

\begin{figure}
  \centering
  \subfigure[quark condensate for $0\le V\le 10$]
  {\includegraphics[scale=0.4]{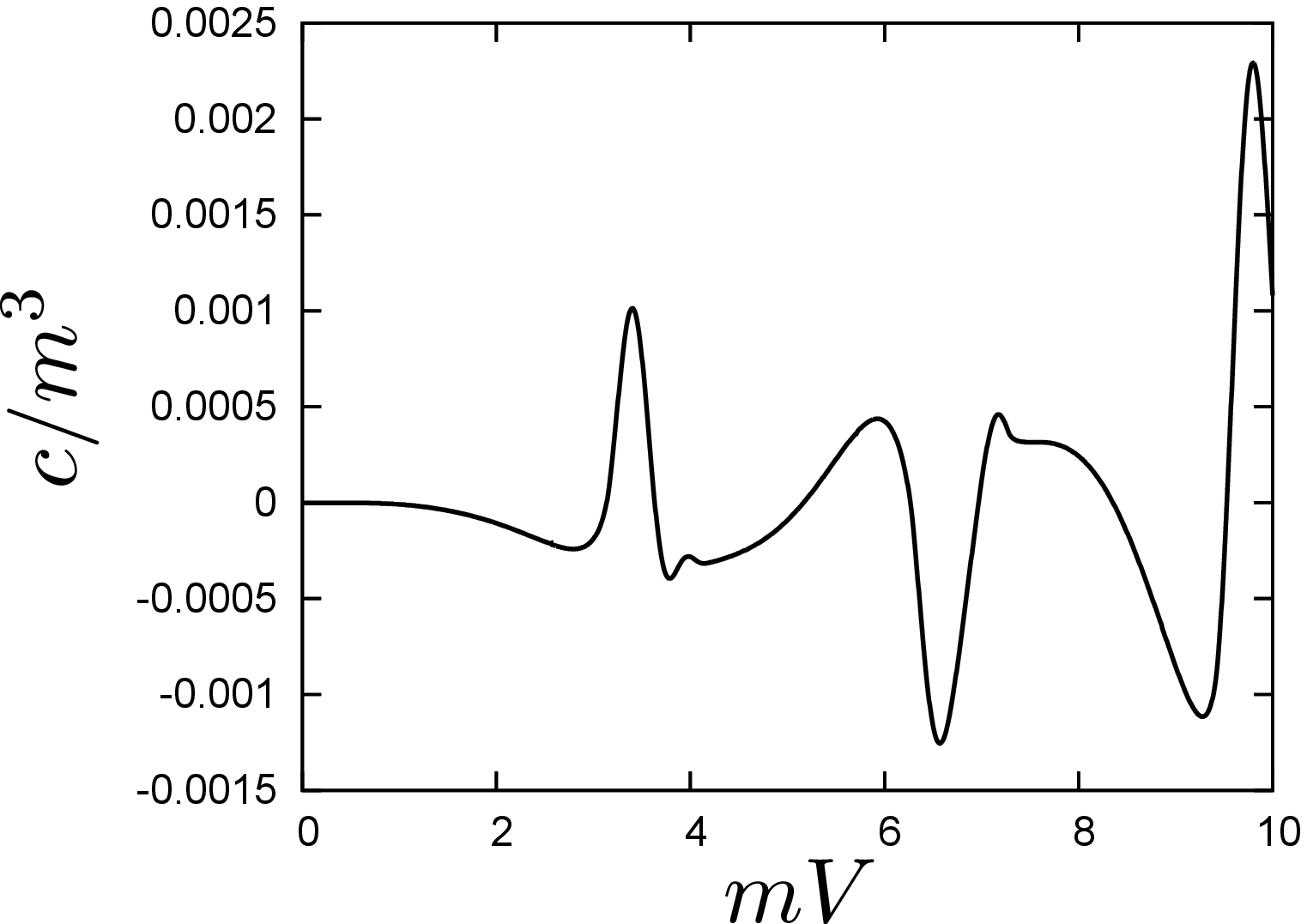}
   }
  \subfigure[electric current for $0\le V\le 10$]
  {\includegraphics[scale=0.4]{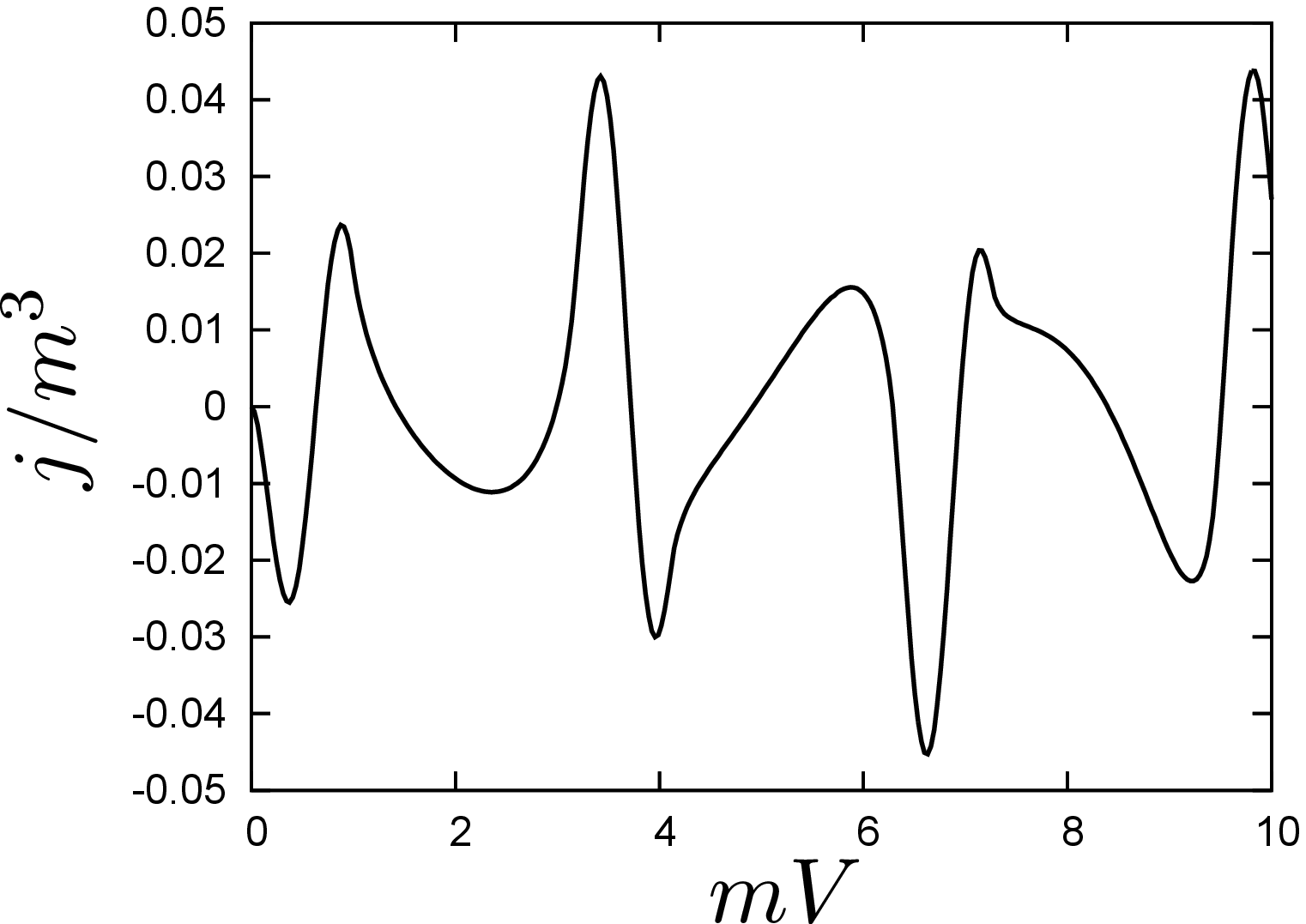} 
  }
  \subfigure[quark condensate for $0\le V\le 620$]
  {\includegraphics[scale=0.4]{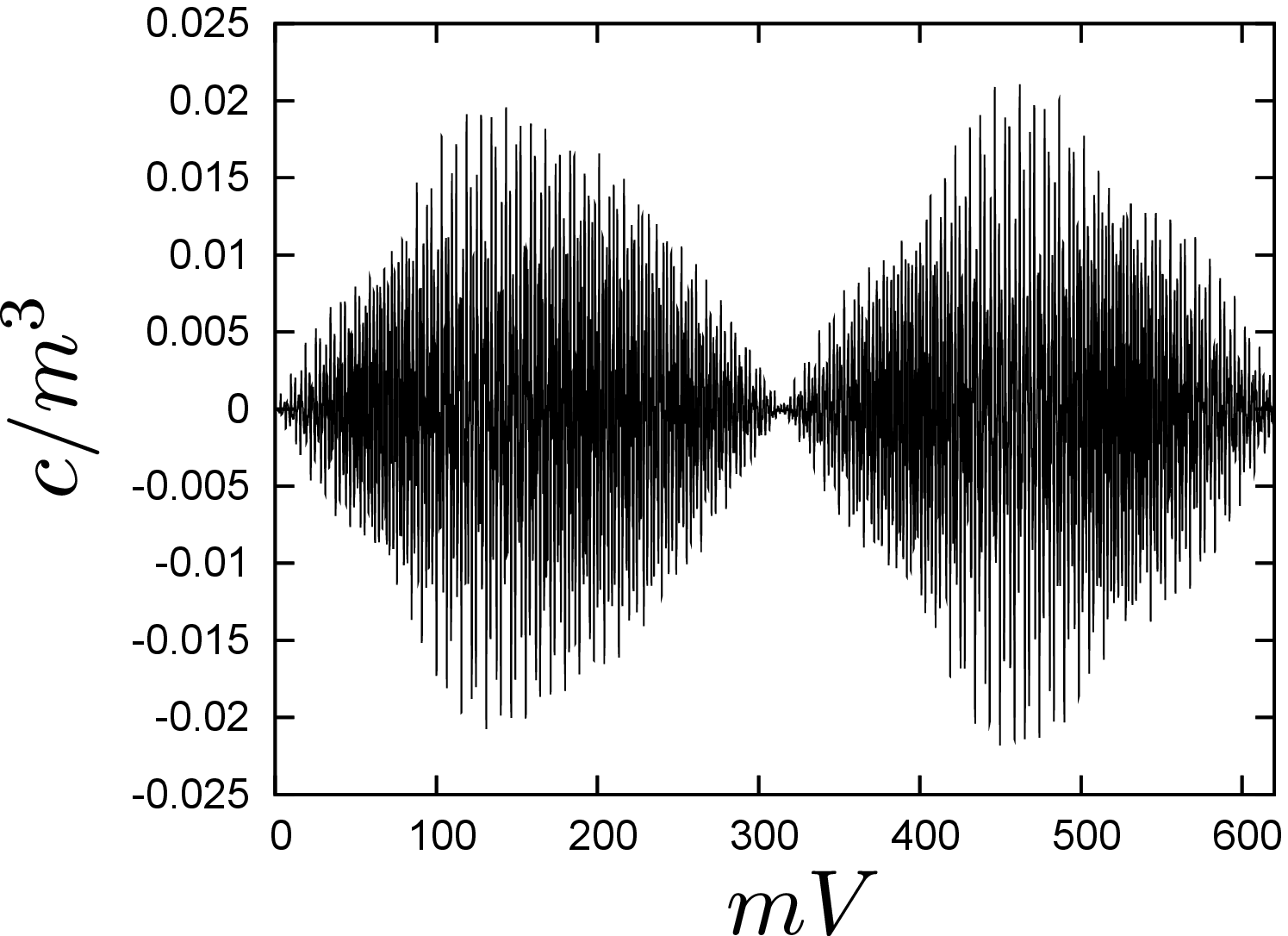}
  }
  \subfigure[electric current for $0\le V\le 620$]
  {\includegraphics[scale=0.4]{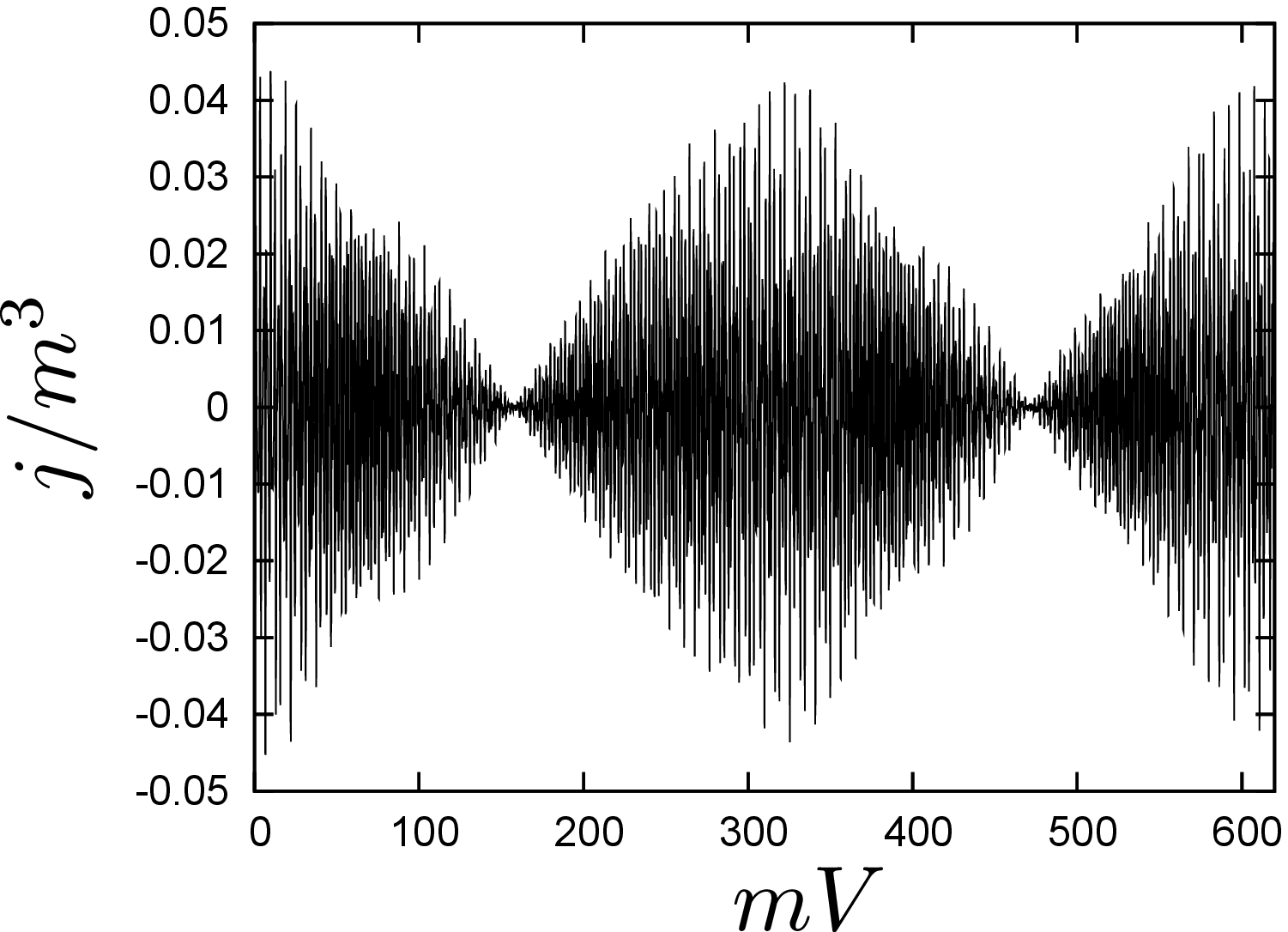} 
  }
  \caption{
Time dependence of quark condensate $c$ and electric current $j$ for
 $E_f/m^2 = 0.01$ and $m\Delta V = 1$.
\label{jc}
}
\end{figure}

In this section, we show numerical results for subcritical electric
fields. 
First, we study the quark condensate and electric current 
for $E_f/m^2=0.01$ and $m\Delta V=1.0$, in which the electric field is sufficiently weak.
Figure~\ref{jc} shows time dependence of the electric current $j$ and
the quark
condensate $c$. 
Figures (a) and (b) show an early stage of the time
evolution $0\le mV\le 10$, while (c) and
(d) show the time evolution over a long time 
$0\le mV\le 620$.

In the static case, only the
Minkowski embedding exists and 
the electric current is exactly zero for the electric field
$E/m^2=0.01$ 
as in Fig.~\ref{cj_rp=0_d=0}(b).
However, in dynamical cases, 
just after turning on the electric field ($V \ge 0$), 
the electric current starts to oscillate with a finite amplitude as well
as the quark condensate.
This corresponds to the oscillation of the bound state of quarks in the
boundary theory, that is polarization current. 
In our setup, this oscillation does not dissipate 
since the energy of the D-brane is conserved within the probe
approximation $N_f\ll N_c$.
This is nothing but a non-linear counterpart of the normal mode in linear perturbations. 

\begin{figure}
 \begin{center}
  \includegraphics[scale=0.6]{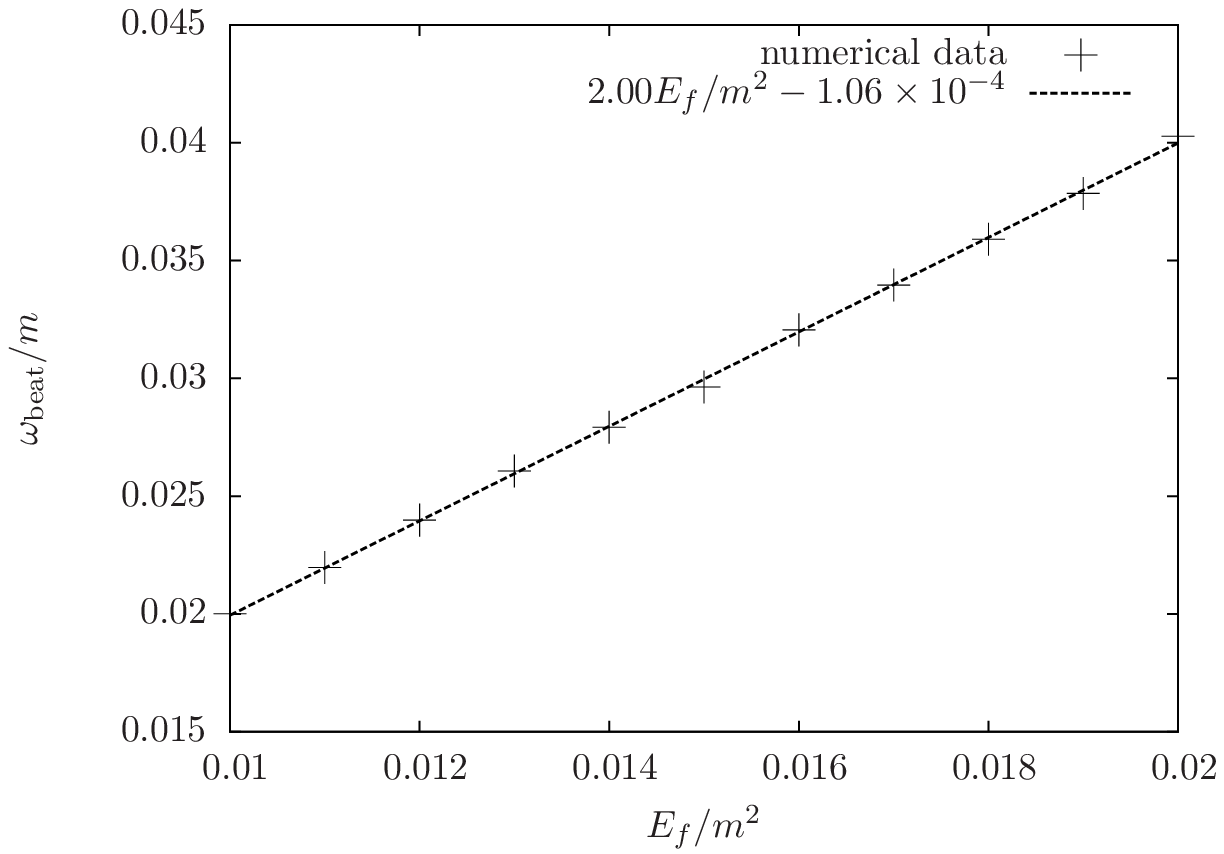}
 \end{center}
 \caption{Relation between the beat frequency $\omega_\mathrm{beat}$ and
 final value of the electric field $E_f$. Dashed line is obtained by
 linear fitting as $\omega_\mathrm{beat}/m = 2.00E_f/m^2 - 1.06\times 10^{-4} $.}
 \label{fig:beat_E_relation}
\end{figure}

The time evolution over a long period shown in Figs.~\ref{jc}(c) and
(d) reveals that there are beats with the oscillations
for both of the quark condensate and the electric current. 
The beat represents the energy exchange between the brane fluctuation 
$\Phi(V,Z)$ and the gauge potential on the brane $a_x(V,Z)$ because the
phase of each beat is opposite.
In the case of zero electric field, they are regarded as coherent
oscillations of scalar and vector mesons,%
\footnote{
From Eq.~(\ref{VZPpara}), we find
$\partial^\mu A_\mu=0$ ($\mu=V,x_1,x_2,x_3$) for $a_u=a_v=0$.
Thus, the oscillation of $a_x$ represents the excitation of a
vector meson.
}
whose mass spectra degenerate~\cite{Kruczenski:2003be}.
Therefore, the beat represents the mixing of scalar and vector mesons 
caused by the presence of the external electric field.
Figure~\ref{fig:beat_E_relation} shows the beat frequency 
$\omega_\textrm{beat}$, which is defined based on period of each node of
the envelope, 
for several values of the electric field $E_f$.
By linear fitting we can find $\omega_\mathrm{beat} \simeq 2.0 E_f /m$. 
This implies the mass spectra for the scalar and vector mesons split
because of the Stark effect and then it 
results in the mass difference $\delta M \simeq \omega_\textrm{beat}$.
In appendix~\ref{Stark},  
we evaluate the shifts of spectra 
for a weak electric field and find $\omega_\textrm{beat}=2E_f/m$, analytically.
This is consistent with our numerical results and the perturbative
calculation is so reasonable at least for $E/m^2\lesssim 0.02$.

\subsection{Deconfinement below the critical electric field}
\label{subsec:Decbelow}

Now, we focus on moderate subcritical electric fields.
In such cases we can observe fascinating phenomena characteristic of
dynamical situations.
As we mentioned, for $E_f<E_\textrm{crit}$, the system will never be
thermalized since no static black hole embedding exists as the final state. 
In fact, in Fig.~\ref{strongE_tthd}(a), 
we have seen that the thermalization time
appears to diverge at the critical electric field.
However, this does not mean that deconfinement is impossible below the
critical electric field.

\begin{figure}
\begin{center}
\includegraphics[scale=0.5]{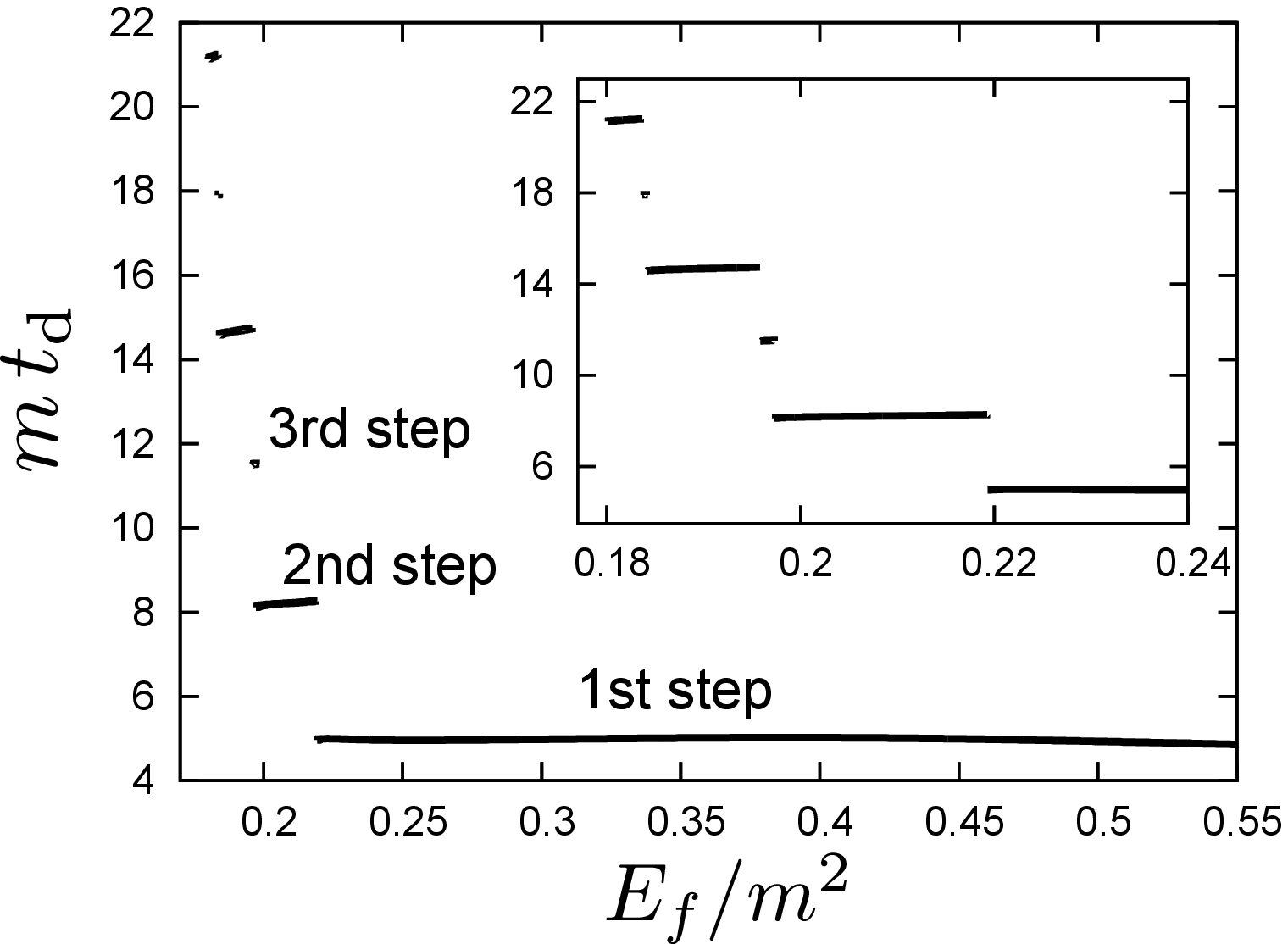}
\end{center}
\caption{
Deconfinement time $t_\textrm{d}$ against electric field $E_f$
for $m \Delta V=2$.
It is a discrete function and  almost constant at each step.
From the bottom, we refer to each step as 1st step, 2nd step, and so on.
}
 \label{td_desc}
\end{figure}

In Fig.~\ref{td_desc}, we plot the deconfinement time $t_\textrm{d}$ as a function of $E_f$
for $m\Delta V=2$ and  $E/m^2 < 0.55$.
We can find that 
the $t_\textrm{d}$ is a discrete function of $E_f$ 
and almost constant at each step, which is referred to as 1st step, 2nd step, and so on.
To understand this curious behavior in the deconfinement time in terms
of the brane dynamics,  
we define a scalar quantity on the brane worldvolume, 
$s \equiv \gamma^{ab} h_{ab} = 4(\gamma_{uv} + h_{uv})/\gamma_{uv}$, and 
investigate the time evolution of $s$ evaluated at the pole $\Phi=\pi/2$.
Note that, since the stress tensor of the brane is proportional to
$\gamma^{ab}$, we can interpret $s|_{\Phi=\pi/2}$ as a rough indication of the energy
density at the pole. 

\begin{figure}
  \centering
  \subfigure[a scalar quantity $s$ at the pole]
  {\includegraphics[scale=0.45]{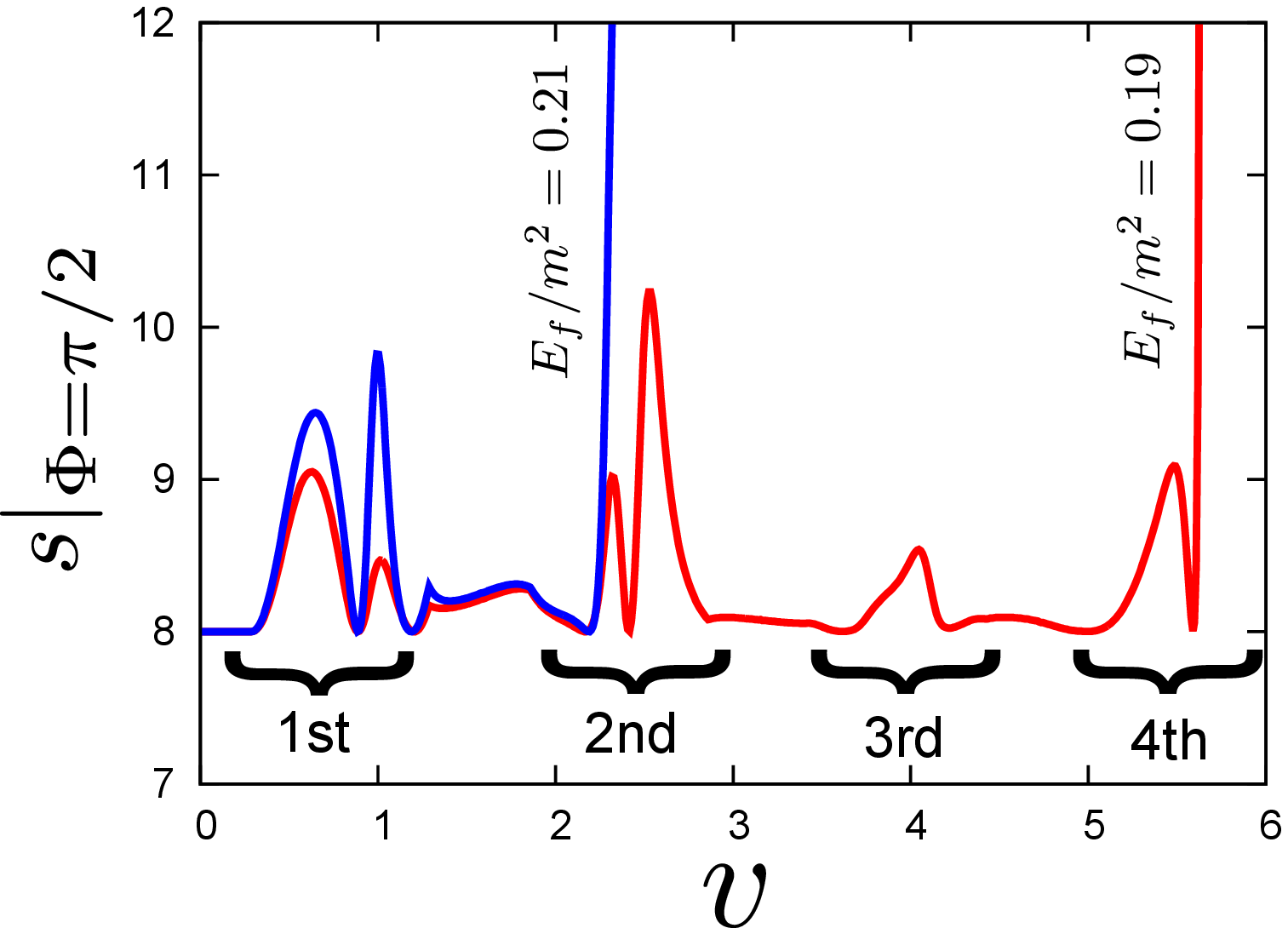}
  }
  \subfigure[redshift factor]
  {\includegraphics[scale=0.47]{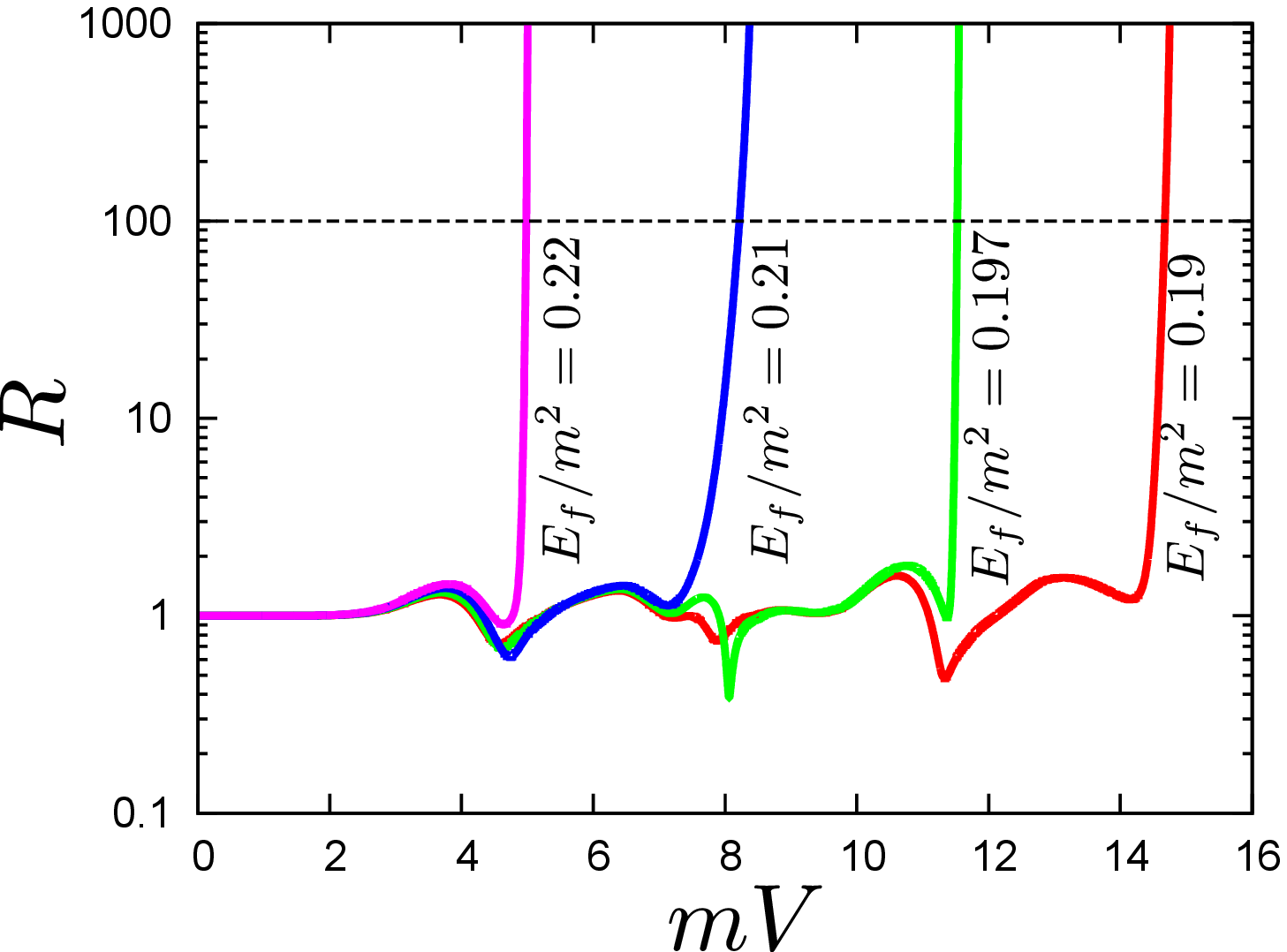} 
  }
  \caption{
(a)Time dependence of the scalar quantity $s$
at the pole for $E_f/m^2=0.21,0.19$ and $m\Delta V=2$.
They are on 2nd and 4th steps, respectively.
(b)Time dependence of the redshift factor $R$ for 
$E_f/m^2=0.22,0.21,0.197,0.19$ and $m\Delta V=2$.
They are on 1st, 2nd, 3rd, and 4th steps, respectively.
}
\label{FF}
\end{figure}

\begin{figure}
\begin{center}
\includegraphics[scale=0.5]{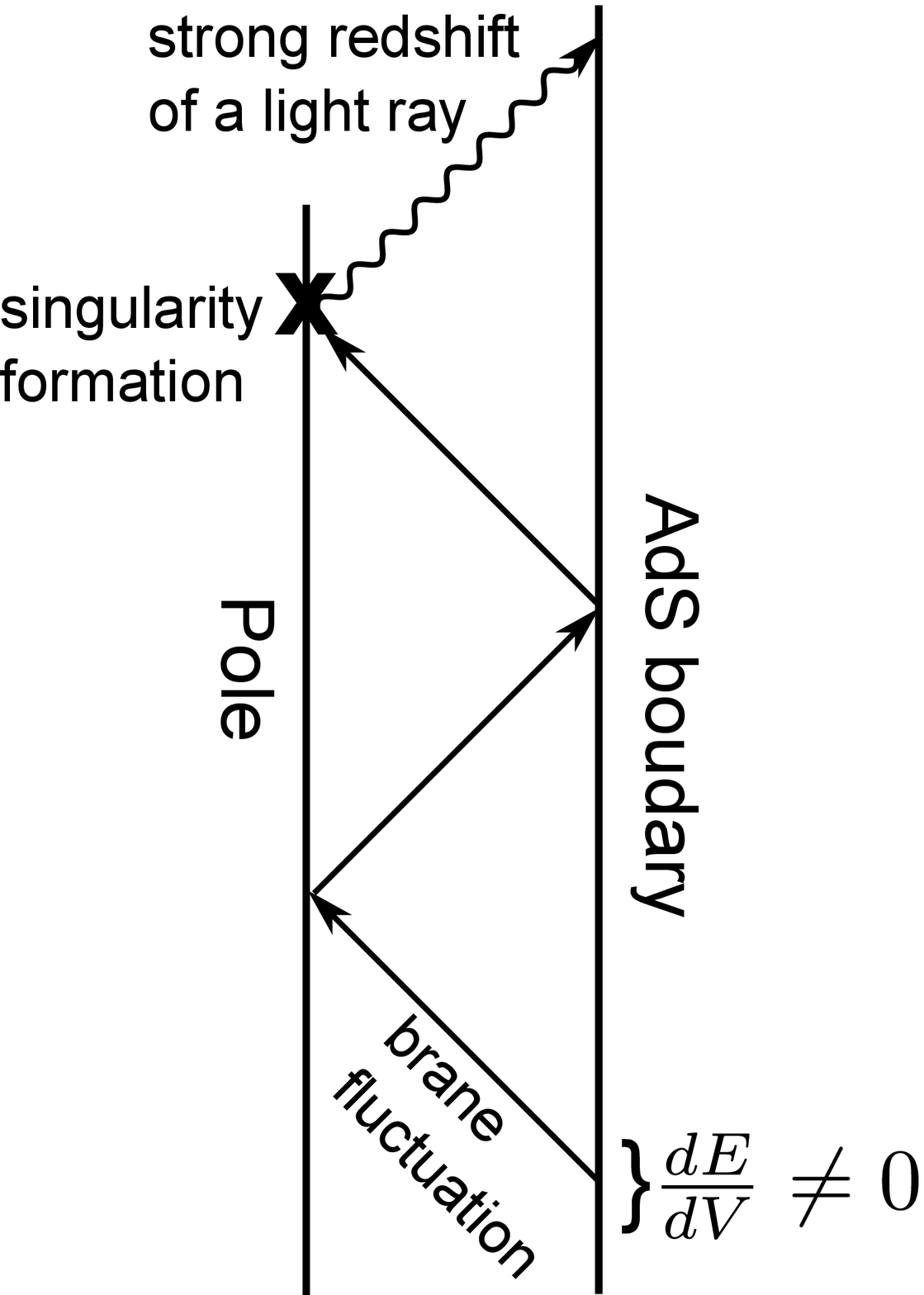}
\end{center}
\caption{
A Penrose diagram of the brane worldvolume. 
A brane fluctuation is injected from the AdS boundary for $0\le V \le \Delta V$
 since the electric field is time dependent there. 
It is reflected at the pole and the AdS
boundary several times.
Eventually, it collapses at the pole and appears to form a naked singularity.
(The number of bounces depends on the parameters, $E_f$ and $\Delta V$.)
A light ray going through near the singularity is strongly redshifted.
}
 \label{zigzag}
\end{figure}

In Fig.~\ref{FF}, we plot $s|_{\Phi=\pi/2}$ as a function of the brane
coordinate $v$ for $E_f/m^2=0.21$, $0.19$ and $m\Delta V=2$. 
These correspond to 2nd and 4th steps in Fig.~\ref{td_desc}.
We can see that pulses are localized in several time intervals which are 
shown by 1st, 2nd, 3rd and 4th in the figure.
This is because the fluctuation on the brane caused by turning on the
electric field is reflected at both sides of the AdS
boundary and the pole. 
It propagates between these boundaries several times.
For $E_f/m^2=0.21$, 
when the wave packet comes to the pole for the second time,
the scalar quantity seems to be diverging. 
On the other hand, for $E_f/m^2=0.19$,
it seems to be diverging when the wave packet comes to the pole for the
fourth time.
The divergence of the scalar quantity implies the appearance of a
singularity on the brane.
Figure~\ref{zigzag} gives a schematic picture of this behavior.
This behavior is similar to the weakly turbulent instability of AdS spacetime:
AdS is non-linearly unstable under arbitrarily small
perturbations~\cite{Bizon:2011gg}.
Detailed analysis of the ``weakly turbulent instability of D-brane'' and
its implication for the field theory 
will be discussed elsewhere.

We can expect that, when the $s$ becomes large, the brane is strongly bended and a
region which causes strong redshift will appear on the brane. 
Actually, in Fig.~\ref{FF}, 
we find that the redshift factor diverges at the same time as the divergence
of $s$ in retarded time.\footnote{
When the scalar quantity $s|_{\Phi=\pi/2}$ diverges at $(u,v)=(u_0,u_0-\pi/2)$, the retarded time is
defined by $V(u_0,u_0)$ at the AdS boundary.
}
It turns out that the divergence of the redshift factor $R$ is extremely
rapid within a finite boundary time rather than the exponential
divergence in the case where the system can be thermalized.
This implies that the singularity is naked and not hidden by an event
horizon.
Since almost only the number of bounces determines the divergence of the
redshift factor, the deconfinement time is discrete and almost constant at each step.
It takes $v=\pi/2$ for one round trip in the worldvolume coordinate. 
If static embeddings with zero electric field, we have $V(v,v)=2v/m$.
Thus, the difference of the deconfinement time for each step can be
roughly estimated as $\Delta t_\textrm{d}\simeq \pi/m$.

\begin{figure}
\begin{center}
\includegraphics[scale=0.6]{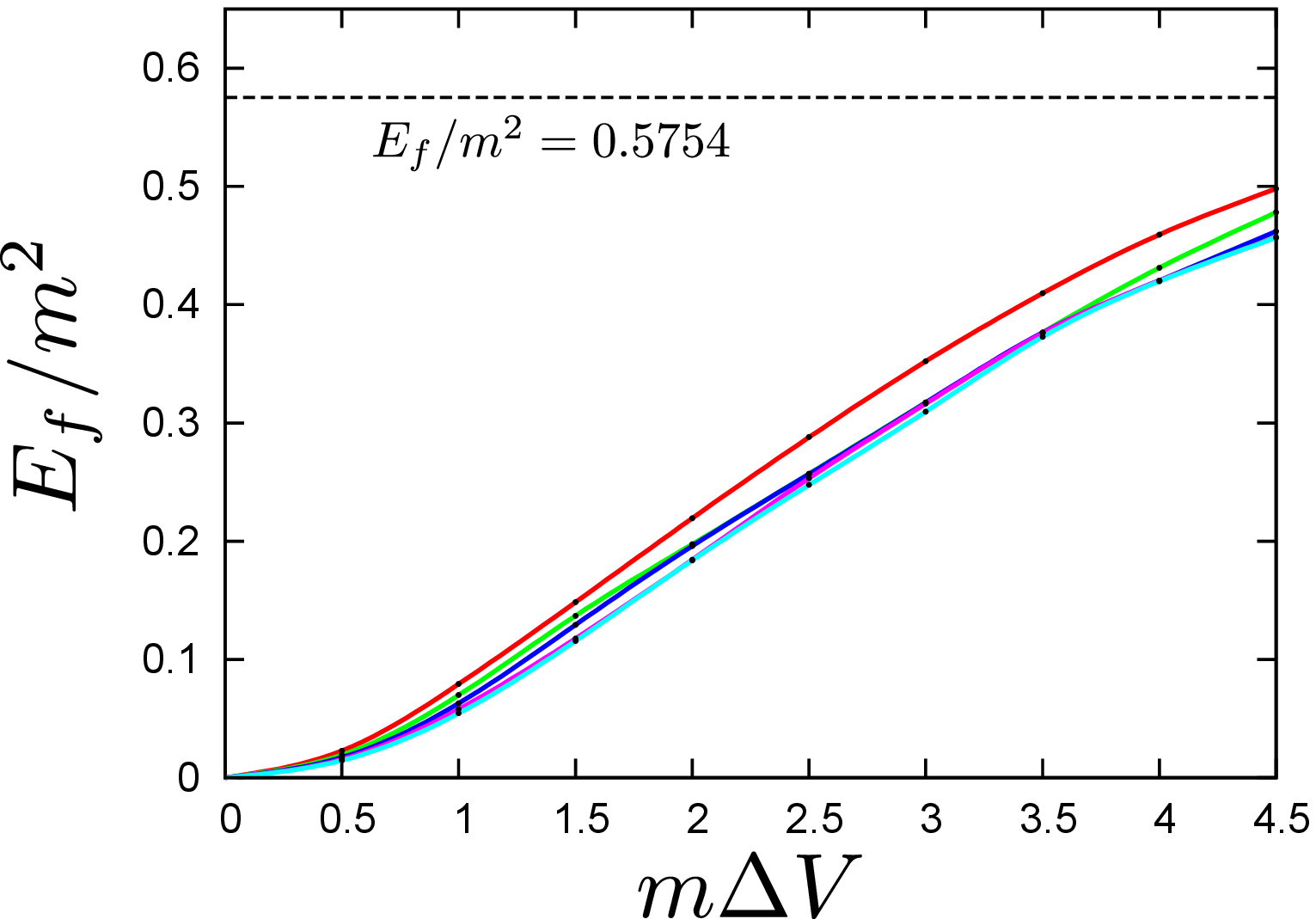}
\end{center}
\caption{
Parameter space of time-dependant solutions of D7-brane.
These curves represent the boundaries of the 
number of bounces needed for the formation of naked singularity.
Black points are our numerical data. 
We have interpolated them by spline curves passing through the origin. 
$E_f/m^2 = 0.5754$ (dashed line) is the critical electric field in the static case.
}
 \label{bouncenum}
\end{figure}

The number of bounces needed for the divergence of $s$ 
depends on parameters $E_f$ and $\Delta V$.
We examine its dependence on two parameters $(E_f/m^2,m\Delta V)$ 
and summarize the result 
in Fig.~\ref{bouncenum}.
Each curve represents the boundary of the 
number of bounces needed for the formation of the naked singularity.
For example, 
above the red curve, the singularity is formed when the wave packet
 reaches the pole at the first time. 
Between the red and green curves, it is formed at the second time. 
(They correspond to the 1st and 2nd steps in Fig.~\ref{td_desc}.)
Note that below the light blue curve we can successively find and draw many
curves.
Practically, we can only perform numerical calculations over a limited
period of time and with a limited resolution.
It is not so clear whether the region below the light blue curve is filled with an
infinite number of the curves or there is a threshold below which the
instability does not occur.%
\footnote{If one takes into account backreactions beyond the probe
approximation, the energy of the oscillations on the
brane will dissipate via emitting closed strings over a long period.
In such case, the instability after a huge number of the bounces may be
physically irrelevant.
}
(For the parameters shown in Fig.~\ref{jc} of the previous subsection,
we have not observed any evidence that a singularity forms at least
within $0\le mV\le 620$.)
However, since areas surrounded by those curves become too narrow to
distinguish each one, we have omitted drawing them in the figure.
As $m\Delta V$ becomes larger, the curves will approach asymptotically to the
critical line $E_\mathrm{crit}/m^2 = 0.5754$ which is the critical
electric field in the static case.
This is because large $\Delta V$ means the electric field is introduced
adiabatically and then the deconfinement transition may occur near the
static critical value.
Note that for large $\Delta V$ dynamics of the brane will begin to depend on the
profile of $E(V)$.
Although precise orbits of the curves shown in the figure might not be universal,  
qualitative behavior should not change.

Using the observables in boundary theory, 
vertical and horizontal axis of Fig.~\ref{bouncenum} are
written as
\begin{equation}
 \frac{E_f}{m^2}=\sqrt{\frac{\lambda}{2\pi^2}}\frac{\mathcal{E}}{m_q^2}
=8\sqrt{\frac{2\pi^2}{\lambda}}\frac{\mathcal{E}}{m_\textrm{gap}^2}\ ,\qquad
 m\Delta V=\sqrt{\frac{2\pi^2}{\lambda}}m_q \Delta V
=\frac{\sqrt{2}}{4}m_\textrm{gap}\Delta V\ ,
\end{equation}
where $m_\textrm{gap}\equiv 4\pi m_q \lambda^{-1/2}$ is mass gap in
$\mathcal{N}=2$ SQCD~\cite{Kruczenski:2003be}. 
In according to the RHIC experiment, 
we set parameters as 
$\mathcal{E}/m_\textrm{gap}^2\sim 0.02$ and 
$m_\textrm{gap} \Delta V\sim 0.4$~\cite{Voronyuk:2011jd}.
Then, we obtain $E_f/m^2\sim 0.6/\sqrt{\lambda}$ and $m\Delta V\sim 0.1$.
From Fig.~\ref{bouncenum}, the system can be in deconfinement phase 
at least for $E_f/m^2\gtrsim 0.01$.
Therefore, 
our result indicates that,
if the 't~Hooft coupling satisfies $\lambda\lesssim 10^3$,
the system can be in deconfined phase even though it is not thermalized 
in RHIC experiment.

\begin{figure}
  \centering
  \subfigure[quark condensate]
  {\includegraphics[scale=0.4]{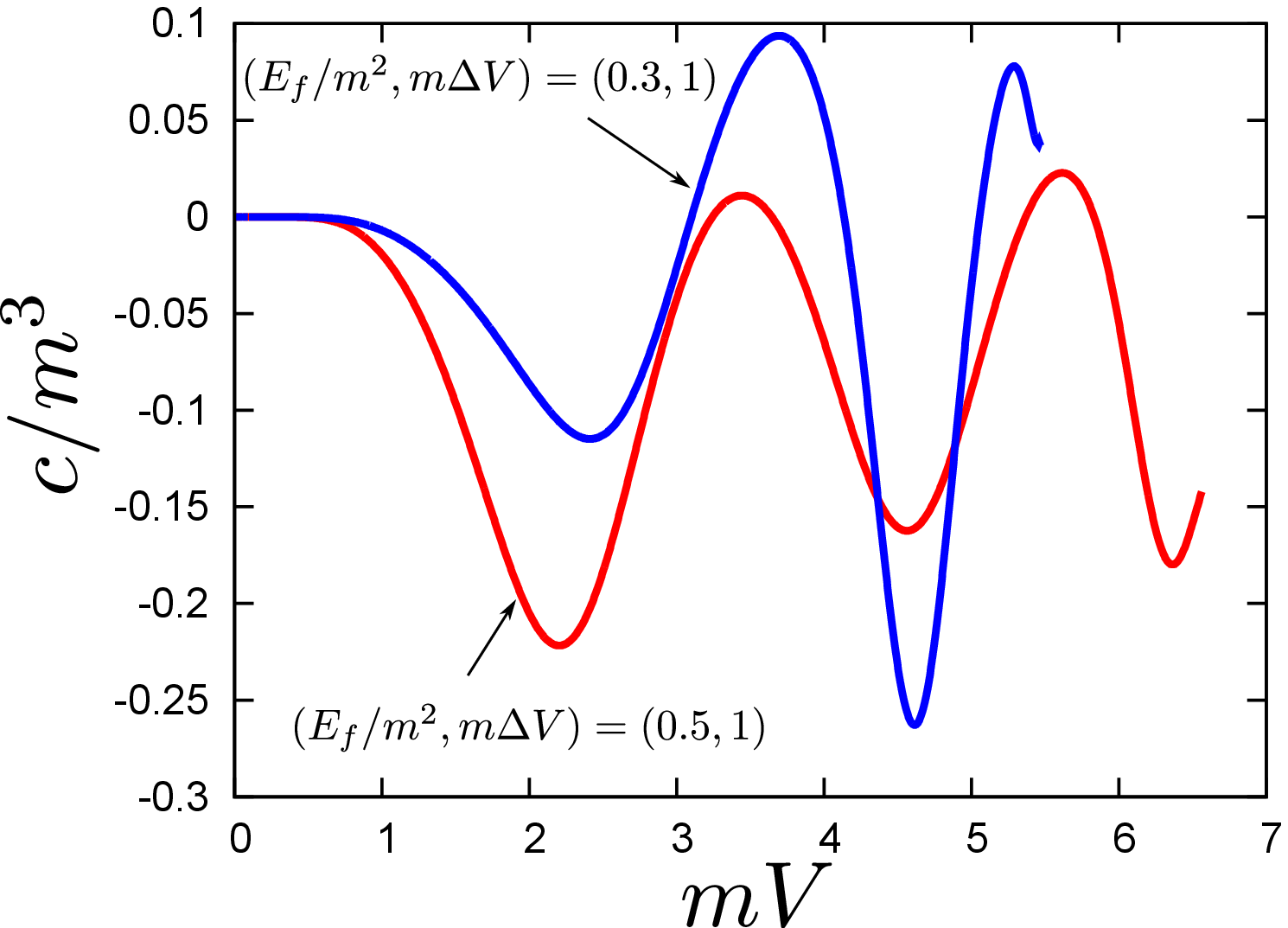}
  }
  \subfigure[electric current]
  {\includegraphics[scale=0.4]{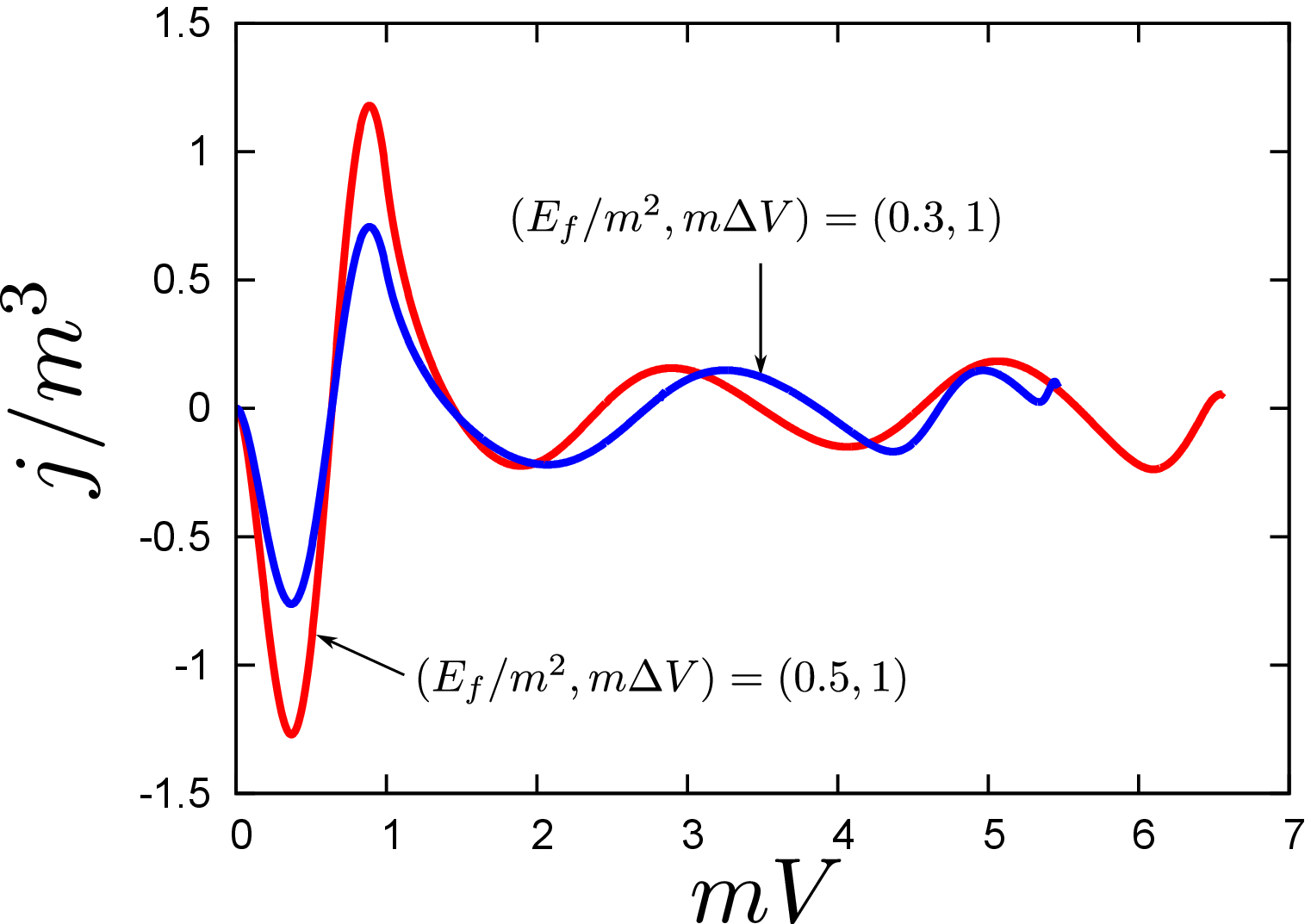} 
  }
  \caption{
Time dependence of quark condensate and electric current $j$
for $(E_f/m^2,m\Delta V)=(0.5,1)$ and $(0.3,1)$.
\label{cjisami}
}
\end{figure}

Now, we set parameters as $E_f/m^2=0.3, 0.5$ and $m\Delta V=1$.
For these parameters, the system will be deconfined although the
electric field is below the critical value. 
Figure~\ref{cjisami} shows time dependence of the electric current $j$ and quark
condensate $c$. 
They oscillate and does not converge. 
As shown before, 
the scalar quantity $s$ tends to be diverging within a finite time, 
while, in $c$ and $j$, we do not find any singular behavior. (The
right ends of the curves in Fig.~\ref{cjisami} correspond to the time of
divergence of $s$ in retarded time.)
This is, presumably, because the singularity near the pole will form
suddenly (the divergence of the redshift factor is extremely rapid).
As a result, these observables in the boundary theory does not respond
and remain finite.
However, the singularity, in which we have to take into account the
various effects beyond the current probe approximation, is naked, namely
visible from the AdS boundary.\footnote{The fast oscillation of the brane in the
target spacetime is T-dual to a D-brane with electric field on it. The open string
metric on the D-brane with large electric field shows a peculiar property (emergent
Carrollian metric) where the light cone collapses \cite{Gibbons:2002tv}, 
and it would be related to our deconfinement.}
We can expect to observe interesting phenomena such as quantum effect on the brane,
backreaction to the bulk spacetime, and so on.

For example, in order to estimate quantum effect on the brane, 
let us consider minimally coupled massless field on
the ($1+1$)-dimensional part of the brane effective metric.
We introduce two kinds of null coordinates $u$ and $U$, which are 
retarded times 
to define positive frequency modes in a final state and an initial
state, respectively.
Then, assuming the initial state does not have any out-going flux,
the expectation value of the stress tensor of the massless field is given by
\begin{equation}
\langle T_{uu} \rangle \sim - \frac{1}{24\pi}\{U,u\}
 = \frac{1}{48\pi} 
\left[
\left(\frac{U''}{U'}\right)^2 - 2\left(\frac{U''}{U'}\right)'
\right] , 
\end{equation}
where $\{U,u\}$ is the Schwarzian derivative and the prime denotes
$u$-derivative.
(See Ref.\cite{BIRRELL}, for example.)
Recalling the fact that the redshift factor is relation between the
initial time and the final time associated with out-going null geodesics, one can find $R(u) = 1/U'(u)$ and
$\kappa(u) = - U''(u)/U'(u)$. 
Thus, we have out-going flux of particle creation as 
$\langle T_{uu} \rangle \sim (\kappa^2 + 2\kappa')/(48\pi)$.
When the effective horizon is formed ($\kappa(u)\sim \mathrm{const.}$), 
this out-going flux leads to thermal flux of the final steady state.
On the other hand, when the naked singularity emerges, it may blow up
because the divergence of $R(u)$ is extremely rapid.

\section{Conclusion and discussions}
\label{sec:conclusion}

In this paper, we analyzed response of the strongly coupled gauge theory against
an electric field quench, by using the AdS/CFT correspondence. The system is ${\cal N}=2$
supersymmetric QCD with ${\cal N}=4$ super Yang-Mills as a gluon sector,
and has a confining spectrum for the meson sector (while the gluon sector is
always deconfined). We turn on the electric field in a time-dependent manner,
and find that the system develops to a deconfinement phase of mesons.

We have studied time-dependent behavior of various observables such as
electric current carried by the quarks and the quark condensate. We
have defined the thermalization time scale and the deconfinement time
in terms of the gravity dual side: the thermalization is with the Hawking temperature,
and the deconfinement is with the strong redshift. 

Among our findings, the most interesting is the fact that the deconfinement transition
of the mesons occurs even with a small electric field once it is applied time-dependently.
In the static electric field, there exists a critical value of the electric field
beyond which the electric current flows and the system is deconfined. 
In our time-dependent quench, if the quench is made sufficiently fast,
even with a final electric field which is smaller than the critical value, the system
goes to a deconfinement phase --- there appears a strong red shift region in the
gravity dual. See section \ref{sec:resultsub} for details.

In the dual gravity picture, this phenomena can be understood as the
D-brane version of the weakly turbulent instability~\cite{Bizon:2011gg}:
The wave packet on the D-brane is getting sharp as time increases and,
eventually, collapses into the naked singularity.
Accordingly, we also found a curious behavior of the deconfinement time --- the
time scale when a strong redshift region appears on the D7-brane. The deconfinement 
time takes only discrete values, see Fig.~\ref{td_desc}. 

We also found that when the applied electric field is small enough, the deconfinement 
transition does not occur within a practical time-scale , but there appears a beat frequency which dictates 
the energy inflow-outflow between the chiral condensate and the electric current, 
see Fig.~\ref{jc} (c) and Fig.~\ref{jc} (d).
Each corresponds to the scalar fluctuation and the gauge fluctuation on the D7-brane.
The beat frequency is found to be proportional to the electric field value.
This fact can be well explained by the analytic formula of the mass
splitting for the Stark effect. 

Our findings are of course a consequence of the analyses performed in the gravity dual side,
and they wait for possible interpretation in the gauge theory side. It is encouraging that
even with a small electric field, if it is applied sufficiently fast, it leads to a deconfinement
phase. Its implication to heavy ion collision experiment would be important.

\begin{figure}
  \centering
\includegraphics[scale=0.4]{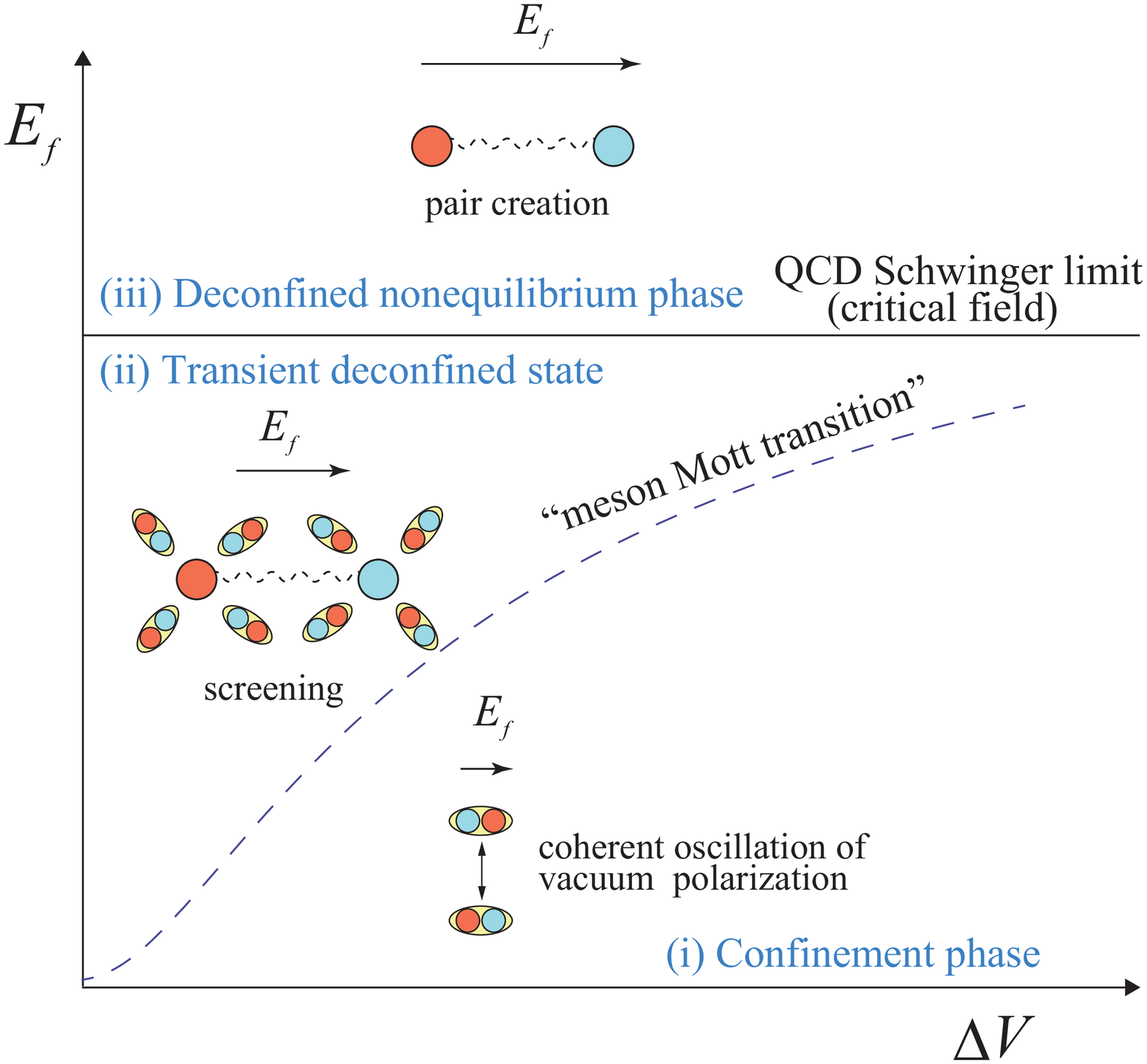}
  \caption{
Schematic "dynamical phase diagram" of states realized in the 
present study by a static electric field $E_f$
following an initial ramp (parametrized by the time parameter $\Delta V$). 
See text for details.
}
\label{summary}

\end{figure}

Furthermore, the potential implication of the present study of 
nonequilibrium dynamics in QCD to strongly correlated 
electron system is suggestive. 
In condensed matter, nonequilibrium dynamics 
of correlated electrons induced by strong electric fields 
is being intensively studied 
experimentally \cite{Taguchi:2000,Iwai:2003,Wall:2011,Suzuki:2012}
and theoretically \cite{Oka:2003,Eckstein:2009,Oka:2012,Aoki:2014,Yoshioka:2011,Moskalenko:2000}. 
Strong Coulomb interaction between electrons 
can freeze the electrons' motion leading to an insulating state 
known as the Mott insulator\cite{Imada:1998}. 
Charge excitations, called doublons and holons, 
are energetically forbidden in this phase. 
By applying very strong static \cite{Taguchi:2000} or pulse
\cite{Iwai:2003,Wall:2011} electric fields, one can 
break the insulating state by creation of charge excitations. 
If the field is not strong enough, the created charges
may be bounded by the attractive force and form excitons, i.e., 
pairs of plus and minus charges. 
Excitons do not carry direct electric current 
and the system is insulating. 
However, there is an old and interesting proposal:
``When the density of the excitons exceeds a 
critical value, the attractive force becomes screened and the 
excitons become dissolved leading to a plasma of 
charged particles''. 
This transition is called the exciton Mott transition
(or crossover) \cite{Mott:1961,Mott:1968,Zimmermann1978,Yoshioka:2011}
and was recently observed experimentally \cite{Suzuki:2012}. 
The excitons in condensed matter can be related to mesons
in the present system. Then, it is tempting to speculate that the formation
of naked singularity explained in the previous section
is an indication of the ``meson Mott transition'',
i.e., the QCD version of the exciton Mott transition. 
We plot a schematic phase diagram obtained by this analogy 
in Fig.~\ref{summary} with three regions (i), (ii), and (iii).
\begin{description}
\item[(i) Confinement phase with coherent oscillation]
When the field is weak, the system is always in the confinement phase. 
However, when the ramp speed is fast (small $\Delta V$), 
the field induces a coherent oscillation of
vacuum polarization due to meson excitation. 
The field during the ramp (\ref{Efunc}) 
can be considered as a pulse field with
a frequency parameter $\Omega=2\pi/\Delta V$. 
When $\Omega$ is comparable with the 
meson energy, (multi-)photon absorption process becomes 
possible \cite{Brezin:1970,Popov:1972, Oka:2012} 
and leads to excitation below the critical field. 
We note that a similar oscillation of current was 
observed in a condensed matter model \cite{Eckstein:2009}.

\item[(ii) Transient deconfined phase (``meson Mott transition'')]
This is the speculated ``meson Mott transition'' regime.
When the meson amplitude becomes large, 
the confinement force becomes relatively weak due to screening.  
The quarks become liberated and deconfinement takes place in the 
meson (quark) sector 
\footnote{There is a difference between the 
present situation compared to previous 
theories of exciton Mott transition, e.g., \cite{Yoshioka:2011}. 
The latter is typically considered in a 
static state, i.e., finite density gas of excitons in equilibrium,
while our system experience a coherent oscillation of the mesons. 
The coherent oscillation accelerates the 
deconfinement since the dynamics leads to energy dissipation and heating. 
}.
The dashed line that separates this region with (i) 
corresponds to the infinite bounce limit of Fig.~\ref{bouncenum}. 

Since the field is below the critical field, the static solution 
obtained by adiabatically introducing the field ($\Delta V\to \infty$)
is in the confinement phase. 
Thus, we expect that the plasma state realized by 
the meson Mott transition is transient. In the long time limit, 
pair annihilation of quarks dominates and the plasma disappear.  
Detailed time evolution in this region is still unclear and 
is an interesting future problem. 

\item[(iii) Deconfined nonequilibrium phase above QCD Schwinger limit]
When the electric field is stronger than the 
confining strength (= QCD Schwinger limit), the confinement phase becomes 
unstable against direct pair creation of quark and antiquarks
\cite{Hashimoto:2013mua}.
This state is a static nonequilibrium phase with 
finite current \cite{Karch:2007pd}. 

\end{description}

In summary, by studying the dynamics of 
supersymmetric QCD in strong electric fields,
we observed many interesting, and universal nonequilibrium physics. 
Our finding implies similarities between possible formation mechanism 
of quark gluon plasma in heavy ion collision experiments
to laser induced phase transitions in condensed matter, which 
helps us understand the physics more deeply.

\section*{Acknowledgments}

We are grateful to Takahiro Tanaka for helpful discussions.
K.H. would like to thank Rikkyo university for hospitality.
K.H. and T.O. are supported by KAKENHI 
(Grant No. 23654096, 24224009, 26400350).
This research was
partially supported by the RIKEN iTHES project.


\appendix

\section{Equations of motion from the DBI action}
\label{app:EOM_DBI}

 In this appendix, we will summarize general features of the equations of
 motion from the DBI action.
 
 The DBI action for D$p$-brane is 
 \begin{equation}
  S_{\mathrm{D}p} = \int d^{p+1}\sigma \sqrt{-\det(M_{ab})}, 
 \end{equation}
 where $M_{ab} = h_{ab} + f_{ab}$.
 For convenience we adopt the following abbreviated notation for
 describing matrices: 
 \begin{equation}
  \mathbf{M} = \mathbf{h} + \mathbf{f}, \quad
   \mathbf{M}^{-1} = \mathbf{h}^{-1} - \mathbf{h}^{-1}\mathbf{f}\mathbf{h}^{-1}
    + \mathbf{h}^{-1}\mathbf{f}\mathbf{h}^{-1}\mathbf{f}\mathbf{h}^{-1}
    - \cdots ,
 \end{equation}
 \begin{equation}
  {}^t\mathbf{M} = \mathbf{h} - \mathbf{f}, \quad
   {}^t\mathbf{M}^{-1} = \mathbf{h}^{-1}
   + \mathbf{h}^{-1}\mathbf{f}\mathbf{h}^{-1}
    + \mathbf{h}^{-1}\mathbf{f}\mathbf{h}^{-1}\mathbf{f}\mathbf{h}^{-1}
    + \cdots ,
 \end{equation}
 where $\mathbf{h}$ is symmetric and $\mathbf{f}$ is anti-symmetric.

 The symmetric part of $\mathbf{M}^{-1}$ is 
 \begin{equation}
  \begin{aligned}
   \boldsymbol\gamma^{-1} =& (\mathbf{M}^{-1} + {}^t\mathbf{M}^{-1})/2 \\
   =& \mathbf{h}^{-1}
    + \mathbf{h}^{-1}\mathbf{f}\mathbf{h}^{-1}\mathbf{f}\mathbf{h}^{-1}
   + \mathbf{h}^{-1}\mathbf{f}\mathbf{h}^{-1}\mathbf{f}\mathbf{h}^{-1}
   \mathbf{f}\mathbf{h}^{-1}\mathbf{f}\mathbf{h}^{-1} + \cdots\\
   =& (\mathbf{I} + \mathbf{X}^2 + \mathbf{X}^4 + \cdots)\mathbf{h}^{-1},
  \end{aligned}
 \end{equation}
 where we have defined $\mathbf{X} \equiv \mathbf{h}^{-1} \mathbf{f}$.
 Shortly, we obtain  
 \begin{equation}
  \begin{aligned}
   \boldsymbol\gamma = \mathbf{h} ( \mathbf{I} - \mathbf{X}^2)
   = \mathbf{h} - \mathbf{f}\mathbf{h}^{-1}\mathbf{f}, 
  \end{aligned}
 \end{equation}
 and some relations between determinants of them as 
 \begin{equation}
  \det \boldsymbol\gamma = \det \mathbf{h} \det (\mathbf{I} - \mathbf{X}^2)
   = \det \mathbf{h} [\det (\mathbf{I} + \mathbf{X})]^2 ,
 \end{equation}
 \begin{equation}
  \det \mathbf{M} = \det \mathbf{h} \det (\mathbf{I} + \mathbf{X})
   =  \det \mathbf{h} \det (\mathbf{I} - \mathbf{X}) .
 \end{equation}
 As a result, we have 
 \begin{equation}
  \det \mathbf{M} = \det \boldsymbol{\gamma} [\det (\mathbf{I} + \mathbf{X})]^{-1}
 \end{equation}
 If the matrix $\mathbf{f}$ has rank $3$, 
 $\det (\mathbf{I} + \mathbf{X}) = 1 - \frac{1}{2} \mathrm{Tr}\mathbf{X}^2$.

 The anti-symmetric part of $\mathbf{M}^{-1}$ is 
 \begin{equation}
  \begin{aligned}
   (\mathbf{M}^{-1} - {}^t\mathbf{M}^{-1})/2 
   =& - \mathbf{h}^{-1}\mathbf{f}\mathbf{h}^{-1}
    - \mathbf{h}^{-1}\mathbf{f}\mathbf{h}^{-1}\mathbf{f}\mathbf{h}^{-1}
   \mathbf{f}\mathbf{h}^{-1}
   + \cdots \\
   =& - (\mathbf{X} + \mathbf{X}^3 + \mathbf{X}^5 +
   \cdots)\mathbf{h}^{-1}\\
   =& - \boldsymbol{\gamma}^{-1} \mathbf{f} \mathbf{h}^{-1}
   = - \mathbf{h}^{-1} \mathbf{f} \boldsymbol{\gamma}^{-1},
  \end{aligned}
 \end{equation}

 Now, we will derive the equations of motion from the DBI action.
 Variation of the Lagrangian is  
 \begin{equation}
  \begin{aligned}
   2 \delta \sqrt{-\det M} =& 
   \delta M_{ab} (M^{-1})^{ba}\sqrt{-\det M} 
   = (\delta h_{ab} + \delta f_{ab})(M^{-1})^{ba}\sqrt{-\det M}\\
   =& \left[
   (2 g_{\mu\nu}(X) \partial_a X^\mu \partial_b \delta X^\nu 
   + \partial_\nu g_{\alpha\beta}(X) \partial_a X^\alpha\partial_b X^\beta
   \delta X^\nu)(M^{-1})^{(ab)}\right. \\
   &\left. - 2 \partial_a \delta a_b (M^{-1})^{[ab]}
   \right] \sqrt{-\det M}
  \end{aligned}
 \end{equation}
 Using the formulas previously shown, the equations of motion are 
 \begin{equation}
  \begin{aligned}
   -\partial_b (\omega\sqrt{-\gamma} g_{\mu\nu}(X) \gamma^{ab} 
   \partial_a X^\mu) 
   + \frac{1}{2}\omega\sqrt{-\gamma}\gamma^{ab}\partial_\nu g_{\alpha\beta}(X) 
   \partial_a X^\alpha\partial_b X^\beta = 0, \\
    2 \partial_a (\omega\sqrt{-\gamma}\gamma^{ab}f_{bc}h^{cd}) = 0
  \end{aligned}
 \end{equation}
 As a result, we have 
 \begin{equation}
  \begin{aligned}
   \hat{D}^2 X^\mu
   + \Gamma^\mu_{\alpha\beta}\hat{D}^a X^\alpha \hat{D}_a X^\beta
   + \hat{D}^a X^\mu \hat{D}_a \ln\omega = 0,\\
    \hat{D}^a (\omega f_{ab}h^{bc}) = 0,
  \end{aligned}
 \end{equation}
 where $\hat{D}_a$ denotes the covariant derivative with respect to 
 $\gamma_{ab} \equiv h_{ab} + f_{ac}f_{bd}h^{cd}$ and 
 $\omega \equiv [\det(\mathbf{I} - \mathbf{X})]^{-1/2} = (\det h/\det \gamma)^{1/4}$.
 Thus, we can regard $\gamma_{ab}$ as an effective metric (up to
 a conformal factor).

 Another Lagrangian giving us the above equations of motion can be constructed as  
 \begin{equation}
  \begin{aligned}
   L[X,f,\gamma,h,\omega]
   =& \sqrt{-\gamma} \omega 
   \left(\gamma^{ab}g_{\mu\nu}(X)\partial_a X^\mu \partial_b X^\nu
   + \frac{1}{2}f_{ac}f_{bd}\gamma^{ab}h^{cd}
   - \frac{1}{2}\gamma^{ab}h_{ab}
   - \frac{p-1}{2} \lambda_1
   \right) \\
   &+ \sqrt{-h} \omega^{-1} \lambda_2 ,
  \end{aligned}
 \end{equation}
 where $\gamma_{ab}$, $h_{ab}$ and $\omega$ are auxiliary fields.
 Since $\lambda_1$ and $\lambda_2$ are non-zero arbitrary constants, we
 can set $\lambda_1 = \lambda_2 = 1$ for simplicity.

\section{$d\neq 0$ cases}
\label{app:general_EOM}

In this appendix, we summarize equation of motions of the D$7$-brane for
general cases with finite temperature and non-zero baryon number density
($d\neq 0$).
We consider Schwarzschild-AdS$_5\times S^5$ spacetime as the
background solution: 
\begin{equation}
 ds^2=\frac{-F(z)dV^2-2dVdz+d\vec{x}_3^2}{z^2} + d\phi^2 +
  \cos^2\phi d\Omega_3^2+ \sin^2\phi d\psi^2\ ,\quad
F(z)=1-r_h^4z^4\ .
\label{bulkmetric_general}
\end{equation}
The bulk event horizon is located at $z=1/r_h$ in this spacetime.
Then, expression of the D7-brane action is the same as Eq.~(\ref{DBI1}), 
except for components of the induced metric: 
\begin{align}
&h_{uv}=-Z^{-2}(FV_{,u}V_{,v}+V_{,u}Z_{,v}+V_{,v}Z_{,u})+\Phi_{,u}\Phi_{,v}\ ,\notag\\
&h_{uu}=-Z^{-2}V_{,u}(FV_{,u}+2Z_{,u})+\Phi_{,u}^2\ ,\quad
h_{vv}=-Z^{-2}V_{,v}(FV_{,v}+2Z_{,v})+\Phi_{,v}^2\ .
\end{align}

To eliminate $f_{uv}$ from the action~(\ref{DBI1}), we perform a
Legendre transformation as
\begin{equation}
\begin{split}
\hat{S}&\equiv S-\int dudv f_{uv}\frac{\delta S}{\delta f_{uv}}\\
&=-\mu_7g_s^{-1} V_3 \Omega_3 \int dudv
 \Big[\Big(\frac{\cos^6\Phi}{Z^6}+d^2\Big)
\{(h_{uv}+Z^2\partial_u a_x  \partial_v a_x )^2\\
&\hspace{6cm}
-(h_{uu}+Z^2\partial_u a_x^2)(h_{vv}+Z^2\partial_v a_x^2)\}\Big]^{1/2}\ ,
\end{split}
\label{newaction}
\end{equation}
where we have eliminated $f_{uv}$ using Eq.~(\ref{conserve}) at the second equality.
As well as the $d=0$ case, we can impose the same coordinate
conditions $C_1\equiv h_{uu}+Z^2\partial_u a_x^2=0$ and $C_2\equiv h_{vv}+Z^2\partial_v a_x^2=0$.
\begin{equation}
\hat{S}= \mu_7g_s^{-1} V_3 \Omega_3 \int dudv
\Big(\frac{\cos^6\Phi}{Z^6}+d^2\Big)^{1/2}
(h_{uv}+Z^2\partial_u a_x  \partial_v a_x ) .
\end{equation}
From this action, we can obtain evolution equations for $V$, $Z$,
 $\Psi$ and $a_x$, where $\Psi(u,v)\equiv \frac{\Phi(u,v)}{Z(u,v)}$. 
The evolution equations are written as 
\begin{align}
&K_1V_{,uv}=
\frac{3}{2}Z(Z\Psi)_{,u}(Z\Psi)_{,v}
+\frac{3}{2}\tan(Z\Psi)
 \{(Z\Psi)_{,u}V_{,v}+(Z\Psi)_{,v}V_{,u}\}\notag\\
&\hspace{9cm}
+\frac{1}{2}K_3V_{,u}V_{,v}
+\frac{Z^3}{2}K_2a_{x,u}a_{x,v}\ ,\\
&K_1Z_{uv}
=
-\frac{3}{2}ZF(Z\Psi)_{,u}(Z\Psi)_{,v}
+\frac{3}{2}\tan(Z\Psi)\{(Z\Psi)_{,u}Z_{,v}+(Z\Psi)_{,v}Z_{,u}\}
\notag\\
&\hspace{1cm}
-\frac{1}{2}K_3
(FV_{,u}V_{,v}+V_{,u}Z_{,v}+V_{,v}Z_{,u})
+\frac{1}{Z}(6-K_2)Z_{,u}Z_{,v}
-\frac{FZ^3}{2}K_2a_{x,u}a_{x,v}\ ,\\
&K_1\Psi_{,uv}=
\frac{3}{2}\left(\Psi F + \frac{\tan(Z\Psi)}{Z}
 \right)(Z\Psi)_{,u}(Z\Psi)_{,v}
\notag\\
&\hspace{2cm}
+ \frac{1}{2Z^2}\{K_2-3 Z \Psi \tan(Z\Psi)\}
\{(Z\Psi)_{,u}Z_{,v}+(Z\Psi)_{,v}Z_{,u}\}
\notag\\
&
\hspace{3cm}
+ \frac{\Psi}{2Z}\left(K_3
+ \frac{3\tan(Z\Psi)}{Z^2\Psi}
\right)
(FV_{,u}V_{,v}+V_{,u}Z_{,v}+V_{,v}Z_{,u})\notag\\
&\hspace{4cm}
- \frac{3\Psi}{Z^2}Z_{,u}Z_{,v}
+\frac{FZ^2\Psi}{2}\left(
K_2-\frac{3\tan(Z\Psi)}{FZ\Psi}
\right)a_{x,u}a_{x,v}
\ ,\\
&K_1a_{x,uv}
=
\frac{3}{2}\tan(Z\Psi)\{(Z\Psi)_{,u}a_{x,v}+(Z\Psi)_{,v}a_{x,u}\}
+\frac{1}{2Z}K_2(Z_{,u}a_{x,v}+Z_{,v}a_{x,u})
\ .
\end{align}
where functions $K_1$, $K_2$ and $K_3$ are defined as
\begin{equation}
\begin{split}
 &K_1=1+d^2\frac{Z^6}{\cos^6(Z\Psi)}\ ,\quad
 K_2=1-2d^2\frac{Z^6}{\cos^6(Z\Psi)}\ ,\\
 &K_3=F_{,Z}-5\frac{F}{Z}+d^2\frac{Z^6}{\cos^6(Z\Psi)}\left(F_{,Z}-2\frac{F}{Z}\right)\
 .
\end{split}
\end{equation}
In general cases, conservation of the constraint equations is
slightly modified as 
\begin{equation}
 \partial_u\Big[
\frac{1}{Z^2}\Big(\frac{\cos^6\Phi}{Z^6}+d^2\Big)^{1/2}C_2
\Big]
=
 \partial_v\Big[
\frac{1}{Z^2}\Big(\frac{\cos^6\Phi}{Z^6}+d^2\Big)^{1/2}C_1
\Big]=0\ .
\end{equation}

\section{Stark effect for scalar and vector mesons}
\label{Stark}

In this section, we analytically examine shifts of spectra of scalar and vector mesons 
caused by a weak electric field, i.e. Stark effect. 
We focus only on homogeneous modes in $(x_1,x_2,x_3)$
and $s$-modes of $S^3$. Then, the brane dynamics is described by
$W(t,z)\equiv z^{-1}\sin\Phi(t,z)$ and $a_x(t,z)$.
The first order static solution in the electric field $E$ 
is given by 
\begin{equation}
\bar{W}=m+\mathcal{O}(E^2)\ ,\qquad
\bar{a}_x =-Et\ . 
\end{equation}
We consider the fluctuation of the static solution: 
$W(t,z)=\bar{W}+w$ and $a_x(t,z)=\bar{a}_x+a$. 
Hereafter, we set $m=1$ to simplify the expression.
Then, using the DBI action, up to the first order in $E$ the quadratic
action in the fluctuations $w(t,z)$ and $a_x(t,z)$ is simply written as 
\begin{equation}
 S=\frac{1}{2}\int dt \int^1_0 dz 
\frac{1-z^2}{z}\Big[
\dot{\chi}_+\dot{\chi}_- - (1-z^2)\chi_+'\chi_-'
-2iEz^2(\dot{\chi}_+\chi_- - \chi_+\dot{\chi}_-)
\Big]\ ,
\end{equation}
where we have introduced complex fields, 
$\chi_\pm\equiv w\pm ia$ and omitted irrelevant overall factor of
the DBI action.
Decomposing the fields into Fourier modes as
$\chi_\pm(t,z)=\int^{\infty}_{-\infty} d\omega \chi_\omega^\pm(z) e^{-i\omega t}$, 
we obtain decoupled equations of motion as
\begin{equation}
 \omega^2\chi_\omega^\pm = (\mathcal{H}\pm 4E\omega z^2)\chi_\omega^\pm\
  ,\quad
\mathcal{H}\equiv-\frac{z}{1-z^2}\frac{d}{dz}\frac{(1-z^2)^2}{z}\frac{d}{dz}
\ .
\end{equation}
The eigenfunction $e_n$ and eigenvalue $\omega_n^2$ 
of the operator $\mathcal{H}$ is given by
\begin{equation}
 e_n=N_n F(-n,-n-1,-2n-2;1/z^2)\ ,\qquad
 \omega_n^2=4(n+1)(n+2)\ ,\qquad (n=0,1,2,\cdots)\ ,
\label{eigenf}
\end{equation}
where $F$ is the Gaussian hypergeometric function and $N_n$ is a normalization factor. 
Defining an inner product as 
$(f,g)\equiv \int^1_0 dz \, z^{-1}(1-z^2)f(z)g(z)$, 
we choose the normalization factor $N_n$ so that $(e_n,e_m)=\delta_{mn}$ is satisfied.
Thus, for $E=0$, meson spectra are given by
$\omega_n^{\pm}=2\sqrt{(n+1)(n+2)}$ for both of the scalar and vector mesons~\cite{Kruczenski:2003be}.
The shifts of the eigenvalues in the presence of the weak electric field 
are given by 
$\delta \omega_n^\pm=\pm 2E(e_n,z^2e_n)$. We can find
$(e_n,z^2e_n)=1/2$ for any $n$. Therefore, we obtain
\begin{equation}
 \delta \omega_n^\pm=\pm\frac{E}{m}\ ,
\end{equation}
where we restored the quark mass $m$.
Note that shifts of spectra do not depend on the mode 
number $n$. So, the beat frequency also does not depend on $n$ and is 
given by $\omega_\textrm{beat}=\delta\omega_n^+-\delta\omega_n^-=2E/m$.
This is consistent with our numerical calculation in section~\ref{subQJ}.

\section{Error analysis}
\label{errorana}

In this section, we estimate the error in our numerical calculations.
We define absolute values of constraints as
\begin{equation}
\begin{split}
&C_u\equiv |-V_{,u}(V_{,u}+2Z_{,u})+Z^2(Z\Psi)_{,u}^2+Z^4 a_{x,u}^2|\ ,\\ 
&C_v\equiv |-V_{,v}(V_{,v}+2Z_{,v})+Z^2(Z\Psi)_{,v}^2+Z^4 a_{x,v}^2|\ .
\end{split}
\end{equation}
Analytically, they have to be exactly zero everywhere once we have
imposed 
$C_u=0$ and $C_v=0$ at the initial surface and the AdS boundary. 
However, in actual numerical calculations, they become non-zero because 
numerical error does exist. 
To check constraint violation in terms of $C_u$ and $C_v$ is one of estimators of our numerical accuracy.
Introducing integer $N$ such that the mesh size is given by 
$\delta u = \delta v = \pi/(2N)$, 
we will see $N$ dependence of the constraints.
As explained in section~\ref{sec:dynamics}, 
we use two numerical methods depending on whether before or after the intermediate
surface. We will refer to the numerical method used after/before the intermediate surface as {\textit{method A/B}}. 
Numerical domains for the method A and B are 
$\{(u,v)|0\leq v \leq u\}$ and $\{(u,v)|0\leq u-v \leq \pi/2, 0\leq v\leq v_\textrm{int}\}$,
respectively. 

\begin{figure}
  \centering
  \subfigure[$C_u$ ($N=400$)]
  {\includegraphics[scale=0.6]{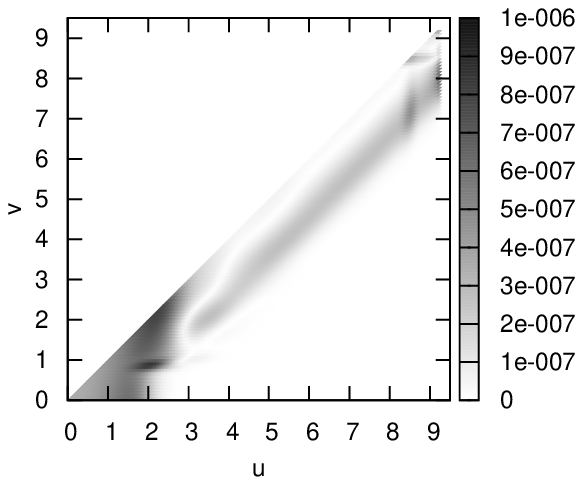}
  }
  \subfigure[$C_u$ ($N=800$)]
  {\includegraphics[scale=0.6]{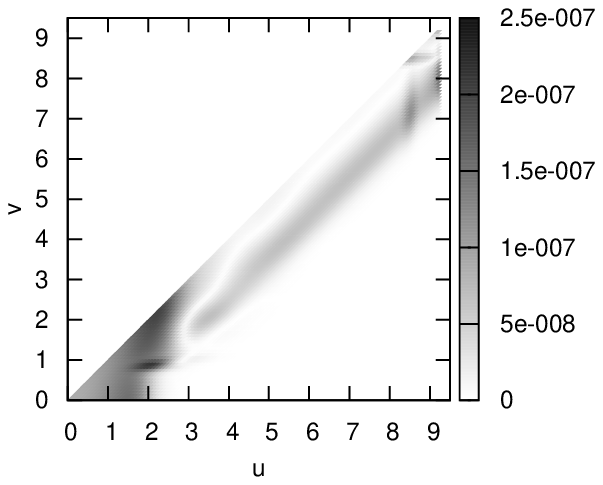} 
  }
  \subfigure[$C_u$ ($N=1600$)]
  {\includegraphics[scale=0.6]{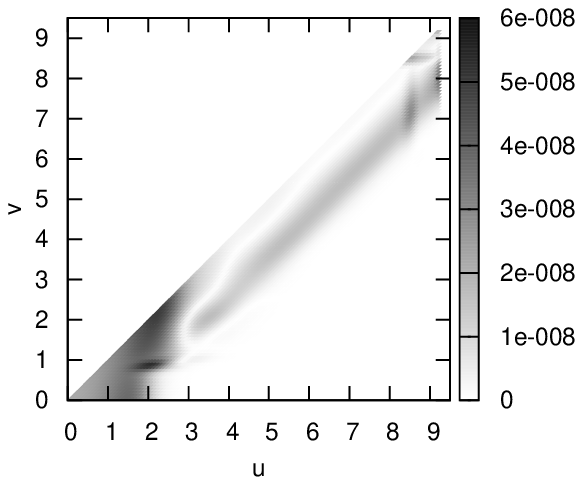} 
  }
  \subfigure[$C_v$ ($N=400$)]
  {\includegraphics[scale=0.6]{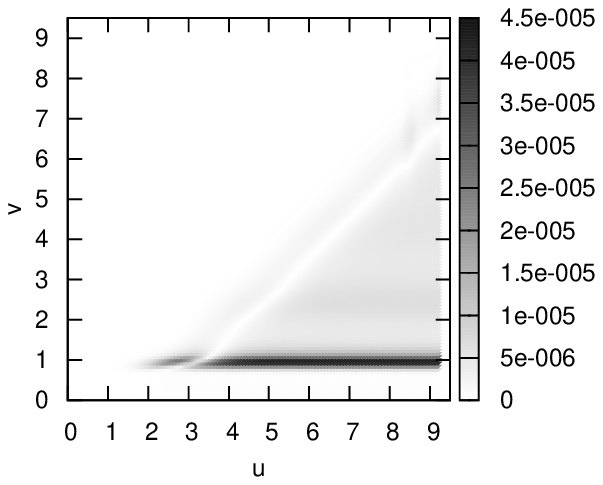}
  }
  \subfigure[$C_v$ ($N=800$)]
  {\includegraphics[scale=0.6]{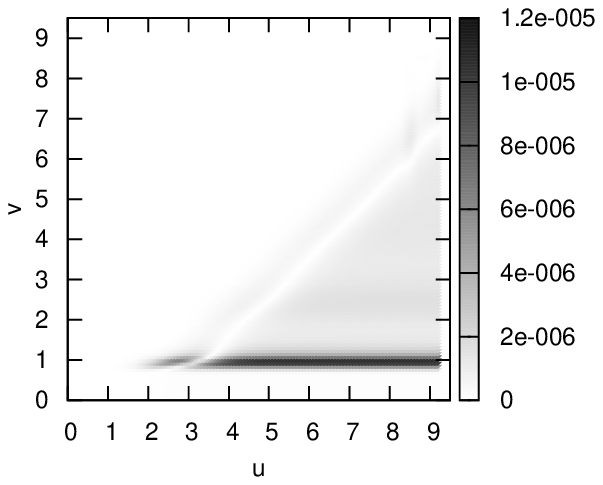} 
  }
  \subfigure[$C_v$ ($N=1600$)]
  {\includegraphics[scale=0.6]{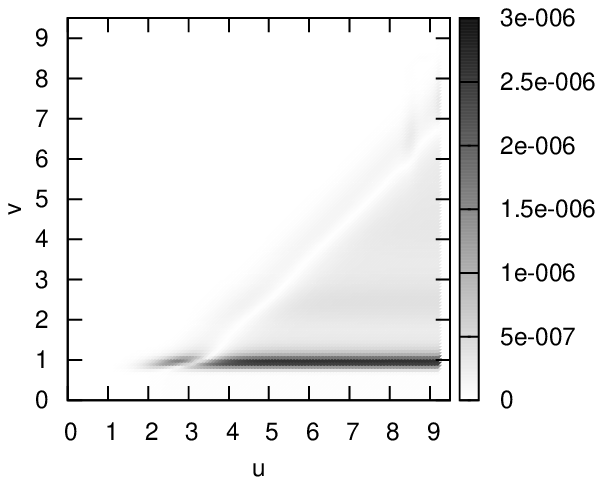} 
  }
  \caption{
Constraint violation for 
$E_f/m^2=1$ and $m \Delta V=0.5$.
}
\label{constraint_super_crit}
\end{figure}

As a typical example of the supercritical electric field, 
we choose the parameter as $E_f/m^2=1$ and $m \Delta V=0.5$.
In this case, the final state of the time evolution is a static black hole
embedding and the effective horizon exists at
the initial surface. 
Thus, we regard the initial surface as the
intermediate surface and use only the method A. 
Figures~\ref{constraint_super_crit}(a), (b), (c) show $C_u$ for $N=400$,
$800$, $1600$. 
We can see that they remain quite small 
(even for $N=400$, we have $C_u<10^{-6}$) and 
decrease as $N$ increases. (See maximum values of color bars.)
Figures~\ref{constraint_super_crit}(d), (e), (f) show $C_v$ for $N=400$,
$800$, $1600$. They share a similar property as $C_u$.

\begin{figure}
  \centering
  \subfigure[$C_u$ ($N=400$)]
  {\includegraphics[scale=0.5]{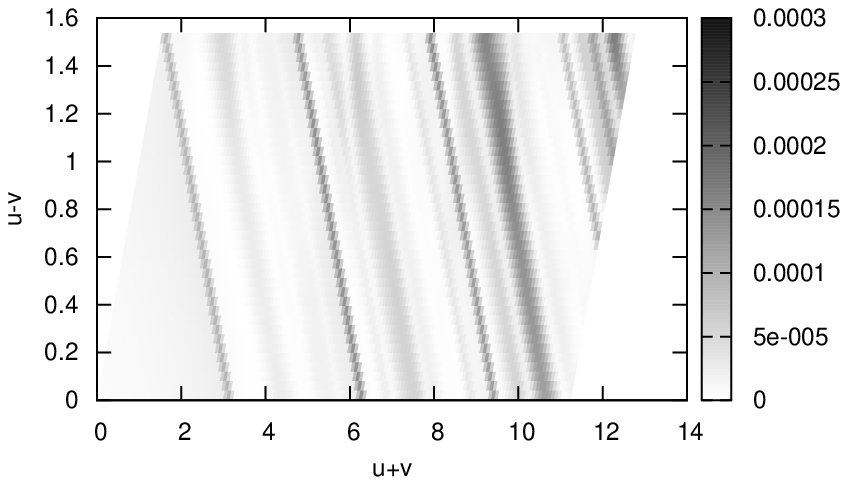}
  }
  \subfigure[$C_u$ ($N=800$)]
  {\includegraphics[scale=0.5]{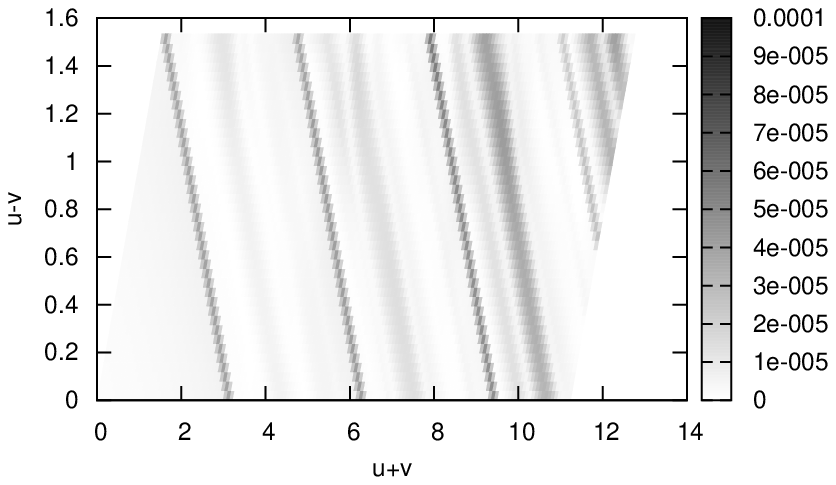} 
  }
  \subfigure[$C_u$ ($N=1600$)]
  {\includegraphics[scale=0.5]{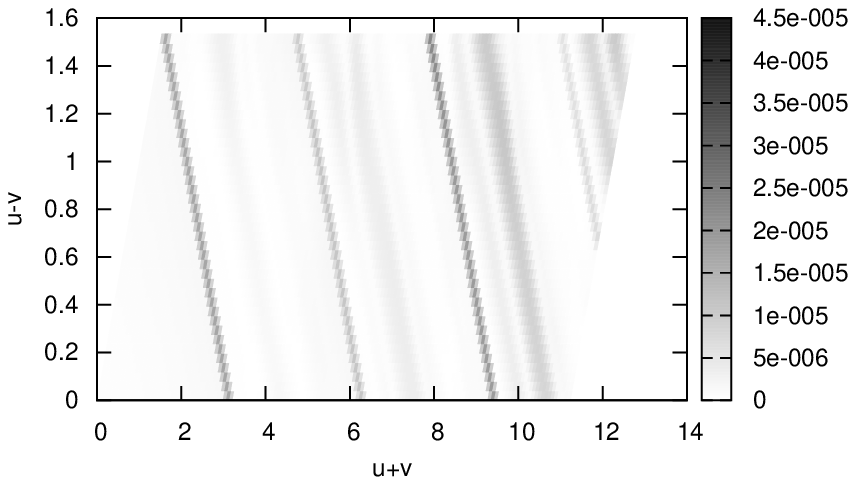} 
  }
  \subfigure[$C_v$ ($N=400$)]
  {\includegraphics[scale=0.5]{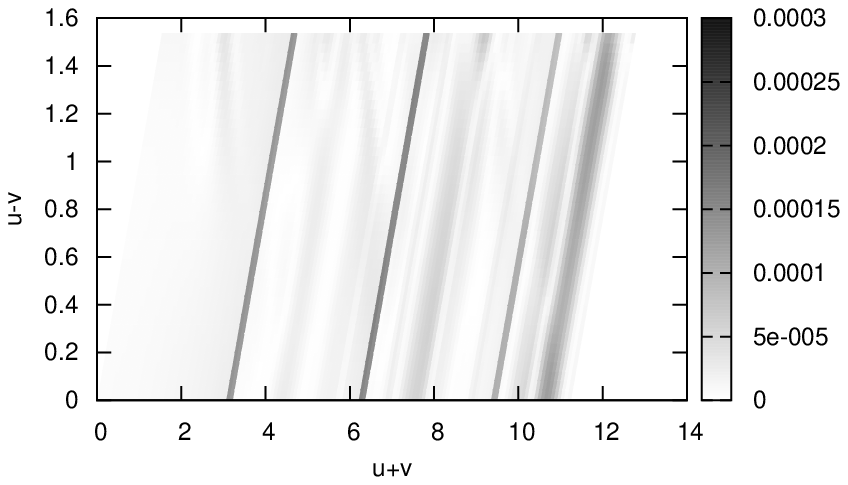}
  }
  \subfigure[$C_v$ ($N=800$)]
  {\includegraphics[scale=0.5]{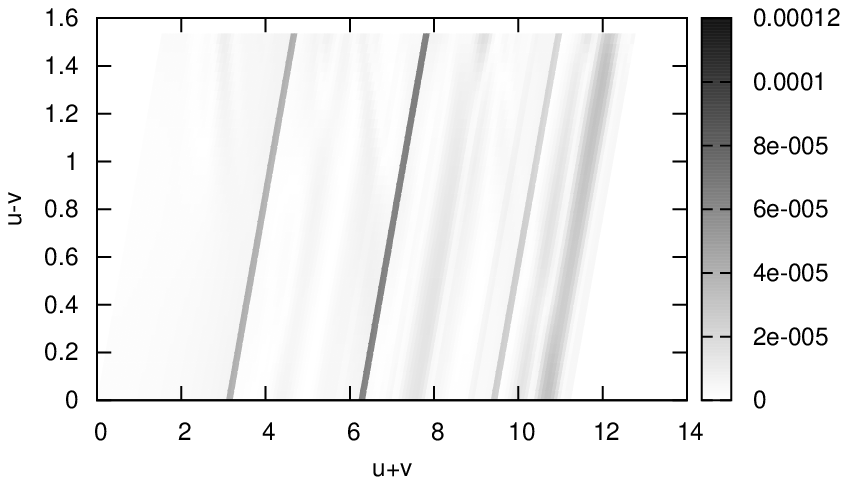} 
  }
  \subfigure[$C_v$ ($N=1600$)]
  {\includegraphics[scale=0.5]{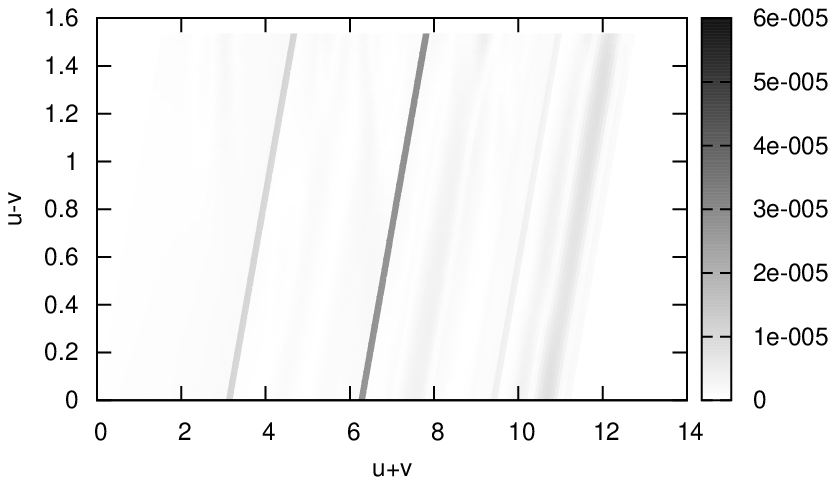} 
  }
  \caption{
Constraint violation for 
$E_f/m^2=0.19$ and $m \Delta V=2$ before the intermediate surface.
}
\label{constraint_sub_crit}
\end{figure}

\begin{figure}
  \centering
  \subfigure[$C_u$ ($N=400$)]
  {\includegraphics[scale=0.5]{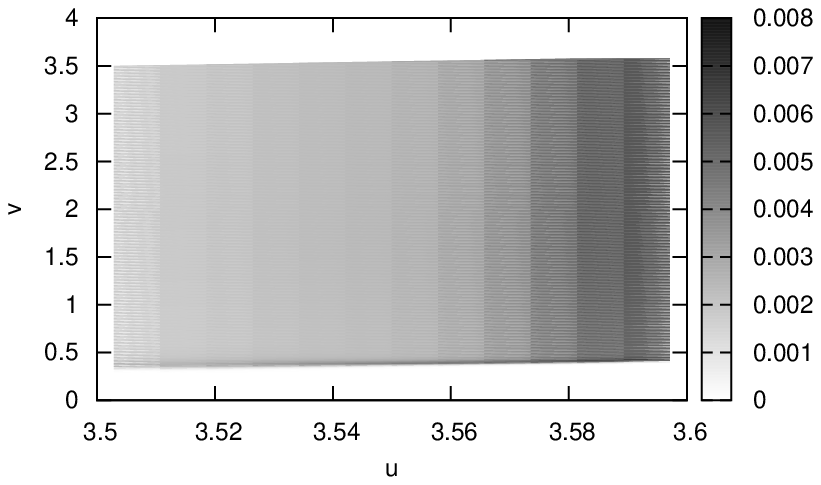}
  }
  \subfigure[$C_u$ ($N=800$)]
  {\includegraphics[scale=0.5]{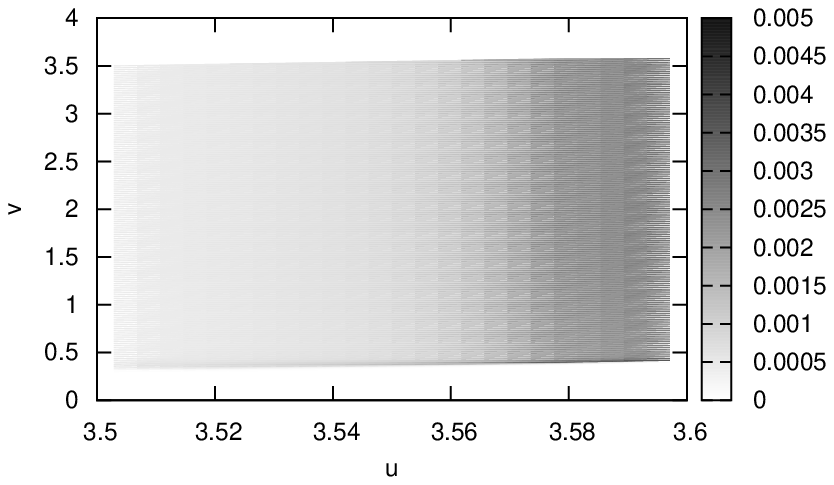} 
  }
  \subfigure[$C_u$ ($N=1600$)]
  {\includegraphics[scale=0.5]{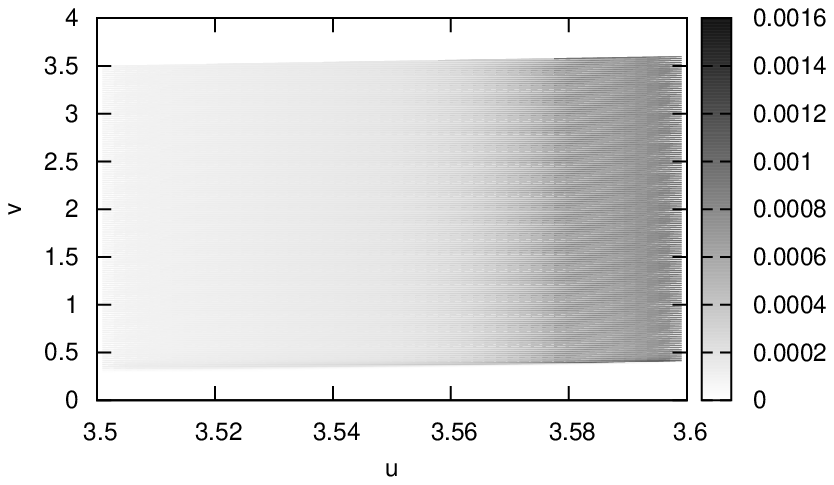} 
  }
  \subfigure[$C_v$ ($N=400$)]
  {\includegraphics[scale=0.5]{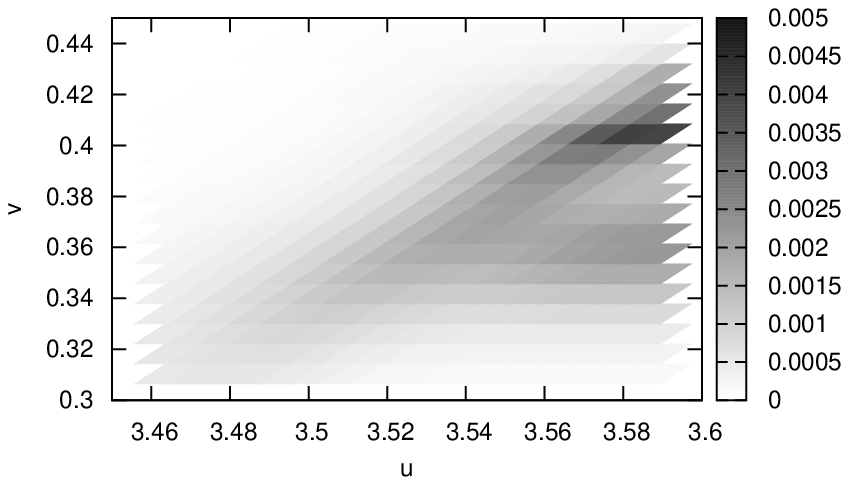}
  }
  \subfigure[$C_v$ ($N=800$)]
  {\includegraphics[scale=0.5]{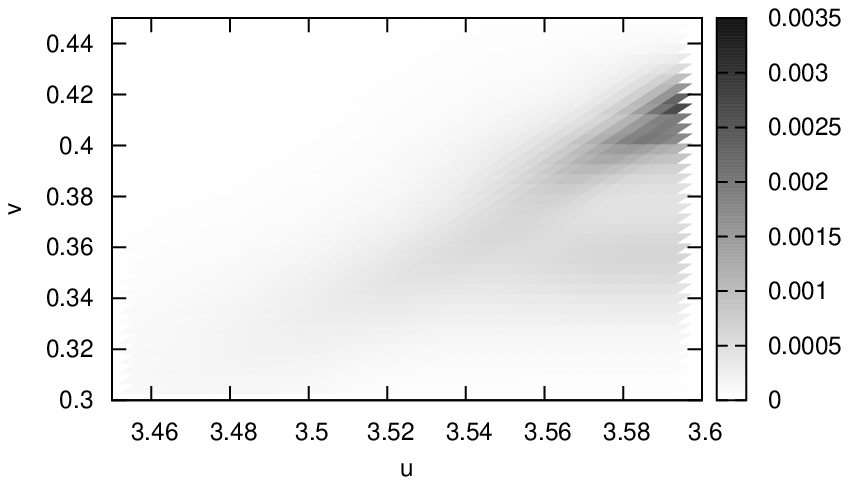} 
  }
  \subfigure[$C_v$ ($N=1600$)]
  {\includegraphics[scale=0.5]{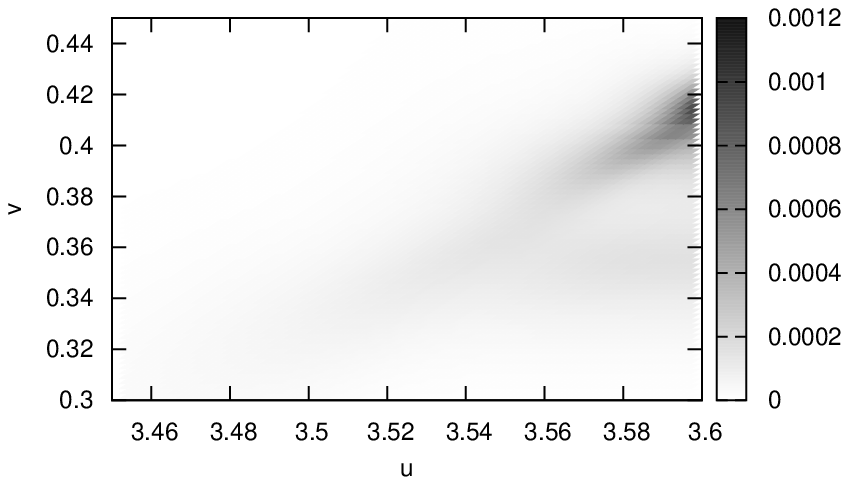} 
  }
  \caption{
Constraint violation for 
$E_f/m^2=0.19$ and $m \Delta V=2$ after the intermediate surface.
}
\label{constraint_sub_crit_B}
\end{figure}

As a typical example of the subcritical electric field, 
we choose the parameters as $E_f/m^2=0.19$ and $m \Delta V=2$.
In this case, the intermediate surface is located at
$v=v_\textrm{int}\simeq 5.5$, namely, numerical computation by the
method B breaks down at $v=v_\textrm{int}$.
Figures~\ref{constraint_sub_crit}(a)-(f) show $C_u$ and $C_v$ for $N=400$,
$800$, $1600$ before the intermediate surface.
Although a sharp noise is generated at the pole on the initial surface 
and propagates between the AdS boundary and the pole,
the constraint violation remains still small ($C_u,C_v<3\times 10^{-3}$ even for $N=400$)
and decreases as $N$ increases.
Figures~\ref{constraint_sub_crit_B}(a)-(f) show $C_u$ and $C_v$ for $N=400$,
$800$, $1600$ after the intermediate surface.
Our numerical calculation by the method A broke down at $u\simeq 3.6$.
In the figures, 
we have focused on $3.5\leq u \leq 3.6$ for $C_u$ and 
$3.45\leq u \leq 3.6$ and $0.3\leq v \leq 0.45$ for $C_v$.
The constraint violation localizes there because 
a singularity is close to the regions. 
We can find that 
the constraint violation remains still small ($C_u,C_v<8\times 10^{-3}$ even for $N=400$)
and decreases as $N$ increases.

\begin{figure}
\begin{center}
\includegraphics[scale=0.6]{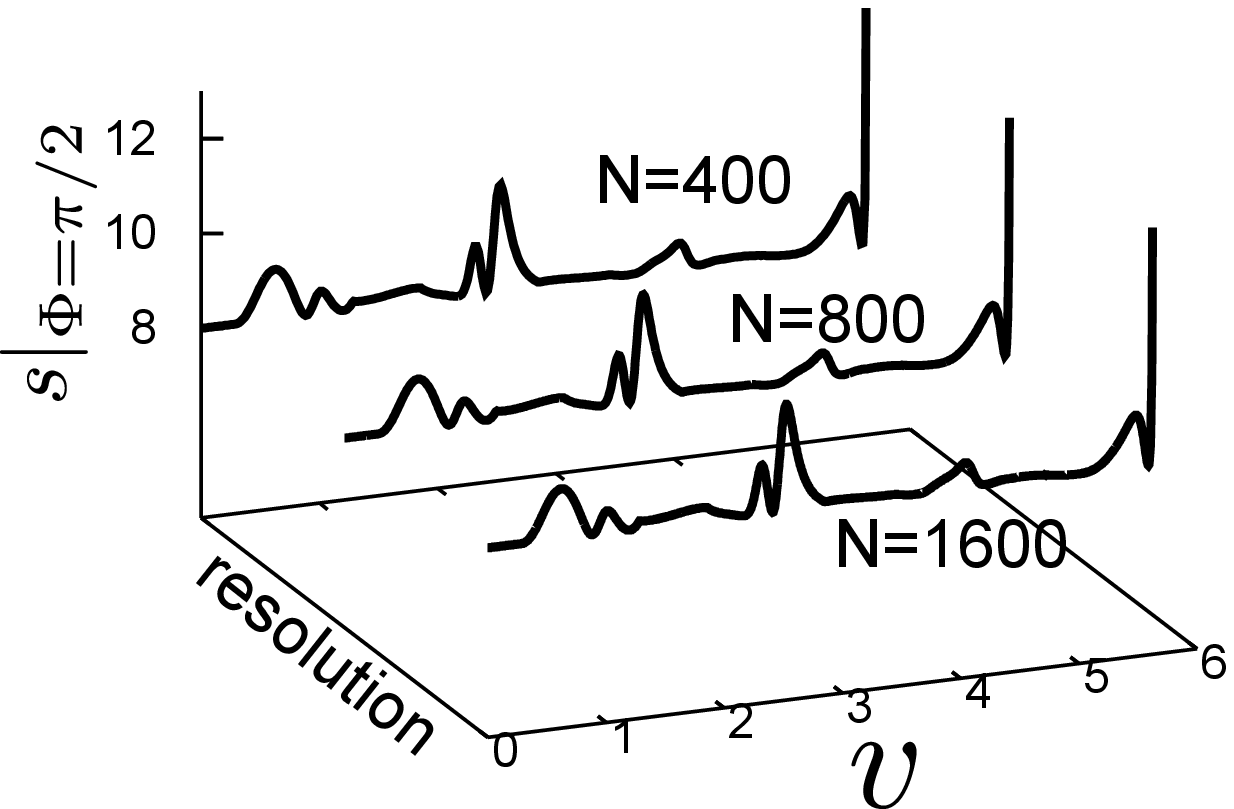}
\end{center}
\caption{
Resolution dependence of the scalar quantity
$s|_{\Phi=\pi/2}$ for $E_f/m^2=0.19$ and $m \Delta V=2$.
}
 \label{fig:trEMconv_s}
\end{figure}

In section~\ref{subsec:Decbelow}, we have inferred that 
a naked singularity appears on the brane for subcritical electric field case
since the scalar quantity
$s|_{\Phi=\pi/2}\equiv \gamma^{ab}h_{ab}|_{\Phi=\pi/2}$ seems to diverge 
within a finite time.
We also found a turbulent-like behavior in brane fluctuations near the singularity.
One may think that 
it is dangerous to treat a singularity by the numerical method and 
our results may be just numerical artifacts. 
Of course, we cannot ``prove'' the existence of the singularity from the
numerical calculation.
We can only show that our results do not depend on the resolution.
Figure~\ref{fig:trEMconv_s} shows the scalar quantity
$s|_{\Phi=\pi/2}$ against the worldvolume
coordinate $v$ and the resolution $N$ for 
$E_f/m^2=0.19$ and $m \Delta V=2$. 
This figure demonstrates that the divergence of the scalar quantity
does not depend on the resolution.

\end{document}